\documentclass[12pt]{article}
\usepackage{setspace}
\usepackage[utf8]{inputenc}
\usepackage{graphicx}
\usepackage{multirow}
\usepackage{multicol}
\usepackage[margin=1in]{geometry}
\usepackage{rotating}
\usepackage{natbib}
\usepackage{setspace}
\usepackage{color}
\usepackage{algpseudocode}
\usepackage{algorithmicx}
\usepackage{enumerate}
\usepackage{subcaption}
\usepackage{adjustbox}
\usepackage{comment}

\usepackage{booktabs}
\usepackage{enumerate}
\usepackage{verbatim}
\usepackage{amssymb}
\usepackage{amsmath}
\usepackage{float} 
\usepackage{bm}
\def\eqx"#1"{{\label{#1}}}
\def\eqn"#1"{{\ref{#1}}}
\usepackage{blindtext}

\newcommand{\Bernoulli}{\operatorname{Bernoulli}}
\newcommand{\Beta}{\operatorname{Beta}}
\newcommand{\Data}{\mathcal D}
\newcommand{\E}{\mathbb E}
\newcommand{\Gam}{\operatorname{Gam}}
\newcommand{\Leaves}{\mathcal L}
\newcommand{\sM}{\mathcal M}
\newcommand{\sQ}{\mathcal Q}

\newcommand{\Tree}{\mathcal T}
\newcommand{\Var}{\operatorname{Var}}

\newcommand{\bt}[1]{\begin{center}\begin{tabular}{#1}\hline}
\newcommand{\et}{\\\hline\end{tabular}\end{center}}

\newcommand{\bx}{\mathbf{x}}

\newcommand{\bfW}{{\bf W}}
\newcommand{\bfX}{{\bf X}}

\newcommand{\bfY}{{\bf Y}}

\newcommand{\bfx}{{\bf x}}
\newcommand{\bfw}{{\bf w}}

\newcommand{\Uniform}{\operatorname{Uniform}}

\usepackage[linesnumbered,ruled,vlined]{algorithm2e}
\SetKwInput{KwInput}{Input}                % Set the Input
\SetKwInput{KwOutput}{Output}              % set the Output

\usepackage[dvipsnames]{xcolor}
\usepackage{soul}

\author{Piyali Basak$^{1}$\thanks{\href{mailto:piyali.basak@merck.com}{piyali.basak@merck.com}}, 
Antonio R. Linero$^{2}$\thanks{\href{mailto:antonio.linero@austin.utexas.edu}{antonio.linero@austin.utexas.edu}},  
Camille Maringe$^{3}$\thanks{\href{mailto:camille.maringe@lshtm.ac.uk}{camille.maringe@lshtm.ac.uk}}, \\ and
F. Javier Rubio$^{4}$\thanks{\href{mailto:f.j.rubio@ucl.ac.uk}{f.j.rubio@ucl.ac.uk}}}

\usepackage[citecolor=blue, colorlinks]{hyperref}

\begin{document}
\title{Relative Survival Analysis Using \\Bayesian Decision Tree Ensembles}
\date{}
\maketitle

\begin{abstract}
  In cancer epidemiology, the \emph{relative survival framework} is used to
  quantify the hazard associated with cancer by comparing the all-cause
  mortality hazard in cancer patients to that of the general population. This
  framework assumes that an individual's hazard function is the sum of a known
  population hazard and an excess hazard associated with the cancer. Several
  estimands are derived from the excess hazard, including the \emph{net
    survival}, which are used to inform decisions and to assess the
  effectiveness of interventions on cancer management. In this paper, we
  introduce a Bayesian machine learning approach to estimating the excess hazard
  and identifying vulnerable subgroups, with a higher excess risk, using
  Bayesian additive regression trees (BART). We first develop a proportional
  hazards extension of the BART model to the relative survival setting, and then
  extend this model to non-proportional hazards. We develop tools for model
  interpretation and posterior summarization and then present an application
  using colon cancer data from England, highlighting the insights our proposed
  methodology offers when paired with state-of-the-art data linkage methods.
  This application demonstrates how these methods can be used to identify
  drivers of inequalities in cancer survival through variable importance
  quantification.

  \vspace{1em}
\end{abstract}

\doublespacing

\section{Introduction}\label{sec:intro}

Reducing the burden of cancer is a top priority for national and international health institutions, such as the World Health Organization and the International Agency for Research on Cancer. To monitor cancer patients at the population level, several complementary indicators are used to assess cancer management, including incidence and prevalence rates, mortality rates, and survival probabilities. The survival probability represents the probability of surviving beyond a time point $t>0$, typically measured from the diagnosis of cancer. The overall survival framework is a common approach for quantifying survival, which aims at estimating survival probabilities associated with any cause of death. In cancer epidemiology, however, the overall survival framework is usually avoided as it does not quantify the survival associated only with the cancer of interest and prevents meaningful comparisons \citep{mariotto:2014,govuk:2024}. In contrast, the ``relative survival'' framework \citep{perme:2012} aims at estimating the survival associated only with the cancer under study, without requiring knowledge of the cause of death (as this may be either unavailable or unreliable at the population level). Consequently, the relative survival framework is the preferred approach for monitoring the effectiveness of the health system in preventing patients dying from cancer, and for comparing cancer survival across countries, regions within a country, or over different time periods \citep{mariotto:2014,govuk:2024,quaresma:2024}. 

The key distinction between relative survival and overall survival methods lies in comparing the observed mortality hazard with an expected mortality hazard in the disease-free population. This approach offers a metric to quantify the mortality associated with a particular disease (such as cancer), removing the need for information on the actual cause of death. Relative survival analysis is a valuable and a widely used tool in population-based cancer studies, providing important insights into cancer prognosis and cancer management, while enabling efficient comparison between different population subgroups of interest. The key assumption in the relative survival framework is that the overall hazard of an individual $\lambda(\cdot)$ decomposes additively into two components: (a) the expected hazard $\lambda_P(\cdot)$ for the general population, and (b) the excess hazard $\lambda_E(\cdot)$ associated with the cancer of interest. That is, 
\begin{equation}
  \lambda(t \mid \bfx) = \lambda_P(\text{age} + t \mid \bfw) + \lambda_E(t \mid \bfx),
  \label{eq:exh_additive_decomp}
\end{equation}
where $\text{age}$ denotes the age at diagnosis. The population hazard $\lambda_P(t \mid \bfw)$ is assumed to be available from population-level life tables for each calendar year and region, based on a subset of the available characteristics $\bfw \in {\mathbb R}^Q$ with $\bfw \subseteq \bfx$; typically, $\bfw$ includes information about sex, age, and measures of societal deprivation. The excess hazard $\lambda_E(t \mid \bfx)$, on the other hand, is estimated using the available data and the patient characteristics $\bfx\in {\mathbb R}^P$, which include those variables in the life table ($\bfw$) along with other socio-demographic and clinical characteristics such as tumor stage. 
 
\subsection{Challenges in Relative Survival Modeling}

One of the main quantities of interest in the relative survival framework is the \emph{net survival}, which is the survival function associated with the excess hazard. For an individual with characteristics $\bfx$, the net survival is defined as $S_E(t \mid \bfx) = \exp\left\{-\int_0^t \lambda_E( r \mid \bfx) \  dr\right\}$. Policy-making and epidemiological reports \citep{mariotto:2014,quaresma:2024} are typically based on the average net survival for the population of interest with characteristics $\bfX_i \sim F_X$:
\begin{equation*}
 S_E(t \mid F_X) = \int S_E(t \mid \bfx) \, F_X(d\bfx).
\end{equation*}
Estimating the net survival presents a variety of challenges for both parametric and nonparametric approaches. Parametric models require restrictive assumptions on the form of the excess hazard function $\lambda_E(t \mid \bfx)$, both in terms of the parametric form of the hazard (\textit{e.g.}, Weibull, log-normal, and so forth) and in terms of how the covariates enter the model and interact with time (\textit{e.g.}, accelerated failure time or proportional hazards models). These assumptions may not always hold in practice, leading to biased estimates or inaccurate predictions if they are not properly validated. In the parametric framework, several methods have been proposed for modeling the excess hazard, such as using general parametric hazard structures \citep{rubio:2019}, or modeling the baseline excess hazard (or cumulative excess hazard) and non-linear effects using splines \citep{giorgi:2003,dickman:2004,charvat:2021,fauvernier:2019,quaresma:2020,eletti:2022}. Other approaches include the four methods described by \cite{dickman:2004}, who assume that the excess hazard is constant within pre-specified intervals to establish a link with Generalized Linear Models (GLM), thus facilitating the estimation of the regression model, as well as approaches utilizing multivariate fractional polynomials (MFP) under an additive hazard structure \citep{lambert:2005}. Each of these modelling frameworks require a careful specification of the role of each covariate as well as the specification of the spline basis.

In practice, structural assumptions like proportional hazards are often unrealistic, and opting for a flexible modeling approach that allows for potential non-linearity and multi-way interactions between the covariates and survival time may be crucial to fully understand the dynamics of the disease. In these cases, fully-nonparametric estimators of the net survival may be preferred. Nonparametric estimators of net survival have a rich and interesting history, with several methods proposed that, in fact, failed to estimate the net survival. These include the so-called \textit{Ederer I} and \textit{Ederer II} estimators \citep{ederer:1961}. We refer the reader to \cite{perme:2012} for a through review of nonparametric estimators of net survival and their corresponding limitations. Nonparametric estimators of net survival are based on estimating the cumulative excess hazard \citep{perme:2012}, in a similar fashion as the Nelson-Aalen estimator of the overall cumulative hazard \citep{aalen:1978}, using counting processes. While often viewed as a gold-standard for estimating the net survival function, Nelson-Aalen type estimators are limited in that (i) estimates of the cumulative excess hazard may be negative (thus leading to non-monotonic net survival functions) and (ii) incorporating covariate information requires stratification by the covariates, which can present challenges when handling continuous variables and may result in a loss of power due to sparse strata. Hence there is a need for methods that combine the flexibility of Nelson-Aalen type estimators with the ability to borrow information across sub-populations offered by parametric methods.

\subsection{Background and Research Questions}

The quality and availability of population-based data have increased in recent years, thus allowing richer information on cancer patients through linkage of various data sources collected at primary and secondary care settings. These efforts have motivated the development of algorithms aimed at deriving key information, not readily available or collected, such as route to diagnosis \citep{elliss:2012}, stage at diagnosis \citep{benitez:2016}, frailty scores \citep{gilbert:2018}, and the presence of comorbidities \citep{maringe:2017}. These variables offer new insights into prognostic factors for cancer survival, at population level. Stage at diagnosis is a key prognostic factor, determining treatment options \citep{nice2020colorectal}, surveillance and follow-up. Efforts in cancer awareness, prevention, and screening aim to detect signs and symptoms early and diagnose cancer at the earliest possible stage, ensuring the best possible outcomes. Nonetheless, despite health systems built on principles of equity of access and use, factors such as patient's age, level of deprivation, education, area of residence as well as clinical characteristics such as frailty or the presence of comorbidities impact cancer outcomes, isolating groups with heightened vulnerability. So far, understanding the impact of comorbidities and other socio-demographic or clinical factors has only been done on the overall survival framework \citep{rubio:2022}, which is a sub-optimal measure for cancer management, or through descriptive studies \citep{michalopoulou:2021}. The proposed methodology provides the tools to highlight the importance of each characteristic on net survival, beyond the impact of the stage at diagnosis. The specific outputs following the use of Bayesian Additive Regressive Trees (BART) for the estimation of net survival nicely complement the accurate estimation of net survival, and provide time-varying effects and variable importance: their interpretation and value are detailed in section \ref{sec:applications}. 

\subsection{Our Contributions}

In this article, we propose a flexible semi-parametric model for estimating excess hazard and the associated net survival function using a log-linear extension of the Bayesian Additive Regression Trees framework \citep{chipman2010bart, hill2019bayesian}. Our approach models the excess hazard of the survival times as a product $\lambda_E(t \mid \bfx) = \lambda_{0E}(t) \, e^{r(\bfx, t)}$ where the baseline hazard $\lambda_{0E}(t)$ is a piece-wise constant function and the non-parametric component $r(\bfx, t)$ is modeled nonparametrically using BART.  We first propose a piecewise constant proportional hazards model $(r(\bfx, t) \equiv r(\bfx))$ and then subsequently extend it to allow the covariates to interact with the time in a completely non-parametric fashion. Using the automatic relevance determination prior of \citet{linero2016bayesian}, we also show how to \emph{shrink} the fully nonparametric model towards the proportional hazards model, allowing us to choose an appropriate structure in a data-adaptive fashion.

Our approach has several desirable properties relative to other approaches to estimating the net survival function:
\begin{enumerate}[i.]
    \item By using BART, we are able to avoid the need to directly specify variable transformations or interactions between covariates. Additionally, our implementation of the proportional hazards model is just as computationally efficient as the original BART algorithm of \citet{chipman2010bart}.
    \item Our nonparametric extension allows for sharing of information across sub-populations of interest, allowing us to maintain the flexibility of Nelson-Aalen type estimators while accounting for continuous characteristics and sparsely-observed sub-populations of interest.
    \item By using Bayesian methods, we are able to combine the predictive power of machine learning with the need to appropriately quantify uncertainty in our inferences.
    \item We also propose several \emph{posterior summarization} techniques to understand the role of covariate interactions and to understand how the effects of covariates change over time. Using these tools, for example, we can quantify and identify variations in the prognostic value of age at diagnosis, route to diagnosis, deprivation and specific comorbidities throughout follow-up, for patients diagnosed with colon cancer in England.
\end{enumerate}
    
The rest of this paper is organized as follows: Section \ref{sec:BART} of this paper provides a brief review of Bayesian Additive Regression Trees and its applications in survival settings. Section \ref{sec:relsurv_BART} provides the details of our proposed semi-parametric piecewise exponential model for excess hazards and net survival estimation in proportional hazard setup, and extends this to the non-proportional scenario. In Section \ref{sec:simulation} we conduct a thorough simulation study to demonstrate the effectiveness of the proposed model and compare the results with those achieved by established methods for estimating net survival. Section \ref{sec:applications} discusses a genuine epidemiological question, illustrating the use of the proposed methodology for calculating standard quantities of interest, such as net survival. It highlights the additional insights the proposed method provides when applied to linked datasets. We close in Section \ref{sec:discussion} with a discussion.

\section{A Brief Review of BART}\label{sec:BART}

Let $T_i$ denote a survival time and $\bfX_i = (X_{i1}, \ldots, X_{iP})^{\top} \in \mathbb R^P$ denote a vector of $P$ covariates for subject $i = 1,\ldots,N$. Our interest is in modeling the survival function $S(t \mid \bfx)$ of $[T_i \mid \bfX_i = \bfx]$ and the associated hazard function $\lambda(t \mid \bfx) = - \frac{\partial}{\partial t} \log S(t \mid \bfx)$. We assume right-censoring of the $T_i$'s, with $C_i$ denoting a non-informative censoring time for subject $i$. The observed data consists of the event time $Y_i = \min(T_i, C_i)$ and the censoring indicator $\delta_i = 1(T_i \le C_i)$.

The Bayesian additive regression trees (BART) framework, proposed by \citet{chipman2010bart}, is a popular Bayesian ensemble method that combines ``weak'' decision trees into a single ``strong'' learner with high predictive accuracy. BART has seen widespread adoption in various research communities, including the causal inference community \citep{dorie2019automated}, due to its ability to perform well under high-noise scenarios and offer uncertainty quantification.

\citet{chipman2010bart} introduced BART in the context of the semiparametric regression model $Y_i = r(\bfX_i) + \epsilon_i$ where $\bfX_i$ is a $P$-dimensional covariate vector, $r: \mathbb{R}^P \to \mathbb{R}$ is an unknown regression function, and $\epsilon_i \sim N(0, \sigma^2)$ is a random error. The BART framework models the unknown function $r(\bfx)$ as a sum of $M$ regression trees $r(\bfx) = \sum_{t = 1}^M g(\bfx; \mathcal \Tree_t, \sM_t)$, where $\Tree_t$ denotes the structure and splitting rules of tree $t$ and $\sM_t = (\mu_{t1}, \ldots, \mu_{tL_t})$ denotes the set of predictions associated with the $L_t$ terminal nodes of the $t^{\text{th}}$ decision tree.

\begin{figure}
     \centering
     \includegraphics[width=.8\textwidth]{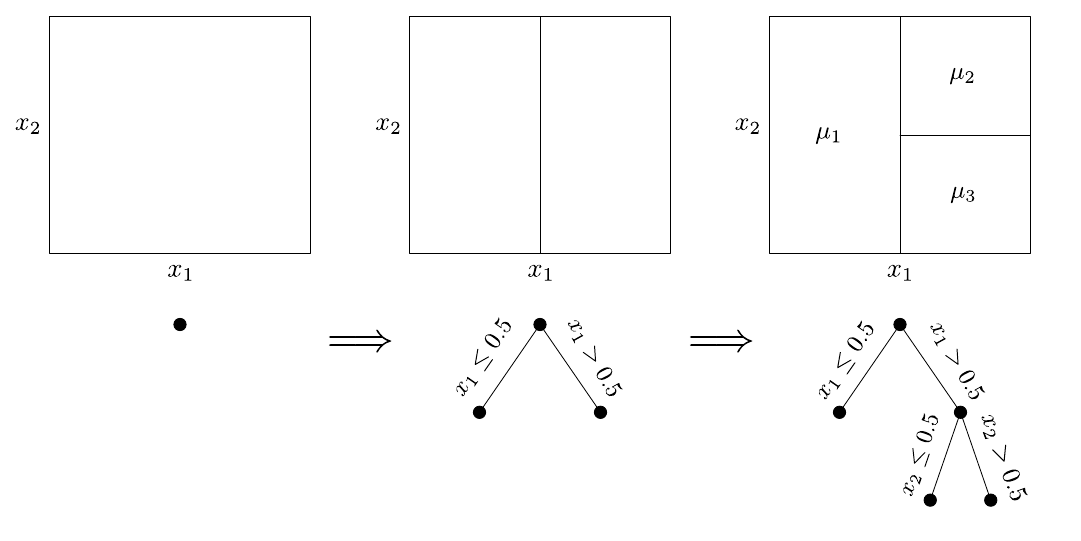}
     \caption{Schematic showing how to sample $(\Tree_t, \mathcal M_t)$. We first determine whether the root node will have a branch, with probability $\gamma$; then sample the splitting coordinate $j = 1$ and the cut-point $C = 0.5$. This process then iterates; the left child node is set to be a leaf node with probability $1 - \gamma/2^\beta$, and the right child is made a leaf with probability $\gamma/2^\beta$. Eventually this process terminates, and we sample a mean parameter $\mu$ for each leaf node.}
     \label{fig:TreeGrow}
\end{figure}

The BART framework uses a \emph{regularization prior} to perform Bayesian inference on $r(\bfx)$; this requires specifying a prior on the tree structures $\Tree_t$ and the leaf node predictions $\sM_t$. Following \citet{chipman2010bart}, $\Tree_t$ is assigned a branching process prior with each node at depth $d$ being non-terminal with probability $\gamma (1 + d)^{-\beta}$, where $\gamma > 0$ and $\beta > 0$ control the shape of the tree. For each branch node $b$, a splitting rule of the form $[x_j \le C_b]$ is assigned with $\bfx$ going left down the tree if the condition is satisfied and right down the tree otherwise. Conditional on $j$, the \emph{cut-point} $C_b$ is assigned a $\Uniform(L_j, U_j)$ prior where $\prod_{k = 1}^P [L_k, U_k]$ is the hyper-rectangle of $\bfx$'s that lead to branch $b$. The splitting variable $j$ is chosen with probability $s_j$, and we refer to the vector $s = (s_1, \ldots, s_P)^{\top}$ as the \emph{splitting proportions}. \citet{chipman2010bart} set $s_j = 1 / P$, however following \citet{linero2016bayesian} we will consider a prior on $s$ that we describe later.     Independent Gaussian priors are designated to the terminal node parameters, with $\mu_{tl} \stackrel{iid}{\sim} N(0,\sigma^2_\mu).$ A schematic showing how the branching process prior generates a sample of a decision tree, and its associated 
partition, is given in Figure~\ref{fig:TreeGrow}.

\paragraph{Inference for Non-Normal Likelihoods}
Beyond semiparametric regression with normal errors, BART has been adapted widely to accommodate non-Gaussian likelihoods. \citet{chipman2010bart} showed how to adapt the BART model to binary classification problems with the probit link using the data augmentation scheme of \citet{albert1993bayesian}. This can be used to form the basis of many other methods, including the survival model of \citet{sparapani2016nonparametric}, which bears some similarity to the models we use here. Beyond this, \citet{murray2020log} showed that the Bayesian backfitting algorithm can be easily generalized to Poisson outcomes without the need for data augmentation, provided that one uses the leaf node prior $\mu_{t\ell} \sim \log \Gam(a, b)$ instead of the normal prior. The Poisson log-linear model spurred further advances, with \citet{linero2018shared} developing a Bayesian backfitting algorithm for gamma-distributed responses, and \citet{linero2021bayesian} showing that Cox's proportional hazard model is also amenable to Bayesian backfitting. The present work also proceeds in this style, using a log-gamma prior for the leaf node parameters.

\paragraph{Inference Via Bayesian Backfitting}
Inference for BART models proceeds via a \emph{Bayesian backfitting} algorithm that alternates between updating $(\Tree_t, \sM_t)$ (conditional on the remaining trees) for $t = 1,\ldots, M$, followed by sampling hyperparameters from their full conditional distributions. As argued by \citet{linero2024generalized}, the main condition required in order for a Bayesian backfitting algorithm to be constructible is that the \emph{integrated likelihood}
\begin{align*}
    L(\bfY, \bm{\eta}) = \int \prod_i f(Y_i \mid \eta_i + \mu) \, \pi_\mu(\mu) \ d\mu,
\end{align*}
have a closed form expression for all $\bfY$ and $\bm{\eta}$, where $f\left(y \mid r(\bfx)\right)$ is the assumed density of $[Y_i \mid \bfX_i = \bfx]$. If this holds, then $L(\mathbf Y, \bm\eta)$ can be used to update $\Tree_t$ via Metropolis-Hastings and then sample $\sM_t$ from its full conditional given $\Tree_t$, with $\eta_i = \sum_{j \ne t} g(\bfX_i; \Tree_j, \sM_j)$. Full details of this approach are given in \citet{linero2024generalized}.

\paragraph{Default Priors}
BART priors have the advantage of being relatively easy to use due to the availability of good default priors that work across a variety of situations. As defaults in this work we set $\gamma = 0.95$, $\beta = 2$, $M \in \{50, 100\}$, and $\sigma_\mu = 1.5 / \sqrt M$, and do not engage in tuning the prior beyond this. It is sometimes possible to get better performance by choosing $M$ via cross-validation, at the expense of more computation time, but the benefits of cross-validation are not consistent \citep{chipman2010bart, linero2017abayesian}.

\section{Relative Survival BART}\label{sec:relsurv_BART}

{The additive relative survival model expresses the overall hazard of an individual as the sum of a \emph{population} hazard and an \emph{excess} hazard, as described in \eqref{eq:exh_additive_decomp}.}
In general, the likelihood of the excess hazard model under non-informative censoring is given by \citep{rubio:2019}:
\begin{align*}
    &\prod_i \left[\{\lambda_P(\text{age}_i + Y_i \mid \bfW_i) + \lambda_E(Y_i \mid \bfX_i)\}^{\delta_i} 
    \times \exp\left\{- \int_0^{Y_i} \lambda_P(\text{age}_i + u \mid \bfW_i) + \lambda_E(u \mid \bfX_i) \ du\right\}\right]
    \\&\propto \prod_i \left[\{\lambda_P(\text{age}_i + Y_i \mid \bfW_i) + \lambda_E(Y_i \mid \bfX_i)\}^{\delta_i} 
    \times \exp\left\{- \Lambda_E(Y_i \mid \bfX_i) \right\}\right].
\end{align*}
where $\Lambda_E(t \mid \bfx) = \int_0^t \lambda_E(u \mid \bfx) \ du$ is the \emph{cumulative excess hazard}, and $\bfW_i \subseteq \bfX_i$.
In the remainder of this section we describe our specific choices for $\lambda_E(t \mid \bfx)$ and describe how to perform posterior inference via Gibbs sampling.

\subsection{Proportional and Non-Proportional Hazards BART Models}

We first propose a Bayesian nonparametric proportional hazards (PH) model for
$\lambda_E(t \mid \bx)$, which is set equal to $\lambda_{0E}(t) \, e^{r(\bx)}$
where $r(\bx)$ is given a BART prior. We then model $\lambda_{0E}(t)$ using a
piecewise exponential model
\begin{align*}
  \lambda_{0E}(t) = \sum_{b=1}^B 1(t_{b-1} \le t < t_b) \, \lambda_b,  
\end{align*}
where $0 = t_0 < t_1 < \cdots < t_{B - 1} < t_B = \infty $. To model the baseline hazard we set $\lambda_b \sim \Gam(1, b_\lambda)$ with a flat prior placed on $b_\lambda$, and generally recommend a modest number of bins; in our illustrations, we take $B = N^{1/3}$ (matching the order of, for example, the Freedman-Diaconis rule for the number bins of a histogram), with the $t_b$'s chosen to be evenly spaced quantiles of the $Y_i$'s. 

This model can be extended to a non-proportional hazards (NPH) model that allows the hazard to vary with the covariates in a fully nonparametric fashion. We set
\begin{align*}
    \lambda_E(t \mid \bx)
    = \sum_{b = 1}^B 1(t_{b-1} \le t \le t_b) \, \lambda_b \exp\{r(\bx, b)\}.
\end{align*}
This model still assumes that the hazard function is piecewise constant, but allows the hazard within each bin to also depend on the covariates. We incorporate $b$ into the decision tree in the same fashion as the other covariates. Additionally, we can induce shrinkage of this model to the proportional hazards model by taking advantage of the automatic relevance determination prior introduced by \citet{linero2016bayesian}. Specifically, let $\omega$ denote the probability that a given splitting rule in $r(\bfx, b)$ makes use of $b$. To induce shrinkage towards the proportional hazards model, we can set $\omega \sim \Beta(P^{-1}, 1)$.

We emphasize that the number of bins $B$ should be kept modest, which lies in contrast to the standard survival setting (i.e., not the relative survival model) where one can instead take $B \to \infty$ and use an improper prior $\lambda_b \sim \Gam(0, 0)$; inferences produced under this later model tend toward inferences based on the Cox partial likelihood for $r(\bx)$, provided that we set $\lambda_b \equiv 0$ in any empty bins \citep{sinha2003bayesian}. We do not recommend setting $B$ large in the relative survival setting for several reasons. First, the population hazards are usually obtained via actuarial tables, and hence are likely themselves computed from a piecewise exponential model with a modest number of bins, and there is little reason to adopt a model for the excess hazard that is more precise than that of the population hazard. Second, for the NPH model, using a smaller number of bins improves the computational efficiency of our Gibbs sampler. Third, a modest choice for $B$ facilitates likelihood-based comparisons with other Bayesian parametric and semiparametric approaches, which would not be possible otherwise.

For the leaf node parameters we set $\mu_{t\ell} \sim \log \Gam(a, b)$ where $a$ and $b$ are chosen so that $\E(\mu_{t\ell}) = 0$ and $\Var(\mu_{t\ell}) = \sigma^2_\mu$ for some user-specified choice of $\sigma^2_\mu$. This choice facilitates computations for our Gibbs sampling algorithm, leading to straight-forward updates for the tree topologies, while mirroring as closely as possible the recommendations of \citet{chipman2010bart}. These constraints imply that $\psi(\alpha) = \log \beta$ and $\psi'(\alpha) = \sigma^2_\mu$ where $\psi(\cdot)$ and $\psi'(\cdot)$ denote the digamma and trigamma functions respectively. In our illustrations we set $\sigma_\mu = 1.5 / \sqrt{M}$, which corresponds to a belief that $r(\bx) \in (-3, 3)$ with prior probability approximately 95\%; it is also straight-forward to place a prior on $\sigma_\mu$.

\subsection{Extracting Quantities of Interest}

We now introduce some tools for interpreting draws from the posterior distribution of the PH and NPH models. We first consider the PH model, where the main object of interest is the function $r(\bx)$, as this determines how the various predictors affect the risk of death. 

\paragraph{Interpreting the Marginal Effects of Variables.}
We use the posterior projection strategy of \citet{woody2020model} to summarize the posterior distribution of $r(\bx)$. This approach defines an interpretable summary of $r(\bx)$ as a functional of the data generating process, which is sampled as we would any other parameter. To summarize $r(\bx)$ we compute its projection onto an interpretable family $\sQ$ as
\begin{math}
    \widetilde r(\bfx) = \arg \min_{q \in \sQ} \|r - q\|,
\end{math}
for some norm $\|\cdot\|$, typically the empirical $L_2$-norm defined by  $\|r\|^2 = N^{-1} \sum_i r(\bfX_i)^2$. Common choices of $\sQ$ include the class  of linear models $q(\bfx) = \bfx^\top \beta_q$, additive models $q(\bfx) = \sum_{j=1}^p  q_j(x_j)$, or single decision trees $q(\bfx) = g(\bfx; \Tree_q, \sM_q)$. We will restrict our attention to the setting of additive models in this work, as additive models happen to work very well as summaries in our illustrations.

\paragraph{Subgroup Identification.}
As an alternative to interpreting marginal effects, we might instead be interested in identifying subgroups of individuals that have different prognoses; this might be useful for informing policy, where we aim to identify relatively coarse subsets of individuals with particularly high risk that we might want to perform some intervention on. To do this, we will use the \emph{virtual twins} approach of \citet{foster2011subgroup}, which amounts to applying the Classification and Regression Tree (CART) algorithm with the Bayes estimator $\widehat r(\bfX_i)$ as the outcome to identify subgroups with different prognoses based on $\bfX_i$.

\paragraph{Summarizing Variable Importance.}
To help quantify the importance of each predictor we use a \emph{predictive variable importance} that measures a given variable's contribution to $r(\bx)$. We do this by computing the posterior distribution of the summary $R^2$ \citep{woody2020model}
\begin{align*}
    R^2 = 1 - \frac{\sum_i \{r(\bfX_i) - q(\bfX_i)\}}{\sum_i \{r(\bfX_i) - \bar r\}}
\end{align*}
where $q(\bfx)$ denotes a projection of $r(\bfx)$ onto some space $\mathcal Q$ and $\bar r = N^{-1} \sum_i r(\bfX_i)$. We do this for several different models:
\begin{enumerate}
\item We first assess the predictive importance simultaneously of all of the
    interactions by considering the summary $R^2$ of a generalized additive model $q(\bx) = \sum_{j = 1}^P q_j(x_j)$ with each $q_j(\cdot)$ estimated using a smoothing spline for numeric predictors.
    \item Next, we assess the predictive performance of the additive components of each individual variable in isolation by considering the summary $R^2$ of a generalized additive model that removes predictor $p$ from the summary, $q_{-p}(\bx) = \sum_{j \ne p} q_j(x_j)$ with each $q_j(\cdot)$ estimated using a smoothing spline for numeric predictors.
\end{enumerate}
The idea behind this procedure is that (i) if there are important interactions, then $q(\bfx)$ should have a poor summary $R^2$, while (ii) if variable $p$ is important, then $q_{-p}(\bfx)$ should have a poor summary $R^2$. Hence, smaller values of summary $R^2$ correspond to higher importance.

To extend these procedures to the NPH setting, we apply the same procedures but to the function $S_E(t \mid \bfx)$ for fixed values of $t$. This allows us to assess, for example, how variable importance changes over time.

\subsection{Posterior Computation}
Computations for the PH and NPH models can be carried out through relatively
simple extensions of existing Gibbs samplers for BART survival models for the
proportional hazards model \citet{basak2021algorithm}. To convert the likelihood
$$
  \prod_i \{\lambda_P(\text{age}_i + Y_i \mid \bfW_i) + \lambda_E(Y_i \mid \bfX_i)\}^{\delta_i}
  \, \exp\left\{-\Lambda_E(Y_i \mid \bfX_i)\right\},
$$
into a more tractable form, we augment a modified censoring indicator $d_i \sim
\Bernoulli(p_i)$ where $p_i = \lambda_E(Y_i \mid \bfX_i) / \{\lambda_E(Y_i \mid
\bfX_i) + \lambda_P(\text{age}_i + Y_i \mid \bfW_i)\}$ for each individual with $\delta_i = 1$, and
set $d_i = 0$ if $\delta_i = 0$. After augmenting these latent indicators, the
likelihood becomes
\begin{equation}
\label{eq:lambda_likelihood}
  \prod_i \lambda_E(Y_i \mid \bfX_i)^{d_i}
  \, \exp\left\{-\Lambda_E(Y_i \mid \bfX_i)\right\},
\end{equation}
which is the standard likelihood form for survival analysis with $d_i$ playing
the role of the censoring indicator. We apply a straight-forward extension of the Bayesian backfitting algorithm of 
\citet{chipman2010bart} to the proportional hazards model; this approach is 
conceptually similar to the approach of \citet{linero2021bayesian}. With the augmented indicators $d_i$ and the likelihood as above, we must update the trees and their leaf parameters $(\Tree_t, \sM_t), t=1, \dots, M$ and the parameters of the baseline hazard function $\lambda_b, b=1, \dots, B.$

We apply a straight-forward extension of the Bayesian backfitting algorithm of 
\citet{chipman2010bart} to the proportional hazards model; this approach is 
conceptually similar to the approach of \citet{linero2021bayesian}. Let $\Data$ denote the observed data and let $\Tree_{-t}$ and $\sM_{-t}$  respectively denote the collection of all tree topologies and leaf parameters except for those associated to tree $t$. Let $\Leaves(\Tree)$ denote the collection of leaf nodes associated to tree $\Tree$.

Our Gibbs sampler alternates between the following steps:
\begin{enumerate}
    \item For $i = 1,\ldots, N$ sample 
    $d_i \sim \Bernoulli\left\{\frac{\delta_i \lambda_{b_i} e^{r(\bfX_i)}}{\lambda_P(\text{age}_i + Y_i \mid \bfW_i) + \lambda_{b_i} \, e^{r(\bfX_i)}}\right\}$.
    \item For $m = 1, \ldots, M$:
    \begin{enumerate}
        \item Sample $\Tree_t$ from a Markov transition function that leaves the conditional posterior distribution
        $\pi(\Tree_t \mid \Tree_{-t}, \sM_{-t}, \Data)$ invariant.
        \item For $\ell \in \Leaves(\Tree_t)$, sample $\mu_{t\ell}$ from its full conditional distribution.
    \end{enumerate}
    \item Sample $\lambda_1, \ldots, \lambda_b$ from 
    $\pi(\lambda_1, \ldots, \lambda_b \mid (\Tree_1, \sM_1), \ldots, (\Tree_M, \sM_M), d_1, \ldots, d_N, \Data)$.
\end{enumerate}
Step 1  is straight forward; below we describe details on Step 2 and Step 3 
below. For notational convenience, we define $Z_{ib} = 1(Y_i \ge t_b) (t_b - t_{b-1}) + 1(t_{b-1} \le Y_i < t_b) (Y_i - t_{b- 1})$, which represents the contribution of bin $b$ to the cumulative excess hazard of observation $i$.

\paragraph{Full Conditional of $\lambda$'s}
Conditional on all other parameters, the likelihood of the $\lambda_b$'s factors across $b$, and hence under the the prior $\lambda_b \sim \stackrel{\text{iid}} \Gam(a_\lambda, b_\lambda)$ the $\lambda_b$'s are  conditionally independent given the other parameters in the model. The full  conditional for each individual $\lambda_b$ is given by
\begin{align*}
    \pi(\lambda_b \mid \text{everything else}) &\propto
    \left\{\prod_i \lambda_b^{d_i \, 1(b_i = b)} 
      e^{-Z_{ib} \lambda_b e^{r(\bfX_i)}} \right\}
      \times \lambda_b^{a_\lambda - 1} \, e^{-b_\lambda \, \lambda_b}
    \\&\propto
    \lambda_b^{a_\lambda + A_b - 1} \, e^{-(b_\lambda + B_b) \, \lambda_b},
\end{align*}
where $A_b = \sum_i d_i \, 1(b_i = b)$ and $B_b = \sum_i Z_{ib} \, e^{r(\bfX_i)}$. This is the full conditional of a gamma distribution, i.e., $\lambda_b \stackrel{\text{indep}}{\sim} \Gam(a_\lambda + A_b, b_\lambda + B_b)$.
It is easy to show that we can compute $B_b$ efficiently via 
recursion by noting that
\begin{align*}
    B_{b+1} = \frac{t_{b+1} - t_b}{t_b - t_{b-1}} \left(B_b - \sum_{i : Y_i \in [t_{b-1}, t_{b+1})} Z_{ib} e^{r(\bfX_i)}\right) + \sum_{i: Y_i \in [t_{b}, t_{b+1})} Z_{i(b+1)} e^{r(\bfX_i)}.
\end{align*}
Rather than requiring $O(NB)$ computations to compute all of the $B_b$'s, 
utilizing this recursion requires only $O(N)$ computations.

\paragraph{Updating the Trees}
To update $(\Tree_t, \sM_t)$ we first make a Metropolis-Hastings proposal to 
update the decision tree $\Tree_t$ and then sample $\sM_t$ from its full 
conditional distribution. Let $Q(\Tree \to \Tree')$ denote a Markov transition 
function (MTF) on the collection of possible tree structures. The MTF is 
typically a mixture of the \texttt{Birth}, \texttt{Death}, \texttt{Swap}, and 
\texttt{Change} proposals introduced by \citet{chipman1998bayesian}; see 
\citet{kapelner2014bartmachine} for a detailed description of these different 
proposals and how to compute the various transition probabilities. After 
sampling a new tree structure from $\Tree \sim Q(\Tree_t \to \Tree)$, we set 
$\Tree_t = \Tree$ with probability
\begin{align*}
    A = \min\left\{\frac{\pi(\Tree) \, L(\Tree) \, Q(\Tree \to \Tree_t)}
         {\pi(\Tree_t) \, L(\Tree_t) \, Q(\Tree_t \to \Tree)}, 1 \right\},
\end{align*}
and leave $\Tree_t$ unchanged otherwise; here $L(\Tree)$ is the 
\emph{integrated likelihood}
\begin{align*}
    L(\Tree) = \prod_{\ell} \int \prod_{i \leadsto \ell}
      \lambda_{b_i}^{d_i} \exp\left(d_i \eta_i + d_i \mu - e^\mu \sum_b Z_{ib} \lambda_b e^{\eta_i}\right) \times \frac{b^a}{\Gamma(a)} \exp(a \mu - b e^\mu) \ d\mu,
\end{align*}
where $\eta_i = r(\bfX_i) - g(\bfX_i; \Tree_t, \sM_t)$. Observing that the term 
inside the integral simplifies to the normalizing constant of a log-gamma 
distribution, this expression simplifies to
\begin{align*}
    L(\Tree) \propto 
    \left\{\prod_\ell \frac{\Gamma(a + A_\ell)}{(b + B_\ell)^{a + A_\ell}}\right\}
    \times \frac{b^a}{\Gamma(a)}
\end{align*}
where $A_\ell = \sum_i d_i$ and  $B_\ell = \sum_{i,b} Z_{ib} \lambda_b e^{\eta_i}$. By similar calculations, the  full conditional for the $\mu_{t\ell}$'s is given by a $\log \Gam(a + A_\ell, b +  B_\ell)$ distribution.

As with the update for the $\lambda_b$'s, it is possible to take advantage of redundancy in $\sum_b Z_{ib} \lambda_b$ to reduce the computational cost of  computing $B_\ell$ from $O(NB)$ to $O(N)$. Note that by definition we have  $Z_{ib} = 1(Y_i \ge t_b) (t_b - t_{b-1}) + 1(t_{b-1} \le Y_i < t_b) (Y_i - t_{b- 1})$. Hence, $\sum_b Z_{ib} \lambda_b = \lambda_{b_i} (Y_i - t_{b_i - 1}) +  \sum_{b < b_i} \lambda_b (t_b - t_{b-1}) = \lambda_{b_i} (Y_i - t_{b_i - 1}) +  F_{b_i}$ where we define $F_b = \sum_{k < b} \lambda_k (t_k - t_{k-1})$, and importantly the $F_b$'s can be computed for $b = 1,\ldots, B$ prior to computing the $B_\ell$'s. Hence, we can write $B_\ell = \sum_i \lambda_{b_i}(Y_i - t_{b_i - 1}) + F_{b_i}$.

Computations for the Gibbs sampler for the NPH model are very similar to those 
for the PH model, and are deferred to the appendix; the main complication is that it is no longer easy to speed up computations via recursion, and so computations for the NPH model are slower by a factor of $B$. Full details for both algorithms are also given in Algorithm~\ref{alg:ph} and Algorithm~\ref{alg:nph} in the appendix.

\section{Simulation Study}\label{sec:simulation}

In this section, we demonstrate the implementation of our proposed method through an extensive simulation study, by mimicking the real dataset \texttt{LeukSurv}, which is available in the \texttt{spBayesSurv} \texttt{R} package. This dataset contains information about $1,043$ patients diagnosed with acute myeloid leukemia, including their age, sex, white blood cell count at the time of diagnosis (truncated at 500) and Townsend score (lower values indicate that the individual is from a less affluent area). We augmented this dataset with an approximation of the baseline survival probabilities of the individuals in the background population, stratified by sex and age at baseline. To make our data generation mechanism as realistic as possible, we fitted four different models incorporating varying baseline hazard functions and incorporating both linear and non-linear  interactions among covariates to the motivating \texttt{LeukSurv} data, and used these fitted models as the ``true model" to generate the necessary survival times for our simulation study.

For each data generating mechanism, we generated $100$ replicates of survival times for 1,043 subjects, and covariates as age, sex, white blood cell count (wbc) and Townsend score (tpi). We first simulated the survival times associated to the excess hazard with the true hazard function $\lambda_E(\cdot)$ modeled as 
\begin{math}
  \lambda_E(t) = \lambda_{0E}(t)\times \exp\{r(\bfx)\}.
\end{math}
We consider the following data generating mechanisms, which were obtained by fitting these models to the data:
\begin{itemize}
    \item Cox-Linear: The baseline hazard is modeled as a piecewise exponential function and the interaction among the covariates $\mathbf{x}$ is modeled in a linear fashion through the exponent is $r(\bfx) = \bfx^{\top}\beta,$ where  $\beta$ is the vector of the regression coefficients.
    
    \item Weibull-Linear: Same as Cox-Linear, except the baseline hazard is a Weibull hazard.
    
    \item Weibull-Spline: Same as Weibull-Linear, except $r(\bfx)$ is modeled additively using splines for the continuous variables.
    
    \item COXPH-BART: Same as Cox-Linear, except $r(\bfx)$ is a decision tree ensemble.
\end{itemize}

We simulated the survival times for the general population  from an exponential distribution using life tables, and introduced a non-informative censoring variable $C$, which was sampled from a uniform distribution. Finally, we obtained the observed event times as the minimum of the population survival, excess survival, and censoring times. 

\begin{figure}
\centering
\begin{subfigure}[t]{0.85\textwidth}
\centering
\includegraphics[width=\textwidth]{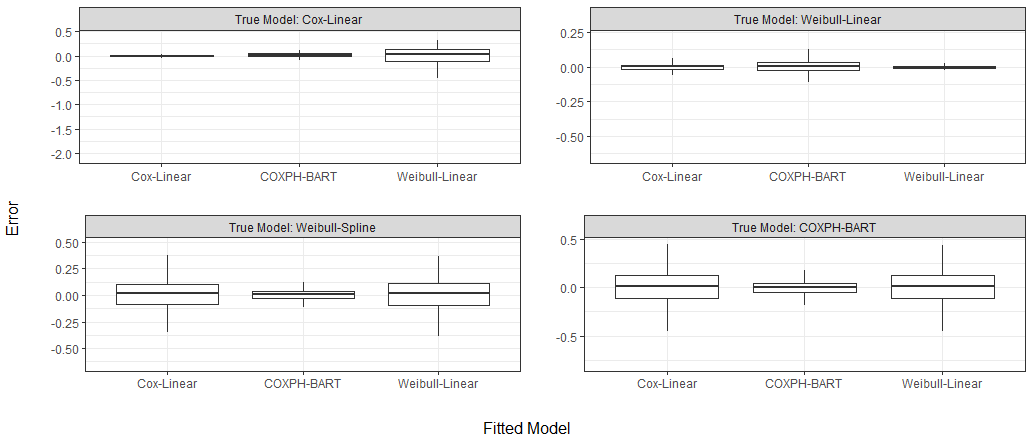} 
\caption{} \label{fig:timing1}
\end{subfigure}

\begin{subfigure}[t]{0.85\textwidth}
\centering
\includegraphics[width=\textwidth]{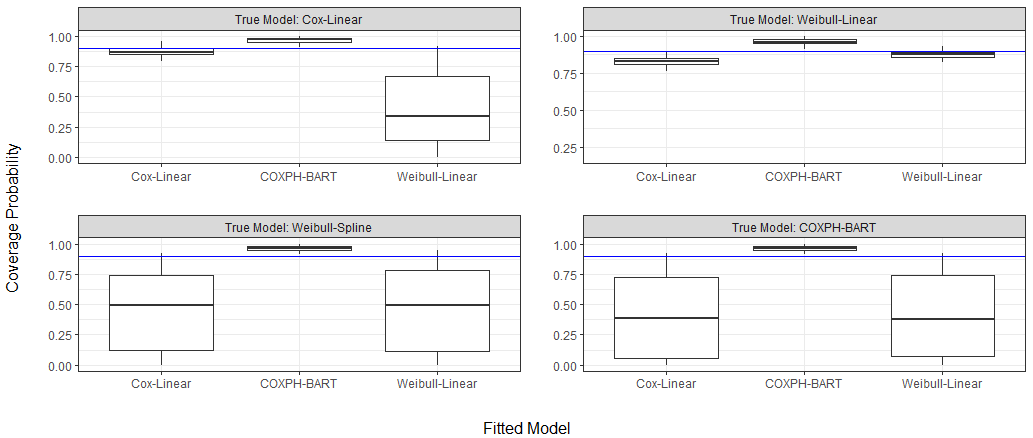} 
\caption{} \label{fig:timing2}
\end{subfigure}

\begin{subfigure}[t]{0.85\textwidth}
\centering
\includegraphics[width=\textwidth]{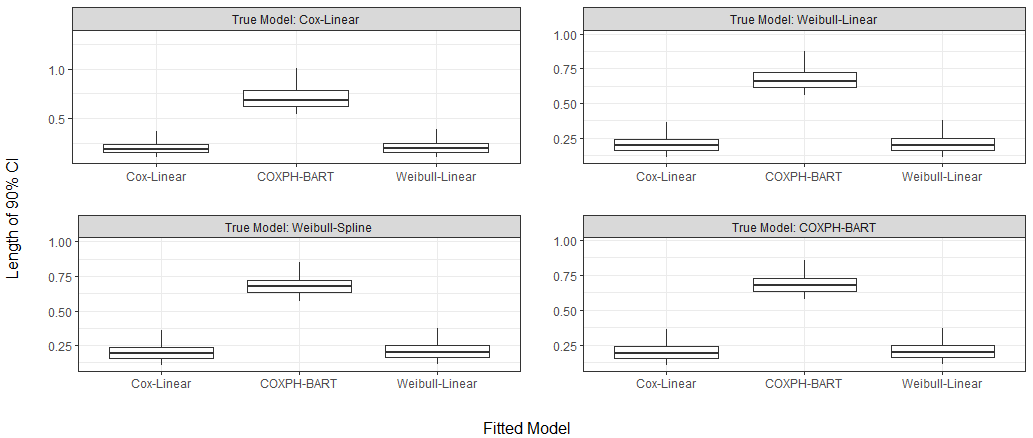} 
\caption{} \label{fig:timing3}
 \end{subfigure}

 \caption{Boxplot comparing the proposed COXPH-BART model with Cox-Linear and Weibull-Linear in terms of the error (panel (a)),  coverage probabilities (panel (b)), and length of the 90\% credible intervals (CI) (panel (c)) in predicting $r(\mathbf{x})$ under data simulated from different true models with proportional hazards.}
\label{fig:combined_boxplot_ph}
\end{figure}

With the survival times generated as above, we fitted our proposed COXPH-BART model and compared model performances with the linear Cox's proportional hazards model (Cox-linear) and the linear Weibull proportional hazards model (Weibull-linear). Monte Carlo approximation of the root mean squared error (RMSE) for the $m^{\text{th}}$ replicated dataset, $m=1,\dots,100$ was obtained as $\text{RMSE}_m = \sqrt{\frac{\sum_{i=1}^{N}\{ r(\mathbf{x}_{i})_m - \widehat{r}(\mathbf{x}_{i})_m\}^2}{N}},$ where $r(\mathbf{x}_{i})_m$ and $\widehat{r}(\mathbf{x}_{i})_m$ denote the true and the estimated values of the exponent term in the excess hazard model, as obtained for the $i^{\text{th}}$ subject from the $m^{\text{th}}$ replicated dataset. Performances from the above models are compared based on the average RMSE, the average proportion of nominal 90\% credible intervals which capture the true value of $r(\mathbf{x})$ and the average length of the 90\% credible intervals (Table \ref{t:table-ph}). 
Figure \ref{fig:combined_boxplot_ph} panel (a) also includes a boxplot of the subject-level prediction error averaged over all $100$ replicated datasets, given by $\frac{\sum_{m=1}^{100}\{\widehat r(\mathbf{x}_i)_m - r(\mathbf{x}_i)_m\}}{100}$ for $i=1, \cdots, N$. Similarly, Figure~\ref{fig:combined_boxplot_ph} panels (b) and (c) provide boxplots of the subject-level proportion of the $90\%$ credible intervals which capture the true value of $r(\cdot)$ and the lengths of these credible intervals, averaged over $100$ replicates, respectively. 

\renewcommand{\arraystretch}{1}
 \begin{table}
	\centering
	\def\~{\hphantom{0}}
		\vspace{0.2in}
		%\resizebox{\textwidth}{!}
  \scalebox{0.7}
  {\begin{tabular}{|c| ccc| ccc| ccc|}
				\hline 
				
				\multirow{3}{*}{\textbf{``True" Model}} &\multicolumn{3}{|c|}{\textbf{avg. RMSE}} &\multicolumn{3}{|c|}{\textbf{avg. coverage probability}}&\multicolumn{3}{|c|}{\textbf{avg. length}}
				\\ \cline{2-10}
                &Cox- &Weibull- &COXPH-&Cox- &Weibull- &COXPH- &Cox- &Weibull- &COXPH-\\
                
                &Linear &Linear &BART &Linear &Linear &BART &Linear &Linear &BART \\
				\hline 
			Cox-Linear&0.068
&0.193
&0.250

&0.874
&0.391
&0.951
&0.212
&0.221
&0.718
 \\
		      Weibull-Linear&0.077
&0.070
&0.170
&0.826
&0.875
&0.954
&0.211
&0.215
&0.676\\
                Weibull-Spline &0.182
&0.179
&0.171
&0.446
&0.461
&0.956 
&0.211
&0.217
&0.690

 \\
                COXPH-BART&0.191
&0.197
&0.164
&0.404
&0.410
&0.958
&0.210
&0.215
&0.686
 \\
			
			\hline
		\end{tabular}}
    		\caption{{Simulation results based on $100$ replicates of data comparing Monte Carlo estimates of average Root Mean Squared Error (RMSE),  average coverage probabilities, and average length of $90\%$ confidence intervals obtained from fitting the proposed COXPH-BART model as compared to Cox-Linear and Weibull-Linear models under different true models with proportional hazards.} \label{t:table-ph}}
	\vspace*{-6pt}
\end{table}

Simulation results show that when the true model is non-linear (Weibull-Spline and COXPH-BART), the proposed COXPH-BART model provides the minimum prediction error. The Cox-BART model also surpasses the nominal coverage rate whereas, other models fail to do so, especially in data generated using the non-linear models. These results suggest that fitting the Cox-Linear and Weibull-Linear  are insufficient in the presence of underlying non-linear associations among the covariates and it will provide improved estimation to adopt an ensemble based model as proposed in such situations. To balance this, COXPH-BART yields wider intervals and has higher RMSE when the parametric models are correctly specified.

Note that each of the data generation mechanisms and the fitted models considered in the above simulation settings assumes proportional hazards, that is, the association and interaction among the covariates (modeled by $r(\mathbf{x})$) is independent of time $t$. To demonstrate performance of the non-proportional relative survival model (COXNPH-BART) proposed in this paper, we perform another simulation study mimicking the same \texttt{LeukSurv} data. We obtain 100 replicated datasets using the same data generation technique as in the proportional hazards simulation, except that in this case, we first fit a non-proportional hazard survival model $\lambda(t) = \lambda_0(t) \times \exp{r( \mathbf{x}, t)}$ to the data and then use this as the ground truth to simulate the survival times for the diseased population. With the non-proportional hazards survival times simulated as above, we fitted our proposed COXPH-BART and COXNPH-BART models and compared model performances based on the prediction accuracy of the relative survival function $S_E(t \mid \bfx) = \exp\left\{-\int_0^t \lambda_{0E}(z) \, e^{r(\bfx,z)} \ dz\right\}.$ As before, Monte Carlo approximations of the RMSE for the $m^{\text{th}}$ replicated dataset at time point $t$ was obtained as $\text{RMSE}_{m,t} = \sqrt{\frac{\{S_E(t, \bfx_i)_m - \widehat S_E(t, \bfx_i)_m\}}{N}}$ where $S_E(t \mid \mathbf{x})_m$ and $\widehat{S}_E(t \mid \mathbf{x})_m$ denotes the true and the estimated relative survival at time point $t$ obtained from the $m^{\text{th}}$ replicated dataset. 

\begin{figure}
     \centering
     \includegraphics[width=\textwidth]{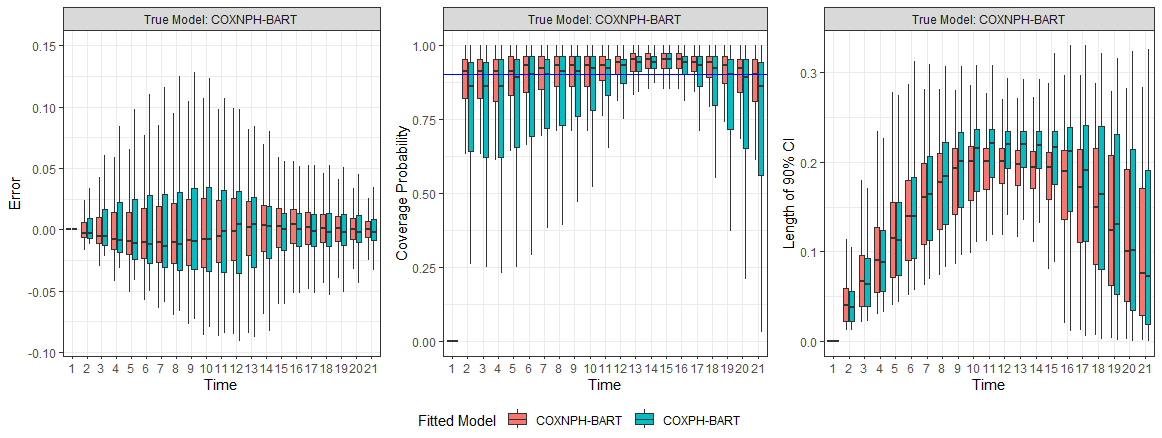}
     \caption{Boxplot comparing the COXPH-BART model with COXNPH-BART model in terms of the error (left),  coverage probabilities of 90\% credible intervals (CIs)  (middle), and lengths of the 90\% CIs (right) in predicting $S_E(t)$ plotted over time under data simulated from COXNPH-BART model. Solid horizontal line in the middle panel is at $0.9$ indicating the $90\%$ nominal coverage probability in predicting $S_E(\cdot).$}
     \label{fig:NPH-Boxplots}
 \end{figure}

\begin{figure}
     \centering
     \includegraphics[width=\textwidth]{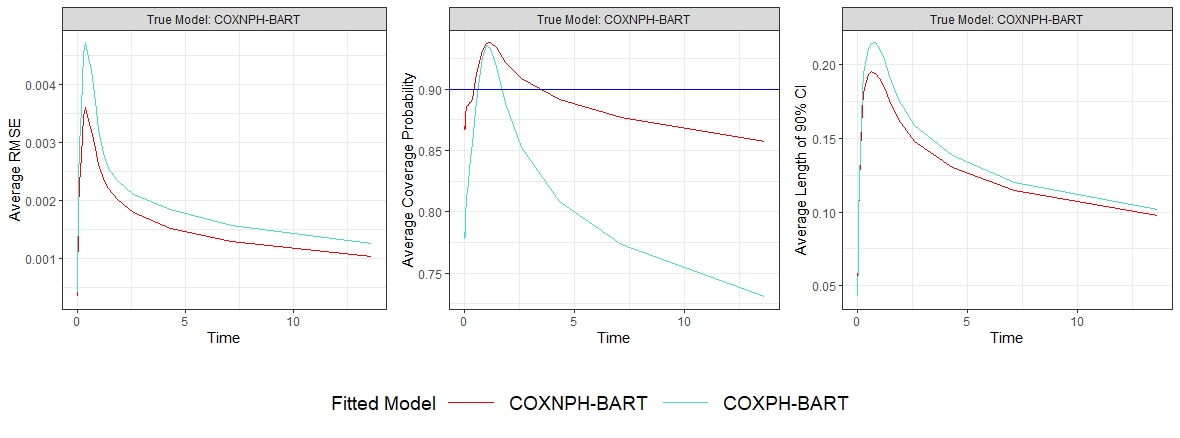}
     \caption{Plot comparing the COXPH-BART and COXNPH-BART models in terms of the average root mean squared error (RMSE) (left),  average coverage probabilities of 90\% credible intervals (middle), and average length of the 90\% credible intervals (right) in predicting $S_E(t)$ over time under data simulated from the COXNPH-BART model.}
     \label{fig:NPH-plots}
 \end{figure}

We report the average RMSE, average coverage probability of 90\% credible intervals and the average lengths of these intervals plotted over time in figure (\ref{fig:NPH-plots}). Figure (\ref{fig:NPH-Boxplots}) also includes boxplot of the prediction error in $\widehat{S}_E,$ the coverage probabilities and length of the 90\% credible intervals plotted over time. Note that we gain somewhat in terms of bias, coverage probability as well as width of the credible intervals due to using COXNPH-BART as opposed to COXPH-BART, however, the two models are quite comparable, especially in terms of the error. 

\section{Application to Colon Cancer Data}\label{sec:applications}

We extracted socio-demographic and clinical information on $18,296$ adults ($9,746$ males and $8,550$ females) diagnosed with colon cancer (ICD-10 C18) in England in 2012, using data from national population-based cancer registries linked to secondary care records. The most recent vital status was confirmed on or before December 31st 2019, providing 7 years of follow-up. 
We used or derived information on the following key individual factors: age at diagnosis; cancer stage at diagnosis, coded according to the Tumor Node and Metastasis (TMN) system; deprivation score (1 = least deprived, 5 = most deprived), derived from income domain of the Index of Multiple Deprivation (IMD) measured at the small-area level of patient's residence at the time of their diagnosis; emergency presentation (EP, binary); presence of comorbidities (including cardiovascular disease, chronic obstructive pulmonary disease, diabetes, and renal disease, see Table \ref{tab:comorbidities} in the appendix). The tumor stages (I - IV) are derived from the size of the tumor (T), the number of nodes involved (N), and the presence of metastasis (M). EP is one of eight routes to diagnosis, and is defined by an algorithm based on linked electronic health records \citep{elliss:2012}: it corresponds to an emergency route via A\&E, emergency GP Referral, emergency transfer, emergency consultant outpatient referral, emergency admission or attendance.

A summary of the data set is presented in Table \ref{table:table1} for males and Table \ref{table:table1_women} in the appendix for females. We observe higher rates of EPs in patients with stage IV cancer tumor compared to those with stage I: from $7.6\%$ and $10.9\%$ in stage I to $39.4\%$ and $42.2\%$ in stage IV, respectively, for male and female patients. Proportions of patients with comorbidities increase with stage at diagnosis in females, but remain stable throughout all stages in males. As expected, mean follow-up time decreases from $5.6$ and $5.8$ years in males and females diagnosed at stage I to $1.4$ years for patients diagnosed at stage IV, with less than $30\%$ of stage I patients dying in the 7 years of follow-up in contrast to over $90\%$ of stage IV patients.

\begin{table}[!ht]
    \centering
    \resizebox{\textwidth}{!}{
    \begin{tabular}{lllllllll}
    \hline
        & \multicolumn{7}{c}{\textbf{Stage at diagnosis}}    \\ \hline
        & \multicolumn{2}{c}{\textbf{I}}  & \multicolumn{2}{c}{\textbf{II}} & \multicolumn{2}{c}{\textbf{III}} & \multicolumn{2}{c}{\textbf{IV}} \\ \hline
         & N  & \% &  N  & \% &  N  & \% &  N  & \% \\
        \textbf{N} &  1,537  &  &  2,801  &  &  2,620  &  &  2,788  &  \\
        \textbf{Mean age (sd)} &  70.1  & (11.2) &  72.4  & (11.7) &  70.5  & (12.2) &  71.7  & (12.0) \\ 
        \textbf{Emergency presentation} &  116  & \textit{7.6} &  684  & \textit{24.4} &  698  & \textit{26.6} &  1,097  & \textit{39.4} \\ 
        \textbf{Comorbidities} &  &  &  &  &  &  &  &  \\ 
        Cardiovascular disease &  191  & \textit{12.4} &  439  & \textit{15.7} &  343  & \textit{13.1} &  408  & \textit{14.6} \\ 
        COPD &  139  & \textit{9.0} &  249  & \textit{8.89} &  243  & \textit{9.3} &  262  & \textit{9.4} \\ 
        Diabetes &  170  & \textit{11.1} &  324  & \textit{11.6} &  283  & \textit{10.8} &  327  & \textit{11.7} \\ 
        Renal disease &  54  & \textit{3.5} &  131  & \textit{4.7} &  84  & \textit{3.2} &  127  & \textit{4.6} \\ 
        \textbf{Deprivation (in quintiles)} &  &  &  &  &  &  &  &  \\ 
        Least deprived &  353  & \textit{23.0} &  613  & \textit{21.9} &  584  & \textit{22.3} &  569  & \textit{20.4} \\ 
        2 &  361  & \textit{23.5} &  671  & \textit{24.0} &  599  & \textit{22.9} &  598  & \textit{21.5} \\ 
        3 &  315  & \textit{20.5} &  553  & \textit{19.7} &  544  & \textit{20.8} &  602  & \textit{21.6} \\ 
        4 &  261  & \textit{17.0} &  517  & \textit{18.5} &  494  & \textit{18.9} &  541  & \textit{19.4} \\ 
        Most deprived &  247  & \textit{16.1} &  447  & \textit{16.0} &  399  & \textit{15.2} &  478  & \textit{17.1} \\ 
        \textbf{Number of deaths} &  420  & \textit{27.3} &  1,132  & \textit{40.4} &  1,409  & \textit{53.8} &  2,595  & \textit{93.1} \\ 
        \textbf{Mean follow-up time, years (sd)} &  5.6  & (1.9) &  5.1  & (2.3) &  4.3  & (2.5) &  1.4  & (1.9) \\ \hline
    \end{tabular}
    }
    \caption{Characteristics of male patients diagnosed with colon cancer}
    \label{table:table1}
\end{table}

Stage at diagnosis is a well-established clinical prognostic factor, with more advanced stages linked to fewer treatment options and lower survival probabilities. In addition, a vast literature has been devoted to understand differences in cancer survival by deprivation \citep{woods:2006}, which has evidenced inequalities in survival for most deprived patients compared to less deprived groups. Deprivation level, however, might be a proxy for other underlying factors such as unhealthy lifestyles and difficulties in healthcare access, which contribute to lower survival. This motivates our epidemiological question of understanding the importance of other prognostic factors on explaining cancer survival, and to understand how they interact with stage and deprivation. With this aim, we fit both COXNPH-BART and COXPH-BART models to the data, separately for males and females diagnosed with colon cancer; results for females are presented in the appendix \ref{sec:appendix_application}. The models included all variables ($\bfx$) discussed above (age, stage, deprivation (dep), cardiovascular disease (CVD), diabetes, renal disease (renal), and emergency presentation (EP)). Life-tables of population mortality rates estimated for each year of age, sex, deprivation level (by IMD quintiles) ($\bfw$), region, and calendar year were linked to patients records, to allow estimation of the excess hazard. 

We fit the models with $M = 100$ trees, a burn-in period of $1,000$ iterations and saved $1,000$ samples after burn-in with a thinning interval of $10$ (for $11,000$ total iterations). We then computed the leave-one-out expected log predictive density (ELPD) for each of these models, which is an omnibus measure of goodness-of-fit for model comparison. Additional details on using this statistic are given by \cite{vehtari2017practical}. The ELPD of the COXNPH-BART model was found to be substantially higher ($-10,285.4$) compared to the COXPH-BART model ($-10,570.9$), indicating a better fit to the data. Below, we present an interpretation of the results and new findings derived from the COXNPH-BART model.

\begin{figure}
\centering
\begin{tabular}{cc}
   \includegraphics[width=0.4\textwidth]{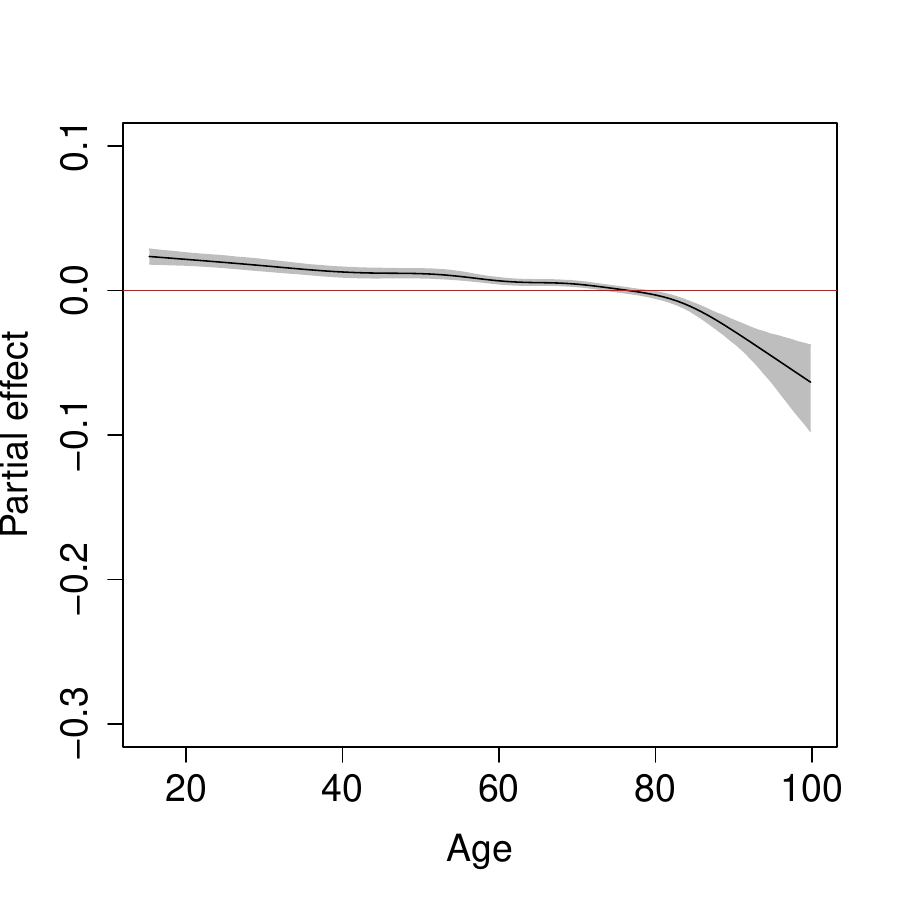}  &
   \includegraphics[width=0.4\textwidth]{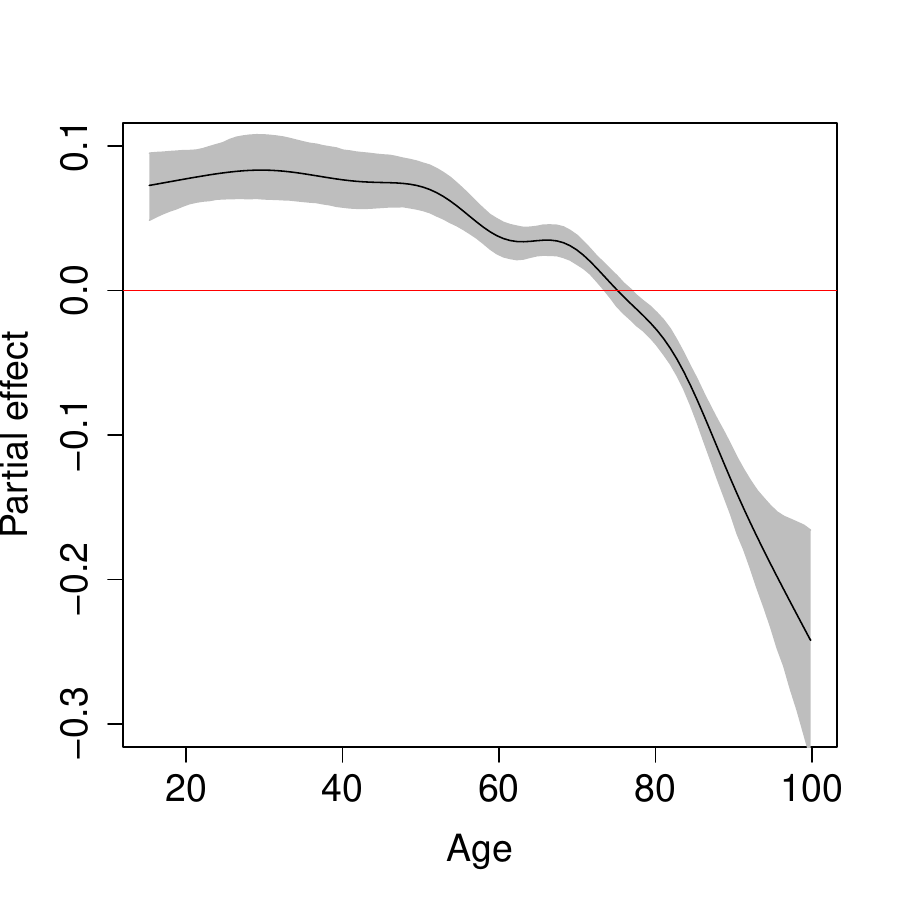} \\
   (a) & (b) \\
   \includegraphics[width=0.4\textwidth]{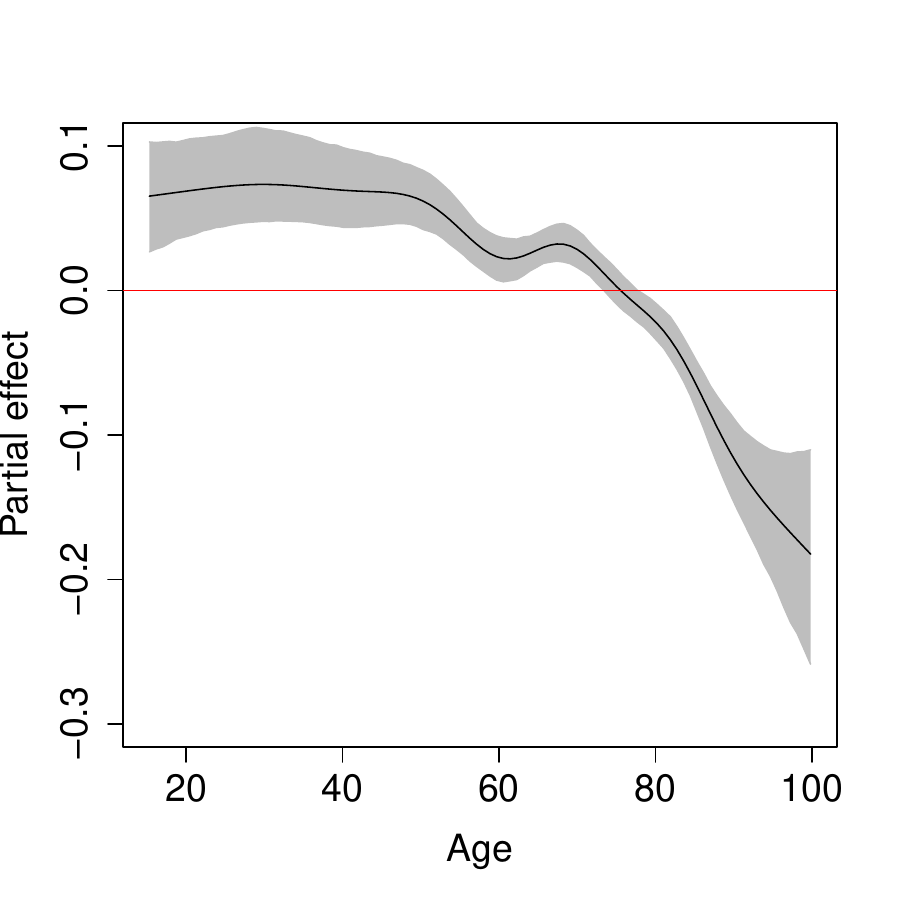} & 
   \includegraphics[width=0.4\textwidth]{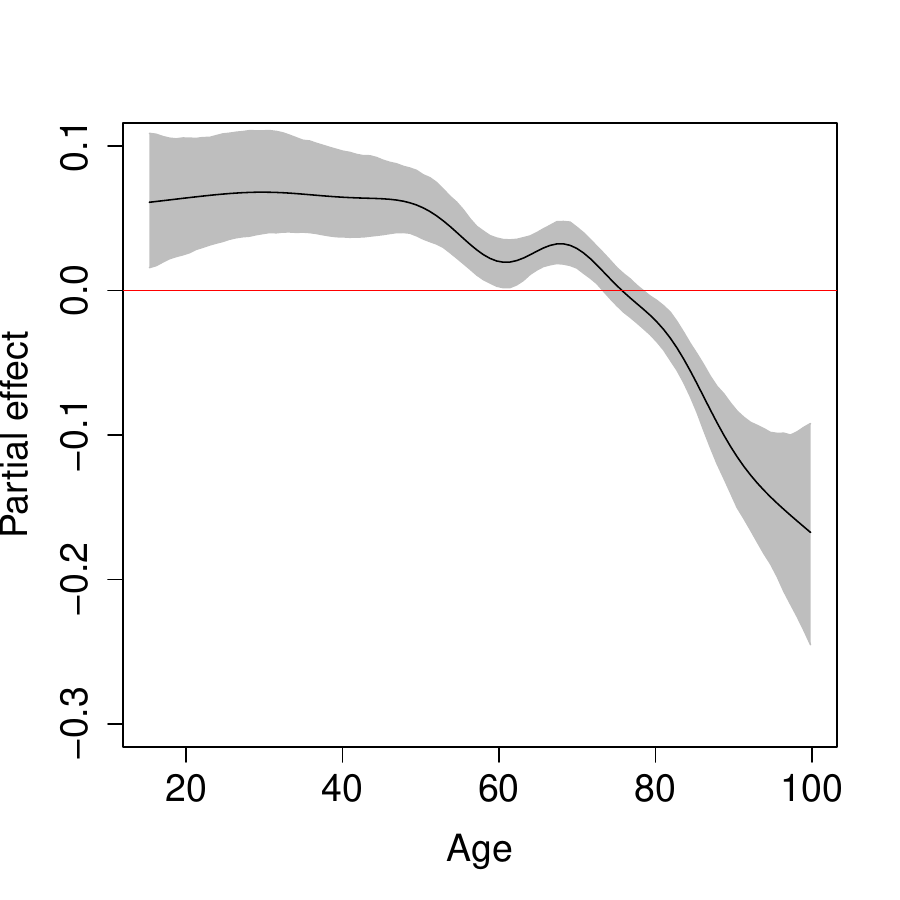} \\
   (c) & (d) \\
   \includegraphics[width=0.4\textwidth]{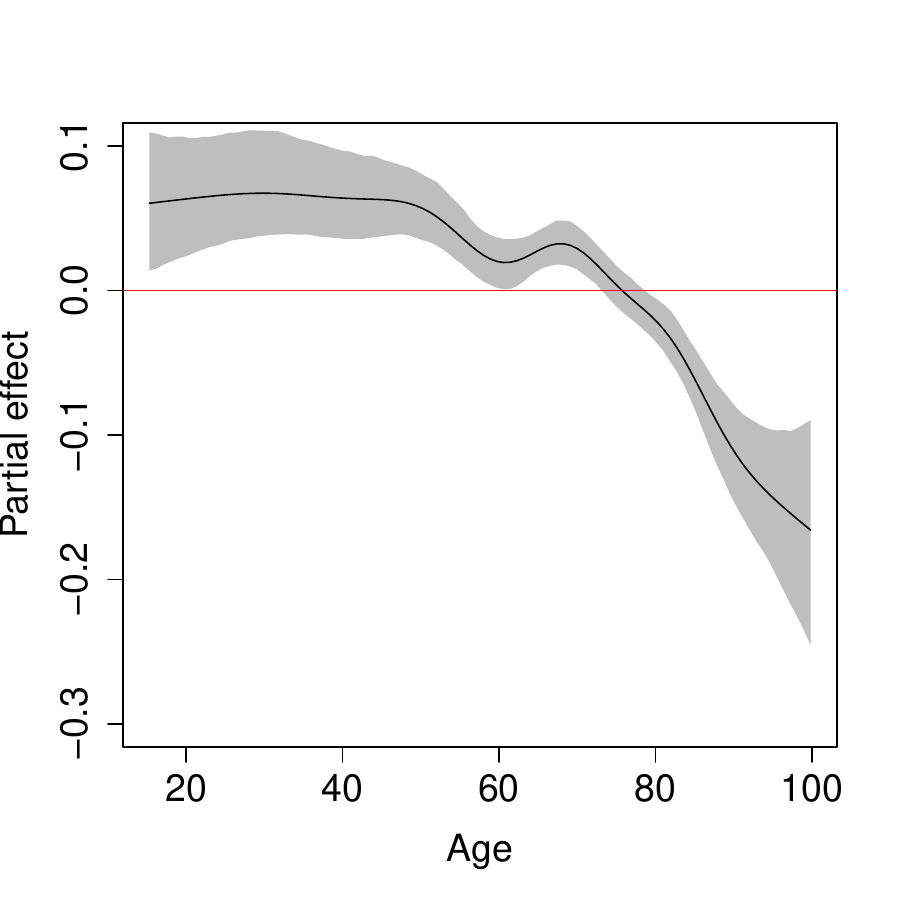} & \\
   (e) 
\end{tabular}
\caption{ Males colon cancer data: Partial effect of age at 2, 20, 40, 60, and 80 months.}
\label{fig:peff_age_males}
\end{figure}

Figure \ref{fig:nsdep_males} in the appendix shows that there is indeed evidence on inequalities in net survival between most and least deprived groups, as suggested in previous literature. Figure \ref{fig:peff_age_males} shows the partial effect of age (as recorded at diagnosis) on net survival at different time points (bins 2, 20, 40, 60 and 80 months). This figure shows a non-linear and time-varying effect of age. It also shows that patients diagnosed at younger ages have increased net survival, up until around 78 years when older ages are associated with lower net survival. The amplitude of the partial effects of age are most extreme at 20 months after diagnosis, and then reduce slightly as follow-up time elapses.

\begin{figure}
\centering
\begin{tabular}{cc}
   \includegraphics[width=0.5\textwidth]{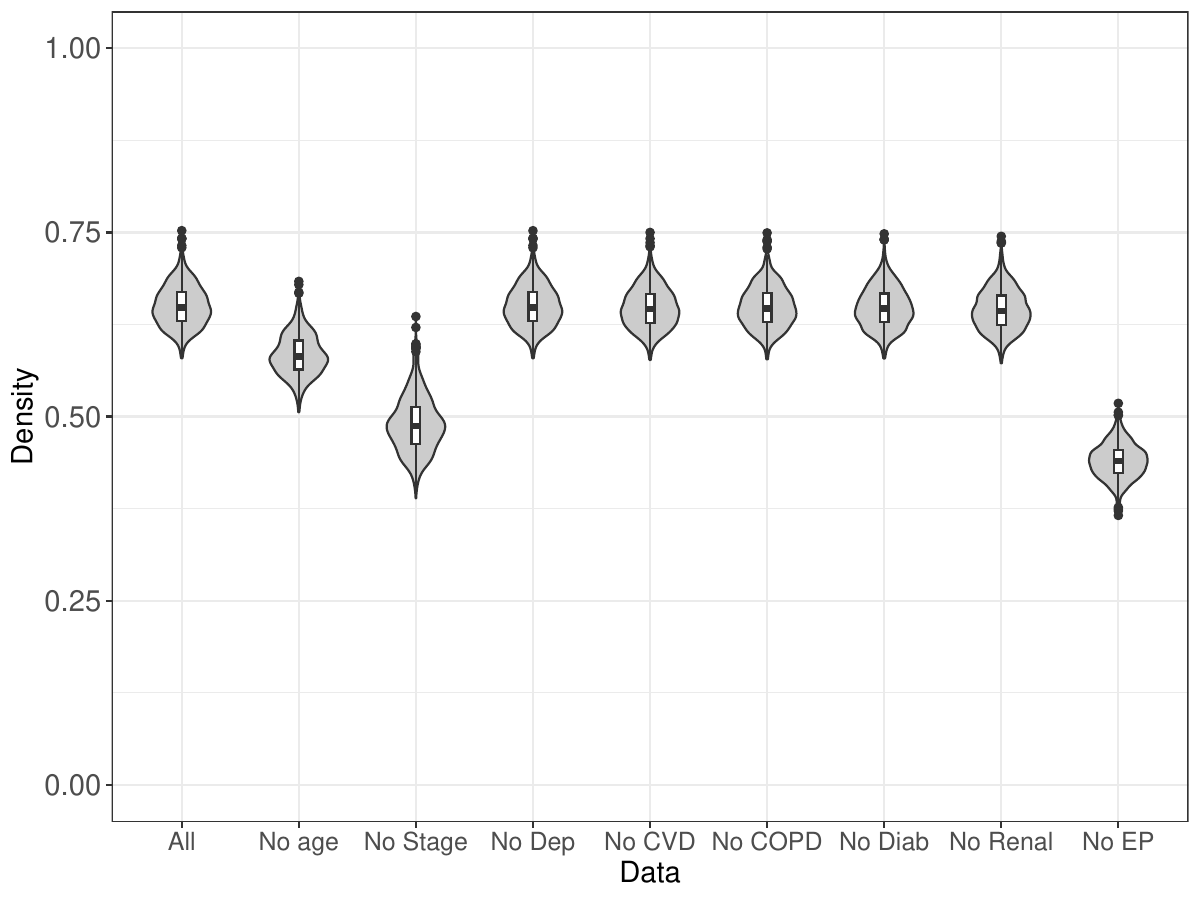}  &
   \includegraphics[width=0.5\textwidth]{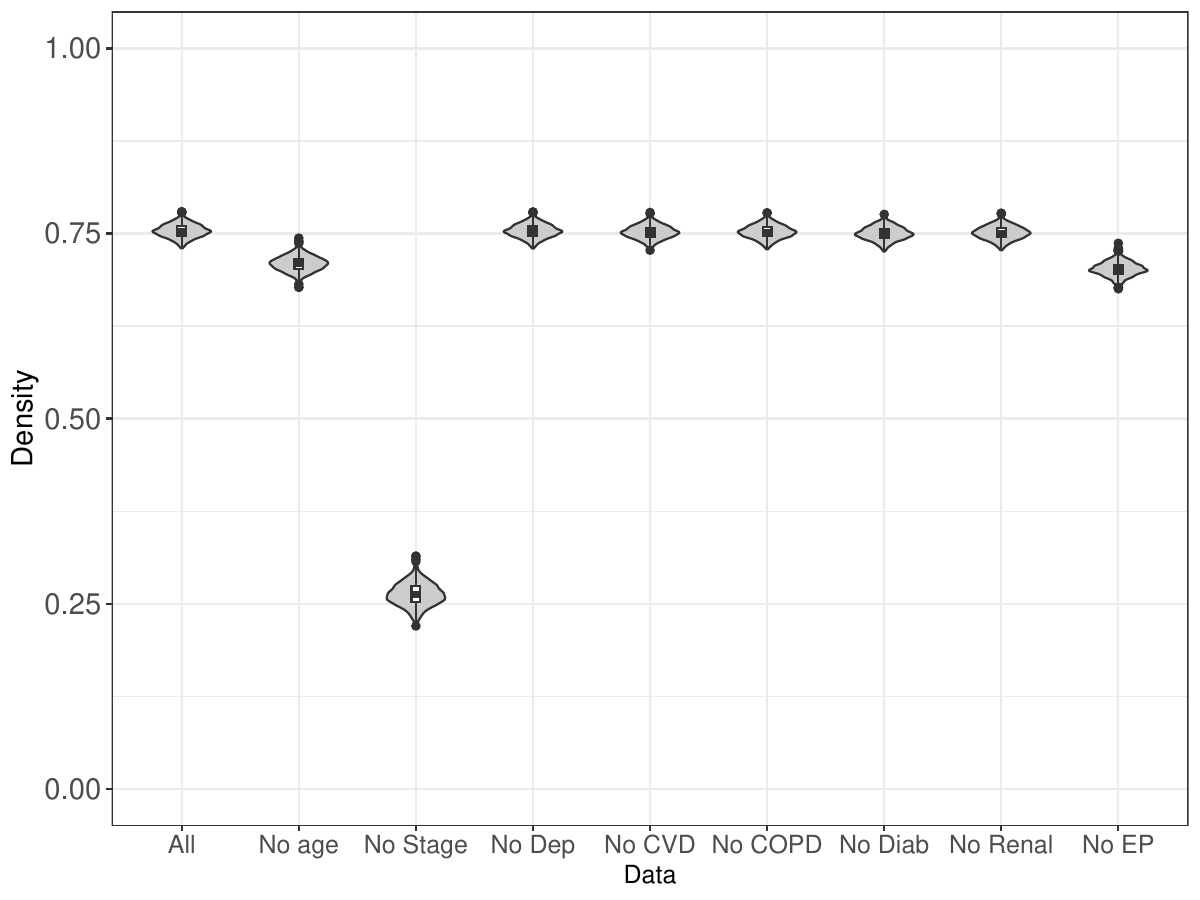} \\
   (a) & (b) \\
   \includegraphics[width=0.5\textwidth]{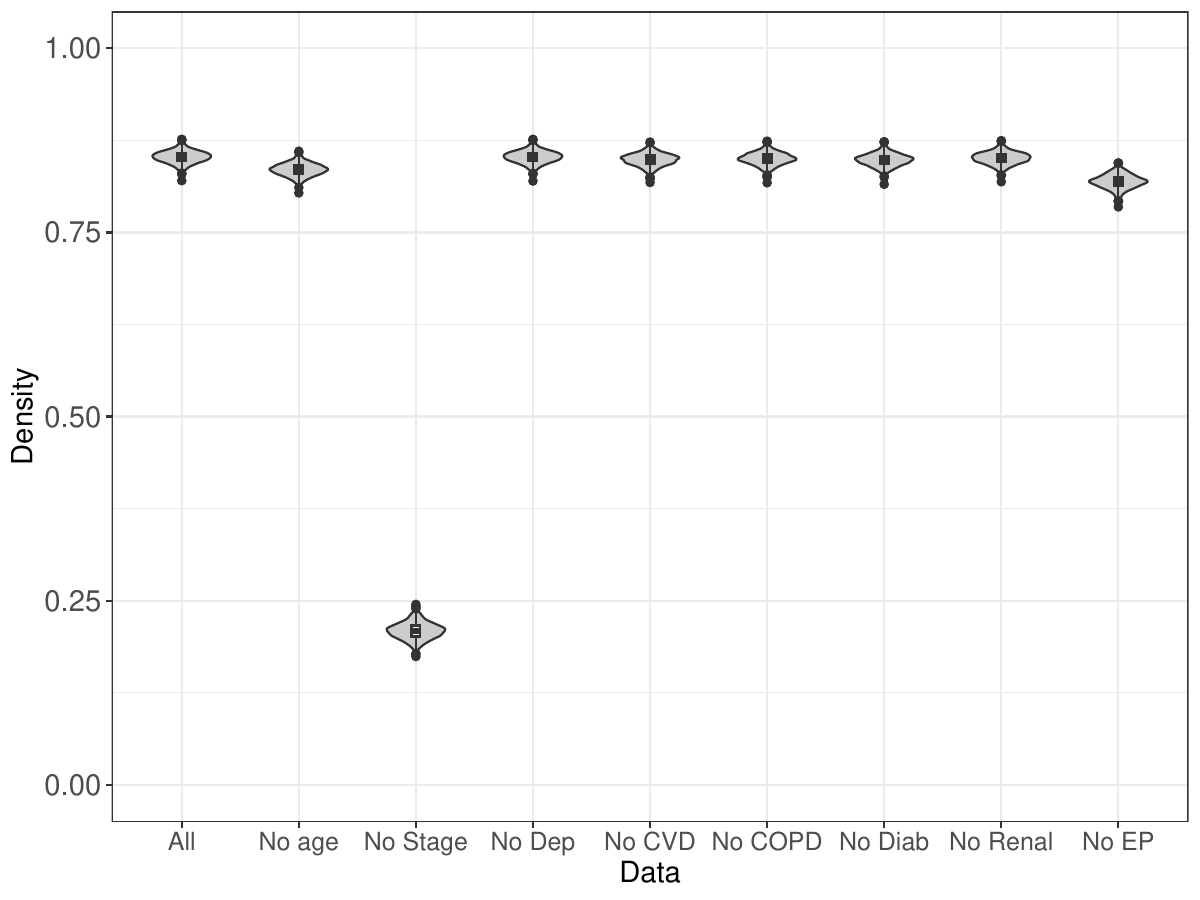} & 
   \includegraphics[width=0.5\textwidth]{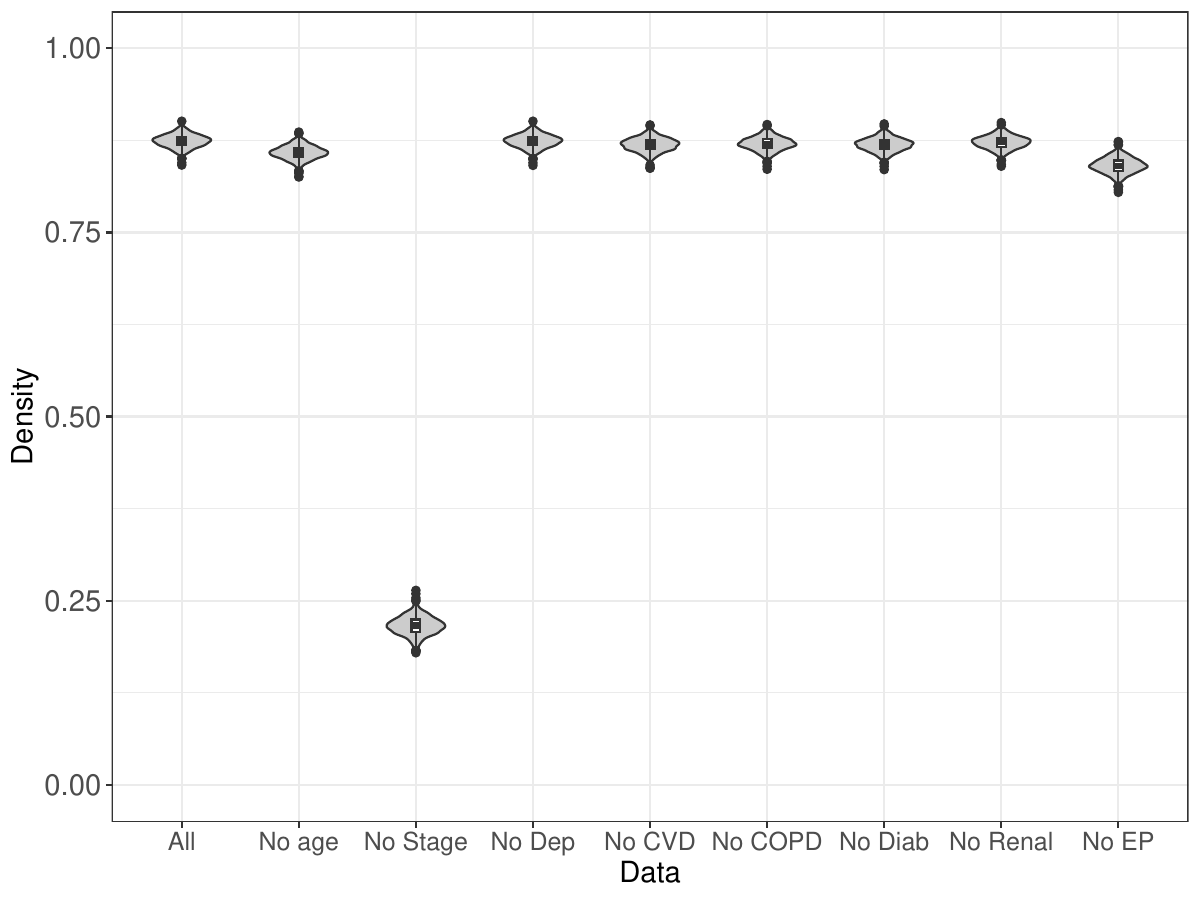} \\
   (c) & (d) \\
   \includegraphics[width=0.5\textwidth]{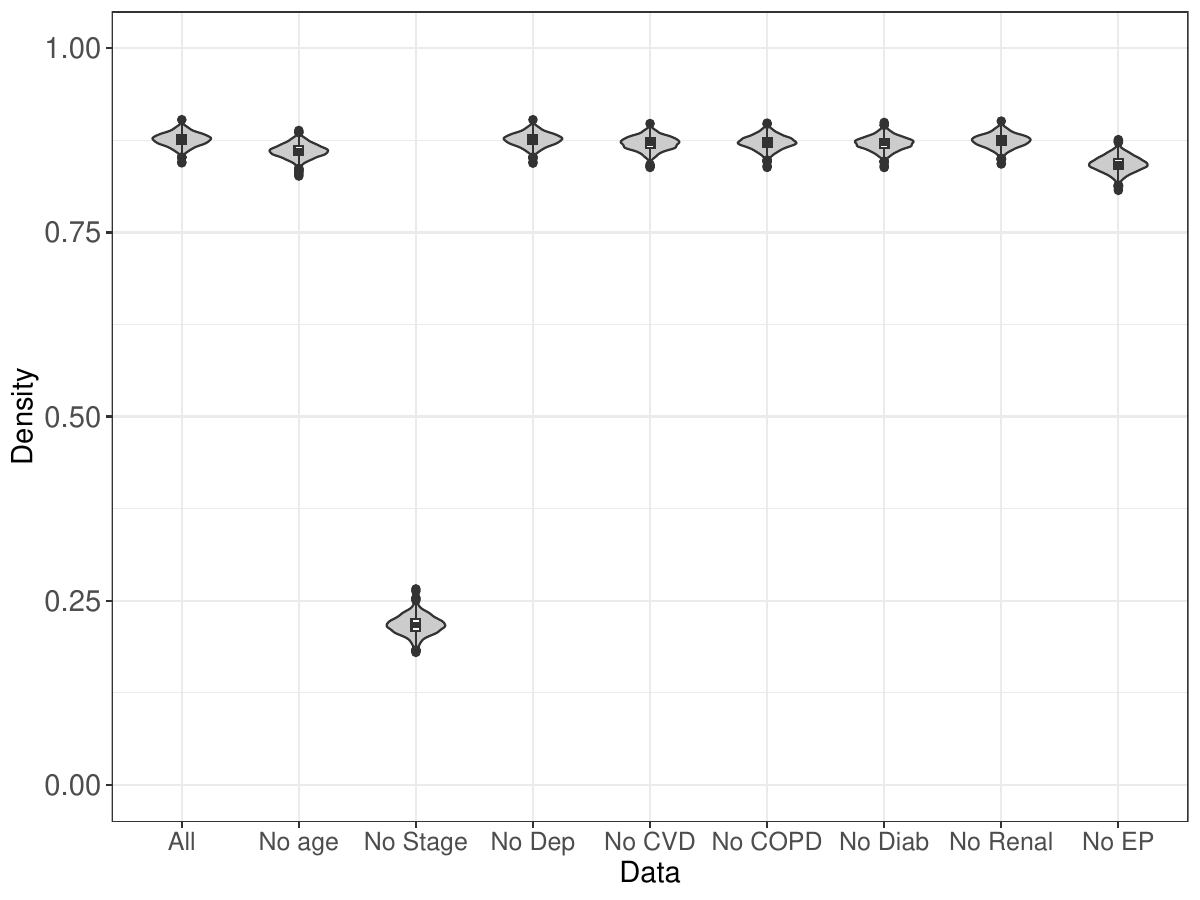} & \\
   (e) 
\end{tabular}
\caption{ Males colon cancer data. Variable importance ($R^2$) distributions at 2, 20, 40, 60, and 80 months. Lower summary $R^2$ deleting each variable indicates \emph{higher} variable importance.\label{fig:vimp_males}}
\end{figure}

The variable importance, measured through summary $R^2$ based on net survival, reflects the contribution of covariates in explaining the levels of net survival and how these contributions vary over the follow-up period (Figure \ref{fig:vimp_males}). At 2 months after diagnosis, the full model explains approximately $68\%$ of the variability in net survival, with the largest contributions coming from the EP status variable, followed by stage and age at diagnosis. Deprivation and comorbidities do not hold much variable importance in addition to the three aforementioned variables. After the acute diagnostic phase, the full-model's importance reaches 75\% and further increases to 87\%, thus explaining even more of the variability in individual net survival. The importance of stage at diagnosis is increased, and while emergency presentation and age at diagnosis remain important prognostic factors, they are not as discriminant as they were shortly after diagnosis. 

\begin{figure}
\centering
\begin{tabular}{cc}
   \includegraphics[width=0.5\textwidth]{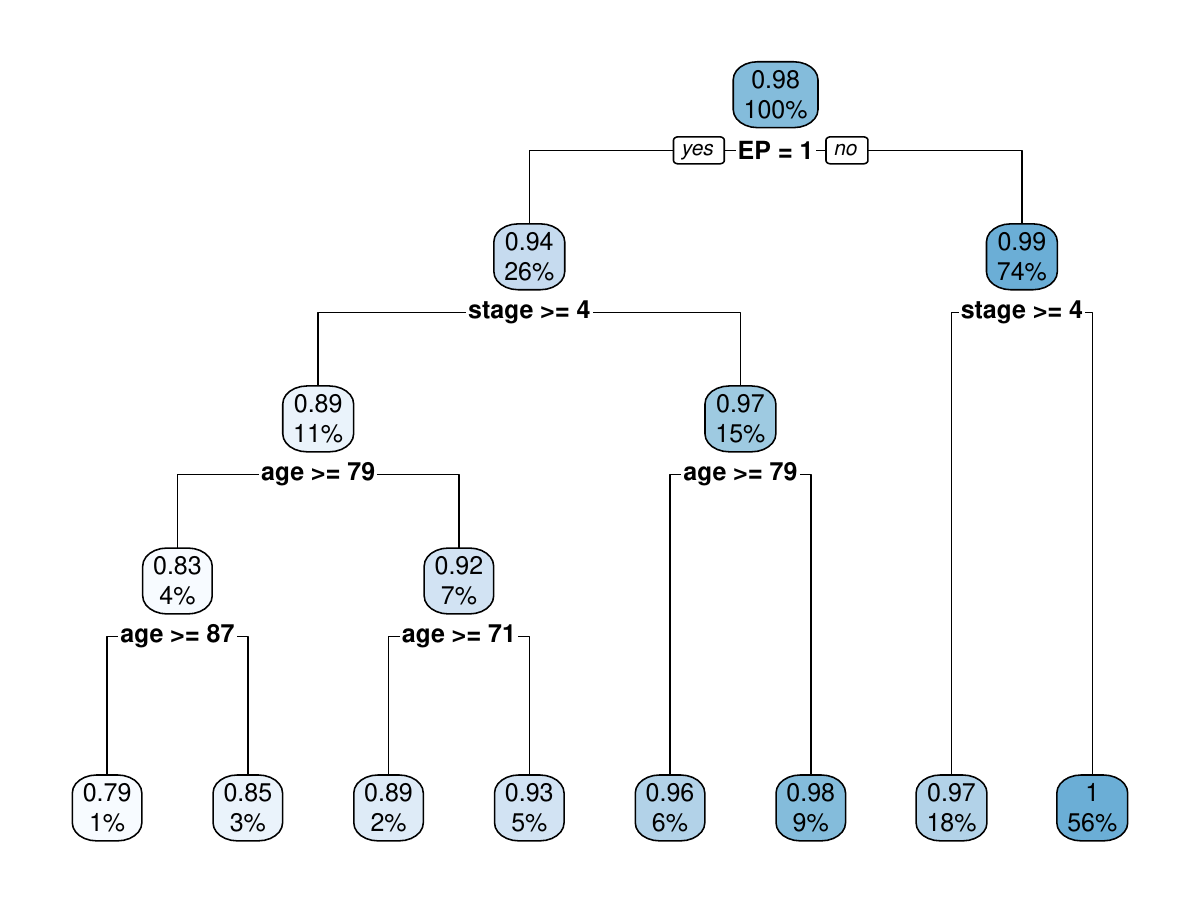}  &
   \includegraphics[width=0.5\textwidth]{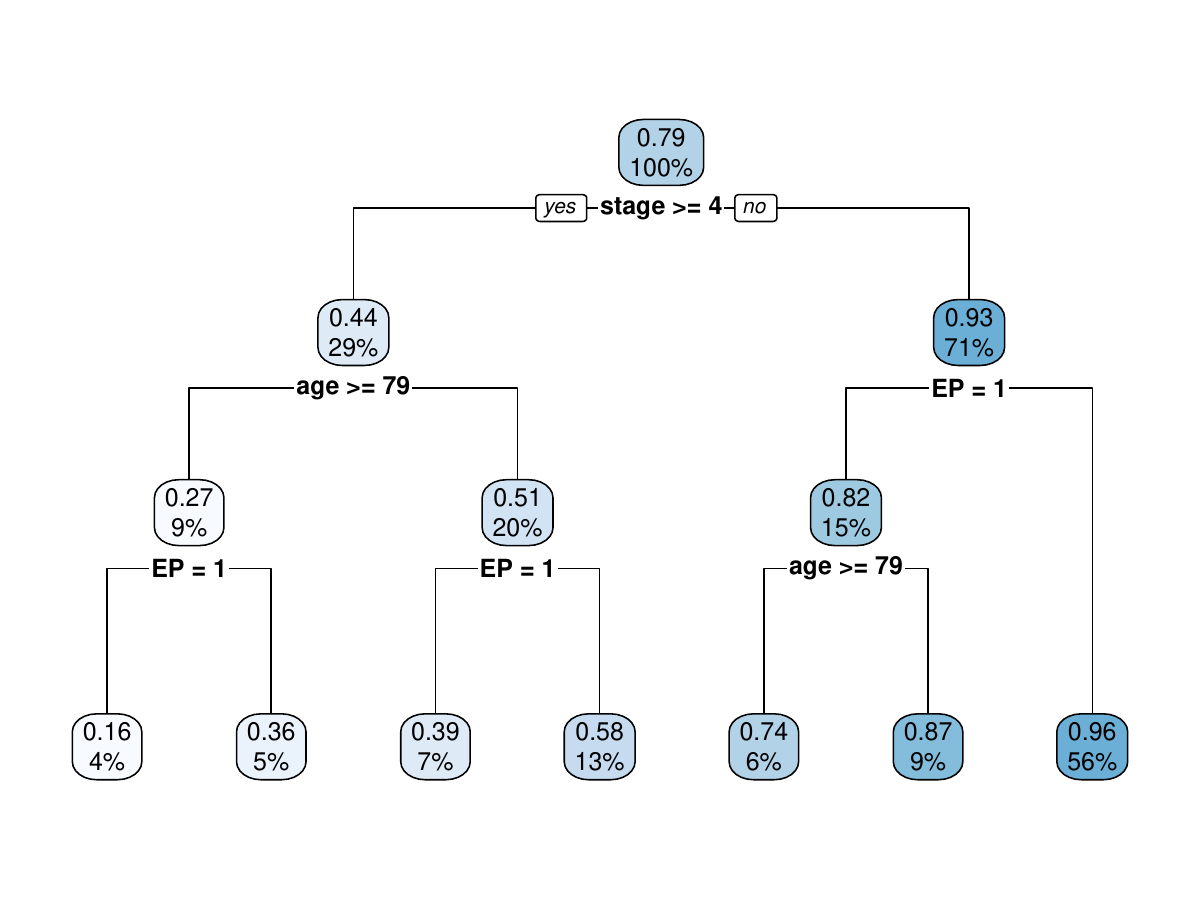} \\
   (a) & (b) \\
   \includegraphics[width=0.5\textwidth]{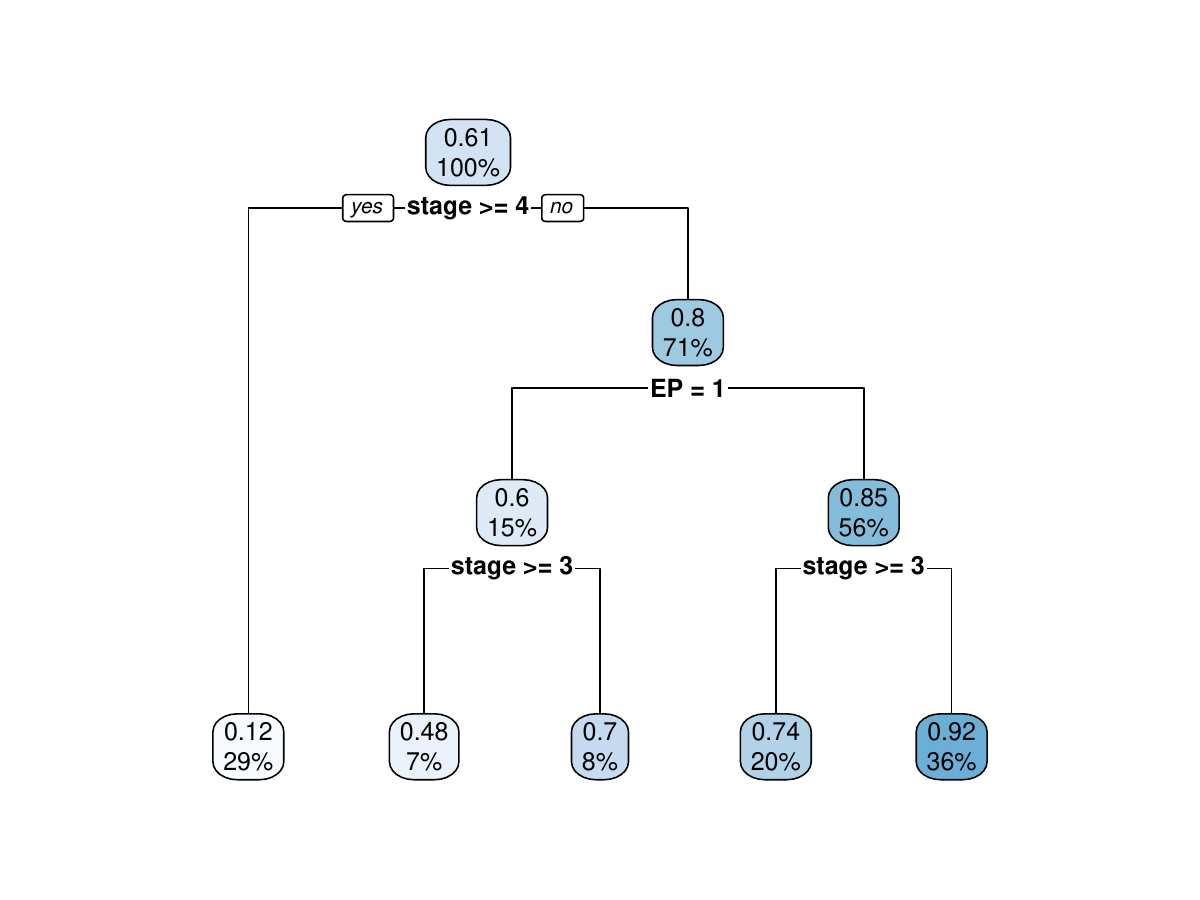} & 
   \includegraphics[width=0.5\textwidth]{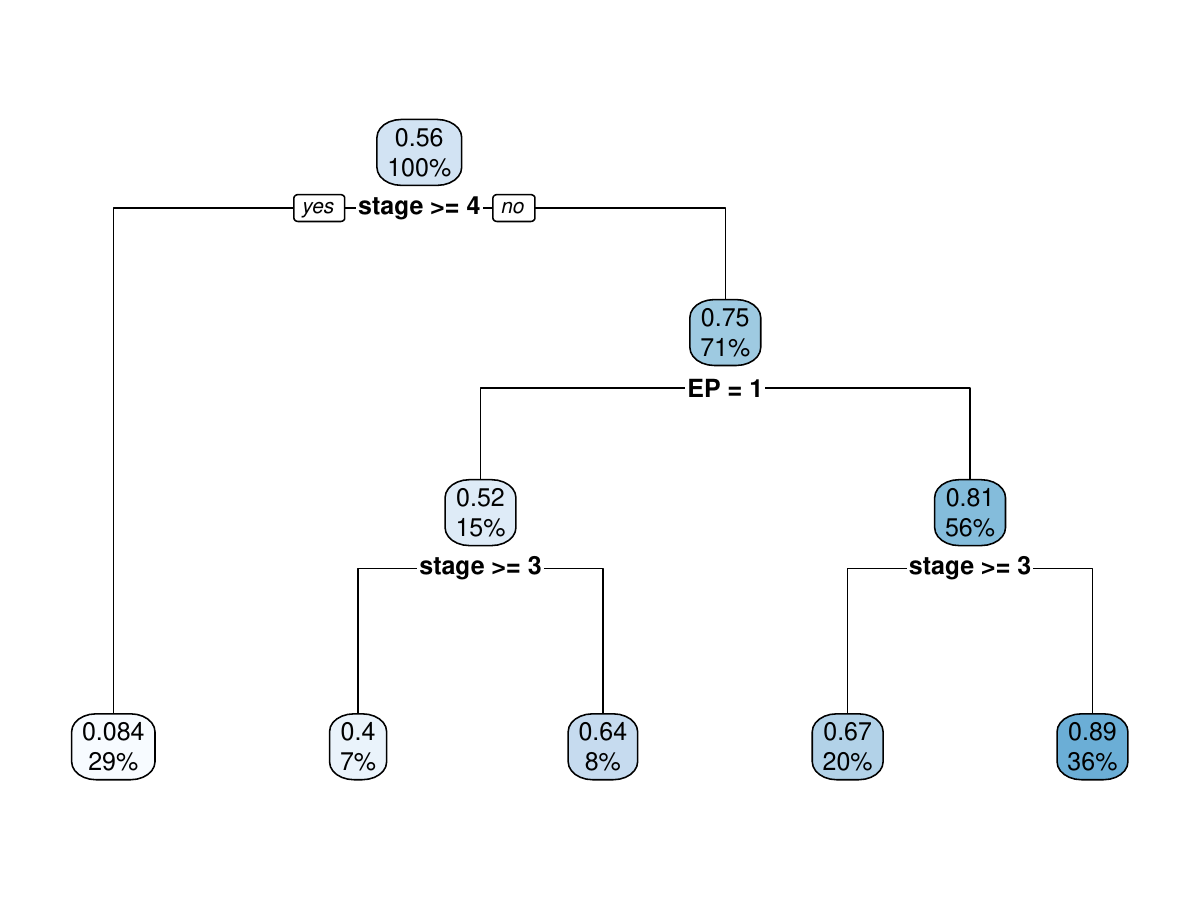} \\
   (c) & (d) \\
   \includegraphics[width=0.5\textwidth]{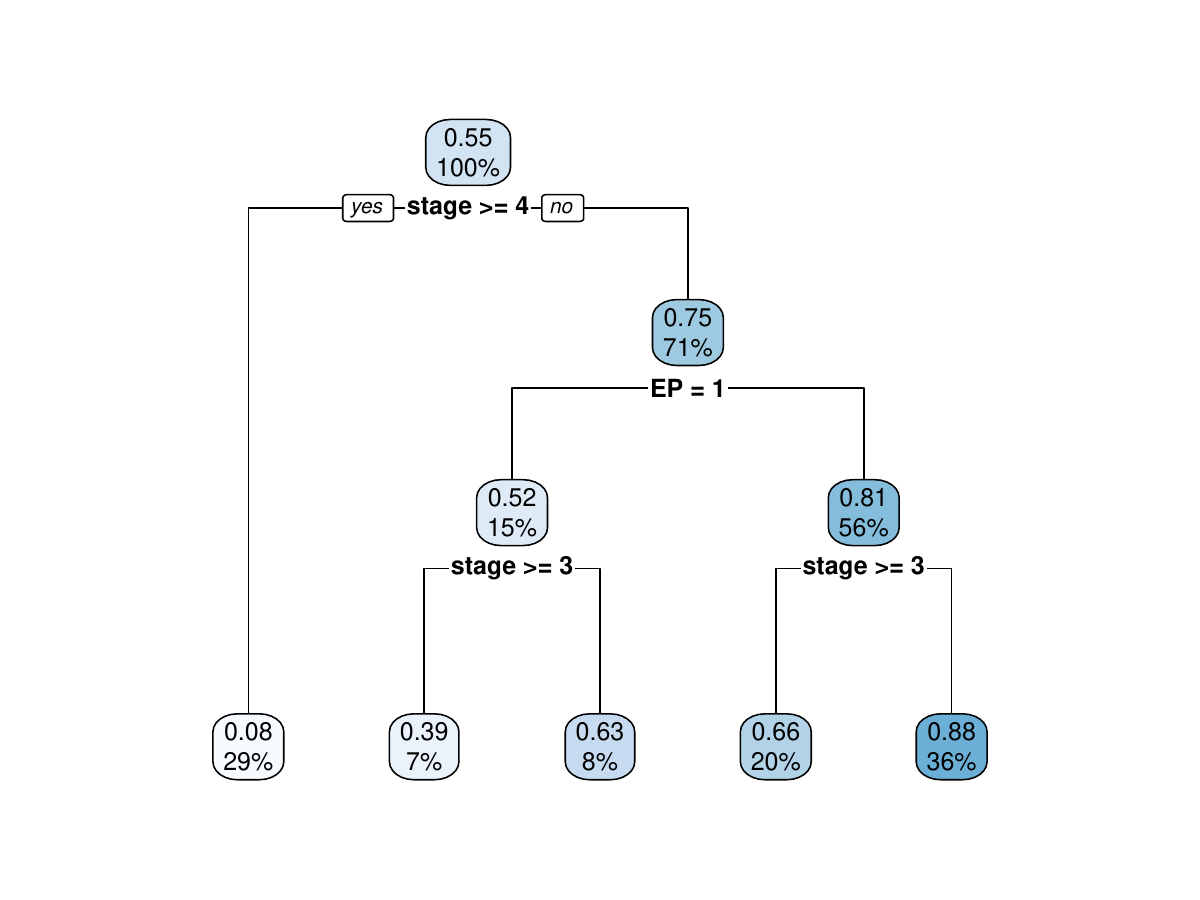} & \\
   (e) 
\end{tabular}
\caption{ Males colon cancer data. Variable importance (net survival) trees at 2, 20, 40, 60, and 80 months.}
\label{fig:tree_males}
\end{figure}

Another way to interpret the fit of the model to the data is through a decision tree, which can be used to find subgroups with different prognosis, as shown in Figure~\ref{fig:tree_males}. At 2 months post-diagnosis, EP status is the first variable to divide the cohort, with 26\% of patients having received an emergency diagnosis. The next division is based on stage, distinguishing 11\% of late-stage (IV, metastatic) from 15\% of early-stage (I–III) patients with $\text{EP} = 1$. Notably, for patients with $\text{EP} = 1$, age at diagnosis further differentiates the group within both stage categories, showing significant variation in net survival (from 79\% to 93\% in stage IV and 96\% to 98\% in stages I-III) across age groups. In contrast, patients with $\text{EP} = 0$ are only divided by stage, with net survival probabilities remaining close to 1. This suggests that an emergency diagnosis has a substantial impact on cancer prognosis across different age groups. However, an emergency diagnosis alone is unlikely to be the primary cause of lower survival probabilities; rather, it may indicate healthcare access challenges or suboptimal care provision. At 20 months post-diagnosis, stage is the first variable to divide the group into early and late stages, with further splits based on age and EP status. Notably, in this time frame, net survival in late-stage patients primarily depends on age, with significant differences observed between groups above and below 79 years. Additionally, EP continues to further divide these age groups, resulting in marked variations in net survival. In contrast, for early-stage patients, EP becomes the next variable to split the group, followed by age group divisions only for those with $\text{EP} = 1$. At later intervals (40, 60, and 80 months post-diagnosis), the decision trees simplify, with only metastatic disease and EP status strongly influencing prognosis. Overall, the BART summary provides valuable insights into the evolution of net survival over time, suggesting that EP may serve as a proxy for healthcare access, which in turn affects patient follow-up and prognosis.

Figure \ref{fig:rpartvimp_males} in the appendix presents the importance of each variable in explaining net survival (on a $[0,1]$ scale), as defined by the R package \texttt{rpart}, at different time points. In this case, variable importance quantifies how much each variable contributes to reducing impurity in the tree. The conclusions are similar to those drawn from the previous variable importance measure, but this method provides an additional, readily available tool for understanding and quantifying variable importance.

\section{Discussion}\label{sec:discussion}

In this article, we have developed a novel flexible Bayesian semi-parametric
approach for estimating excess hazard and associated net survival probability in
the relative survival setting using BART. More specifically, we have modeled the
baseline of the excess hazard function as piecewise exponential and modeled the
complex non-linear associations among the covariates using BART, allowing for
both proportional as well as non-proportional (time-varying) hazard assumptions,
and thus avoiding the need to specify closed-form variable transformations or
covariate interactions. Using a Bayesian ensemble-based model enables us to
effectively merge the predictive capabilities of machine learning with a robust
approach to quantifying uncertainty in our conclusions. Simulation results
validate that our proposed models are conservative in terms of the coverage
provided by credible intervals, even under model misspecification. This method
facilitates the sharing of information across relevant sub-populations, enabling
us to retain the flexibility of Nelson-Aalen type estimators while accommodating
continuous characteristics and sparsely observed sub-groups. 

Applicability of the proposed model has been demonstrated through a case study concerning the analysis of colon cancer data in England, along with the proposed posterior summarization techniques to quantify feature importance and interaction with time. Tumor stage at diagnosis is a well-established clinical prognostic factor, with more advanced stages being associated with fewer treatment options and lower survival probabilities. Additionally, extensive literature has examined disparities in cancer survival linked to deprivation. However, researchers should consider newly available variables, obtained through data linkage, to better identify the underlying causes of these inequalities. Given the flexibility of the COXNPH-BART model and its ability to consider all possible interactions, we were able to provide new insights into factors driving colon cancer net survival, up to 7 years after diagnosis. More specifically, the case study demonstrates the strong prognostic value of emergency presentations (EP), potentially as a proxy for healthcare access and usage. It also illustrates the additional insights epidemiologists and clinicians could gain by considering the effect of factors beyond stage at diagnosis and deprivation level. The outputs of the proposed modeling approach enable the analysis of time-varying variable importance, highlighting different phases in the patients' follow-up. Clinical factors, such as the four comorbidities included here and deprivation do not bear large importance in explaining net survival, throughout follow-up. It is important to emphasize that these conclusions apply at the population level, which is the focus of this work. Certain clinical prognostic factors may still be important at the individual level, but that is beyond the scope of our study. This work opens the door to analyzing and understanding the importance of prognostic factors for different cancer sites, as well as validating our findings in datasets from other countries.

\subsubsection*{Data Availability and Ethical Approval}

The data used for this study are the English National Cancer Registry data 1971–2016. Cancer registration data consist of patient information and as such, it is protected under the Data Protection Act 1998 and GDPR 2018 and cannot be made available as open data. Formal requests for release of cancer registration data can be made to the data custodian NHS Digital (NHSD). The researchers will have beforehand obtained all the ethical and statutory approvals required for accessing sensitive data. Detailed information on the application process can be found at https://digital.nhs.uk/ndrs/. The authors have obtained the ethical and statutory approvals required for this research (PIAG 1-05(c)/2007); ethical approval updated 6 April 2017 (REC 13/LO/0610).

\bibliographystyle{apalike}
\bibliography{mybib}

\appendix

\clearpage

\section{More Details on the Gibbs Samplers} \label{gibbs_ph}

In this section we provide more detail on the computations. First, the full algorithm for the PH model is given in Algorithm~\ref{alg:ph}. Next, we provide details on computations for the NPH model.

The Gibbs sampling updates for the $\lambda_b$'s are nearly identical, except that we instead take $B_b = \sum_i Z_{ib} e^{r(\bfX_i, b)}$ and it is no longer easy to compute the $B_b$'s recursively. Computations for $L(\Tree)$ and $\mu_{t\ell}$ are again similar. We now let $r(\bfX_i, b) - g(\bfX_i, b; \Tree_t, \sM_t) = \eta_{ib}$ and write the integrated likelihood as
\begin{align*}
    L(\Tree) &= \prod_\ell \int \prod_{(i,b) \leadsto \ell}
      \lambda_{b}^{d_i 1(b_i = b)} 
      \exp\left\{d_i 1(b_i = b) (\eta_{ib} + \mu)
      - e^{\mu} Z_{ib} \lambda_b e^{\eta_{ib}}\right\}
      \\&\qquad 
      \times \frac{b^a}{\Gamma(a)} \exp(a\mu - b e^\mu) \ d\mu,
\end{align*}
which evaluates to 
\begin{align*}
    L(\Tree) \propto \left\{\prod_\ell 
      \frac{\Gamma(a + A_\ell)}{(b + B_\ell)^{a + A_\ell}} \right\}
        \times \frac{b^a}{\Gamma(a)},
\end{align*}
where $A_\ell = \sum_{(i,b) \leadsto \ell} d_i 1(b_i = b)$ and $B_\ell = 
\sum_{(i,b) \leadsto \ell} Z_{ib} \lambda_b e^{\eta_{ib}}$; similarly, the 
update for the leaf predictions is given by $\mu_{t\ell} \sim \log \Gam(a + 
A_\ell, b + B_\ell)$. The full details of this algorithm are given in Algorithm~\ref{alg:nph}.

\begin{algorithm}
\caption{Bayesian Backfitting for the Proportional Hazards Model\label{alg:ph}}
\KwInput{Data $\Data = \{(Y_i, \delta_i, \bfX_i, \lambda_P(\text{age}_i + Y_i \mid \bfW_i))\}_{i=1}^N$}
\BlankLine
\KwOutput{Posterior samples of $\{\Tree_m, \sM_m\}_{m=1}^M$ and $\{\lambda_b\}_{b=1}^B$}
\BlankLine
{Initialize the trees $\{\Tree_m\}_{m=1}^M$ and their leaf parameters $\{\sM_m\}_{m=1}^M$}
\BlankLine

\For{each iteration}{
    \For{\textnormal{each individual} $i = 1, \dots, N$}{
        Update $d_i \sim \Bernoulli\left\{\frac{\delta_i \lambda_{b_i} e^{r(\bfX_i)}}{\lambda_P(Y_i \mid \bfW_i) + \lambda_{b_i} e^{r(\bfX_i)}}\right\}$ \;
    }
    \For{\textnormal{each tree} $m = 1, \dots, M$}{
        Propose a new tree structure $\Tree \sim Q(\Tree_m \to \Tree)$\;
        Compute $A_\ell = \sum_{i: X_i \leadsto \ell} d_i$ and $B_\ell = \sum_{b, i: X_i \leadsto \ell} Z_{ib} \, \lambda_b \, e^{\eta_i}$ for each leaf $\ell$ in $\Tree$ and $\Tree_m$\;
        Compute $L(\Tree)$ and $L(\Tree_m)$ as:
        \begin{align*}
            L(\Tree) &\propto \prod_{\ell} \frac{\Gamma(a + A_\ell)}{(b + B_\ell)^{a + A_\ell}} \times \frac{b^a}{\Gamma(a)}.
        \end{align*}
        
        Set $\Tree_m = \Tree$ with probability $A = \min\left\{\frac{\pi(\Tree)\,L(\Tree)\,Q(\Tree \to \Tree_m)}{\pi(\Tree_m)\,L(\Tree_m)\,Q(\Tree_m \to \Tree)}, 1 \right\}$\;
        
        \For{\text{each leaf} $\ell \in \Leaves(\Tree_m)$}{
            Sample $\mu_{m\ell} \sim \log \Gam(a + A_\ell, b + B_\ell)$\;
        }
    }

    \For{\text{each bin} $b = 1, \dots, B$}{
        Compute $A_b = \sum_{i: b_i = b} d_i$\;
        Compute $B_b = \sum_i Z_{ib} e^{r(\bfX_i)}$ using:
        \begin{align*}
            B_{b+1} = \frac{t_{b+1} - t_b}{t_b - t_{b-1}} \left(B_b - \sum_{i: Y_i \in [t_{b-1}, t_{b+1})} Z_{ib} e^{r(\bfX_i)}\right) + \sum_{i: Y_i \in [t_{b}, t_{b+1})} Z_{i(b+1)} e^{r(\bfX_i)}
        \end{align*}
        Sample $\lambda_b \sim \Gam(a_\lambda + A_b, b_\lambda + B_b)$\;
    }
}
\end{algorithm}

\begin{algorithm}
\caption{Bayesian Backfitting for the Non-Proportional Hazards (NPH) Model\label{alg:nph}}
\KwInput{Data $\Data = \{(Y_i, \delta_i, \bfX_i, \lambda_P(\text{age}_i + Y_i \mid \bfW_i))\}_{i=1}^N$}
\BlankLine
\KwOutput{Posterior samples of $\{\Tree_m, \sM_m\}_{m=1}^T$ and $\{\lambda_b\}_{b=1}^B$}
\BlankLine
{Initialize the trees $\{\Tree_m\}_{m=1}^M$ and their leaf parameters $\{\sM_m\}_{m=1}^M$}
\BlankLine

\For{\textnormal{each iteration}}{
    \For{\textnormal{each individual} $i = 1, \dots, N$}{
        Update $d_i \sim \Bernoulli\left\{\frac{\delta_i \lambda_{b_i} e^{r(\bfX_i, b_i)}}{\lambda_P(Y_i \mid \bfW_i) + \lambda_{b_i} e^{r(\bfX_i, b_i)}}\right\}$ \;
    }
    \For{\textnormal{each tree} $m = 1, \dots, M$}{
        Propose a new tree structure $\Tree \sim Q(\Tree_m \to \Tree)$\;
        Compute $\eta_{ib} = r(\bfX_i, b) - g(\bfX_i, b; \Tree_m, \sM_m)$ for all $(i,b)$\;
        
        Compute $A_\ell = \sum_{(i,b) \leadsto \ell} d_i 1(b_i = b)$ and $B_\ell = \sum_{(i,b) \leadsto \ell} Z_{ib} \lambda_b e^{\eta_{ib}}$\;
        
        Compute $L(\Tree)$ and $L(\Tree_m)$ as:
        \begin{align*}
            L(\Tree) &\propto \prod_{\ell} \frac{\Gamma(a + A_\ell)}{(b + B_\ell)^{a + A_\ell}} \times \frac{b^a}{\Gamma(a)}.
        \end{align*}
        
        Set $\Tree_m = \Tree$ with probability $A = \min\left\{\frac{\pi(\Tree)\,L(\Tree)\,Q(\Tree \to \Tree_m)}{\pi(\Tree_m)\,L(\Tree_m)\,Q(\Tree_m \to \Tree)}, 1 \right\}$\;
        
        \For{\textnormal{each leaf} $\ell \in \Leaves(\Tree_m)$}{
            Sample $\mu_{m\ell} \sim \log \Gam(a + A_\ell, b + B_\ell)$\;
        }
    }

    \For{\textnormal{each bin} $b = 1, \dots, B$}{
        Compute $A_b = \sum_{i: b_i = b} d_i 1(b_i = b)$\;
        Compute $B_b = \sum_i Z_{ib} e^{r(\bfX_i, b)}$\;
        Sample $\lambda_b \sim \Gam(a_\lambda + A_b, b_\lambda + B_b)$\;
    }
}
\end{algorithm}

\clearpage

\section{Additional results for the real data application}\label{sec:appendix_application}

%%%%%%%%%%%%%%%%%%%%%%%%%%%%%%%%%%%%%%%%%%%%%%%%%%%%%%%%%%%%%%%%%%%%%%%%%%%%%%%%
%----------------------------------------------------------
\subsection{Comorbidity definitions}
%----------------------------------------------------------
Table \ref{tab:comorbidities} below presents the definition of the comorbidities used in our case study. These definitions are based on ICD-10 codes and the algorithm proposed in \cite{maringe:2017} to derive information on the presence of comorbidities in cancer patients in the UK.
\begin{table}[ht!]
\centering
\begin{tabular}{lc}
\textbf{Comorbidity} & \textbf{ICD-10 codes}  \\ \hline \hline
\textbf{Cardiovascular disease (CVD)} & I21, I22, I252, I70, \\ 
 & I71, I731, I738, I739, \\
 & I771, I790, I792, K551, \\
 & K558, K559, Z958, Z959, \\
 & G45, G46, H340, I60-I69  \\ \hline
\textbf{Chronic Obstructive Pulmonary} & I278, I279, J40-J47, J60-J67, \\
\textbf{Disease (COPD)} & J684, J701, J703 \\ \hline
\textbf{Diabetes} & E10-E14   \\ \hline
\textbf{Renal disease} & I120, I131, N032-N037, \\ 
& N052-N057, N18-N19, N250, \\
& Z490-Z492, Z940, Z992   \\ 
\hline \hline
\end{tabular}
\caption{Comorbidity definitions.}
\label{tab:comorbidities}
\end{table}

%----------------------------------------------------------
\subsection{Male colon cancer patients}
%----------------------------------------------------------
This section provides supplementary figures for the results concerning male colon cancer patients, as presented in the main document. 

\begin{figure}[h!]
\centering
\includegraphics[width=0.5\textwidth]{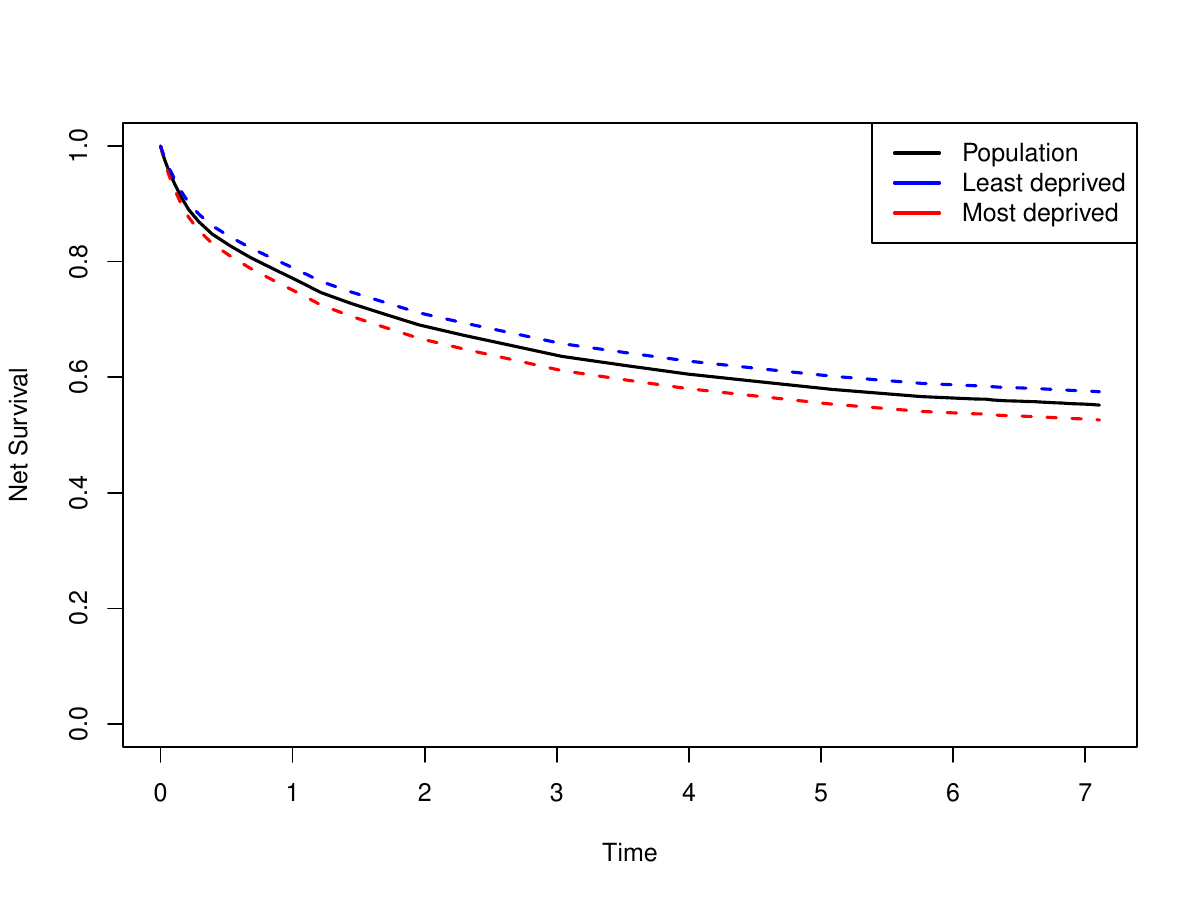} 
\caption{ Males colon cancer data. Net survival. Most deprived vs least deprived vs population}
\label{fig:nsdep_males}
\end{figure}

\begin{figure}[h!]
\centering
\begin{tabular}{c c c}
\includegraphics[width=0.3\textwidth]{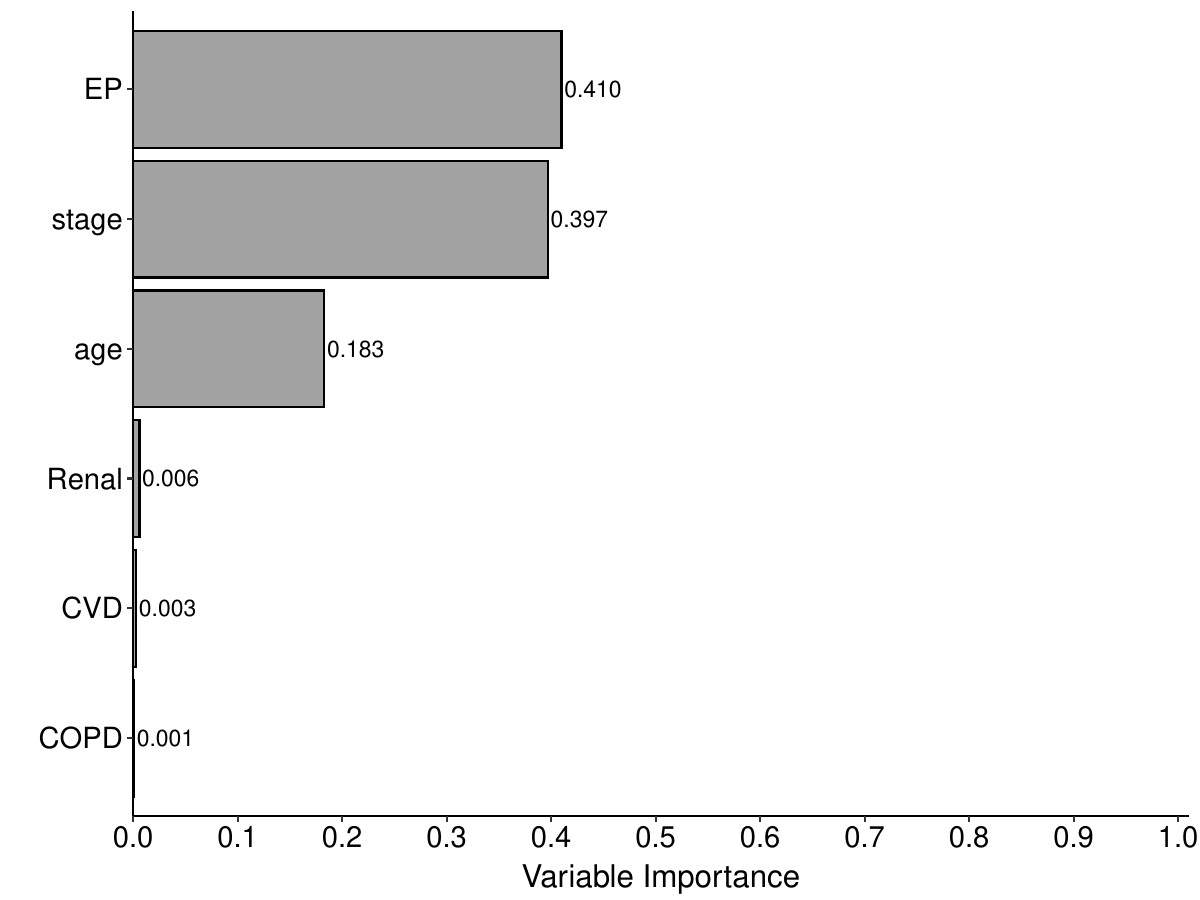} & 
\includegraphics[width=0.3\textwidth]{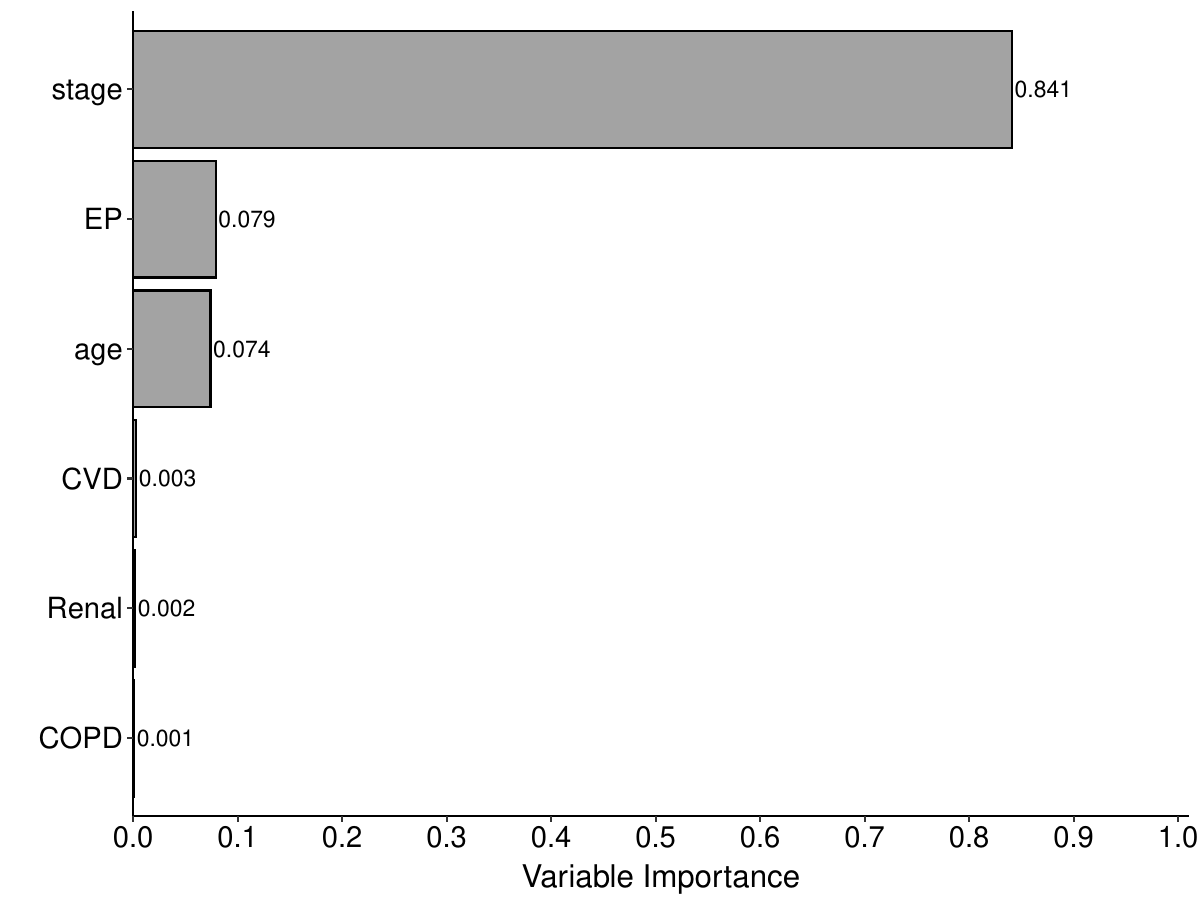} & 
\includegraphics[width=0.3\textwidth]{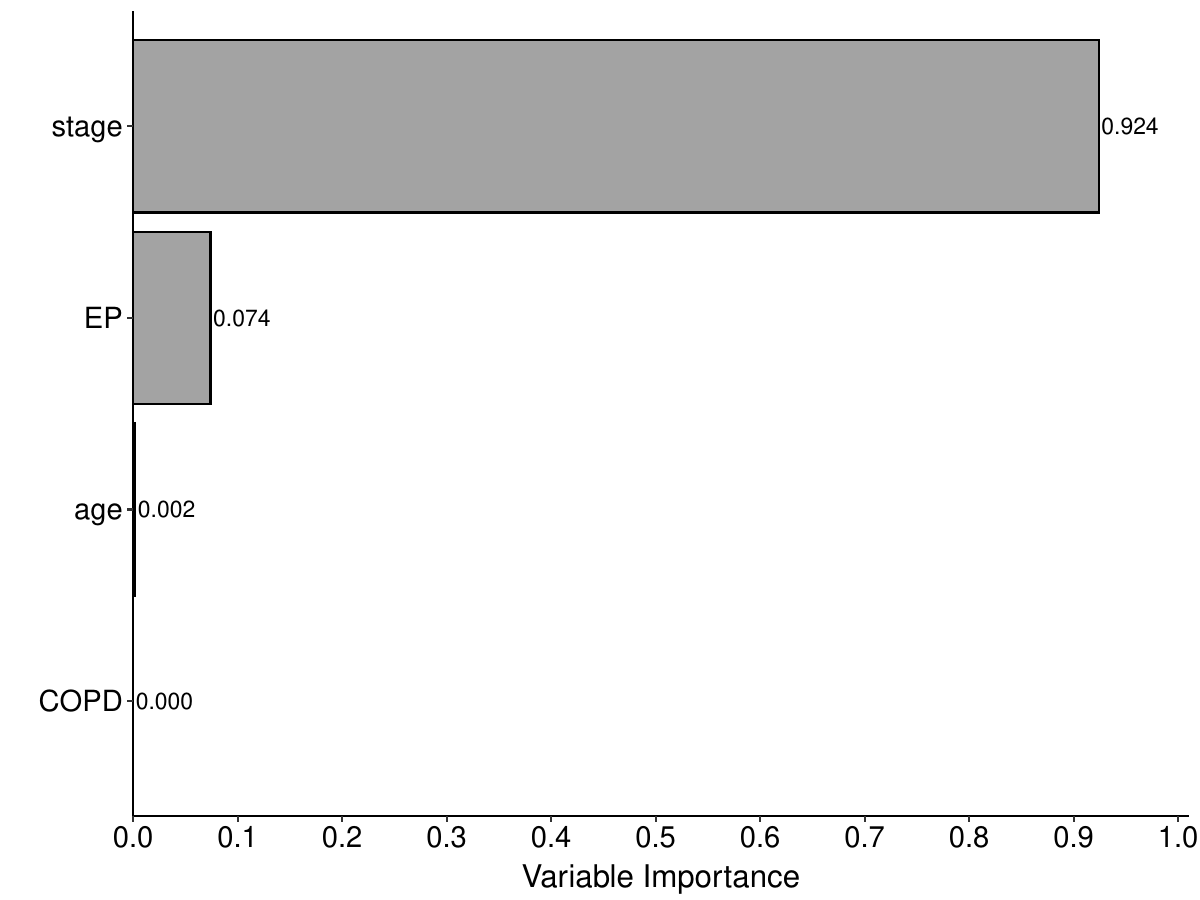}\\
 (a) & (b) & (c) \\
 \includegraphics[width=0.3\textwidth]{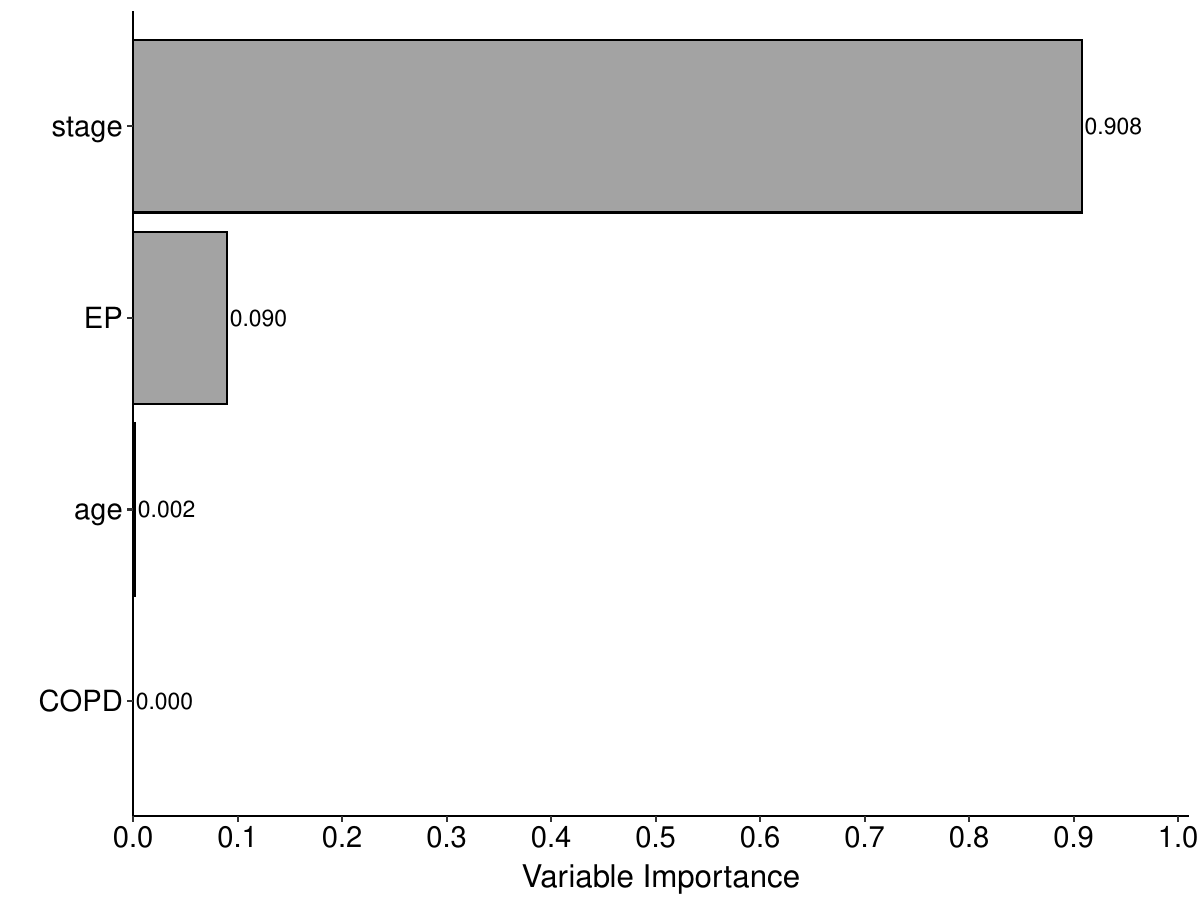} & 
\includegraphics[width=0.3\textwidth]{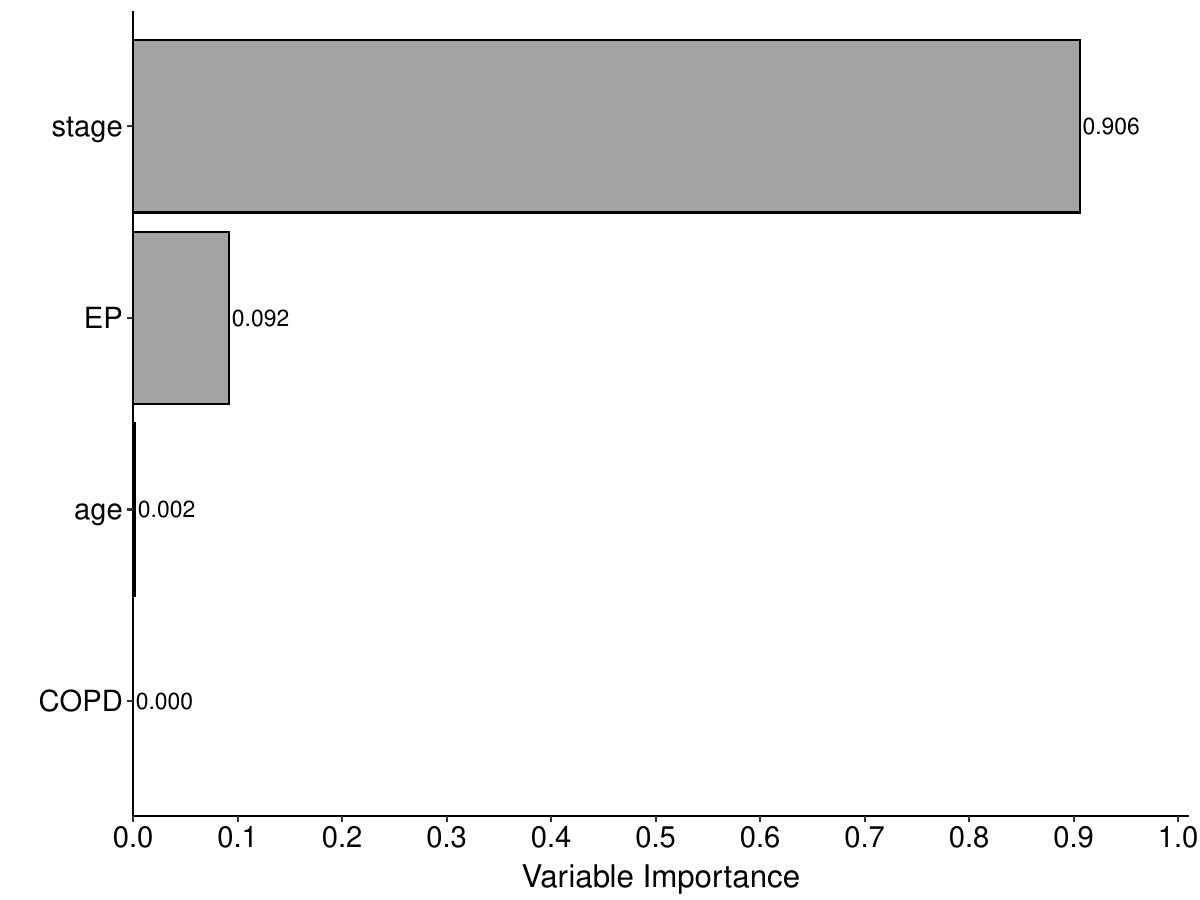} & 
\\
 (d) & (e) & 
 \end{tabular}
\caption{ Males colon cancer data. Variable importance from \texttt{rpart} at 2, 20, 40, 60, and 80 months.}
\label{fig:rpartvimp_males}
\end{figure}

\pagebreak
%----------------------------------------------------------
\subsection{Female colon cancer patients}
%----------------------------------------------------------

We fitted the same BART models to the female colon cancer data than we did for male, and present here their specific results and conclusions.
The ELPD of the COXNPH-BART model was found to be substantially higher ($-8,458.1$) compared to the COXPH-BART model ($-8,716.4$), indicating a better fit to the data. Below, we present an interpretation of the results and new findings derived from the COXNPH-BART model.

\begin{figure}[h!]
\centering
\includegraphics[width=0.5\textwidth]{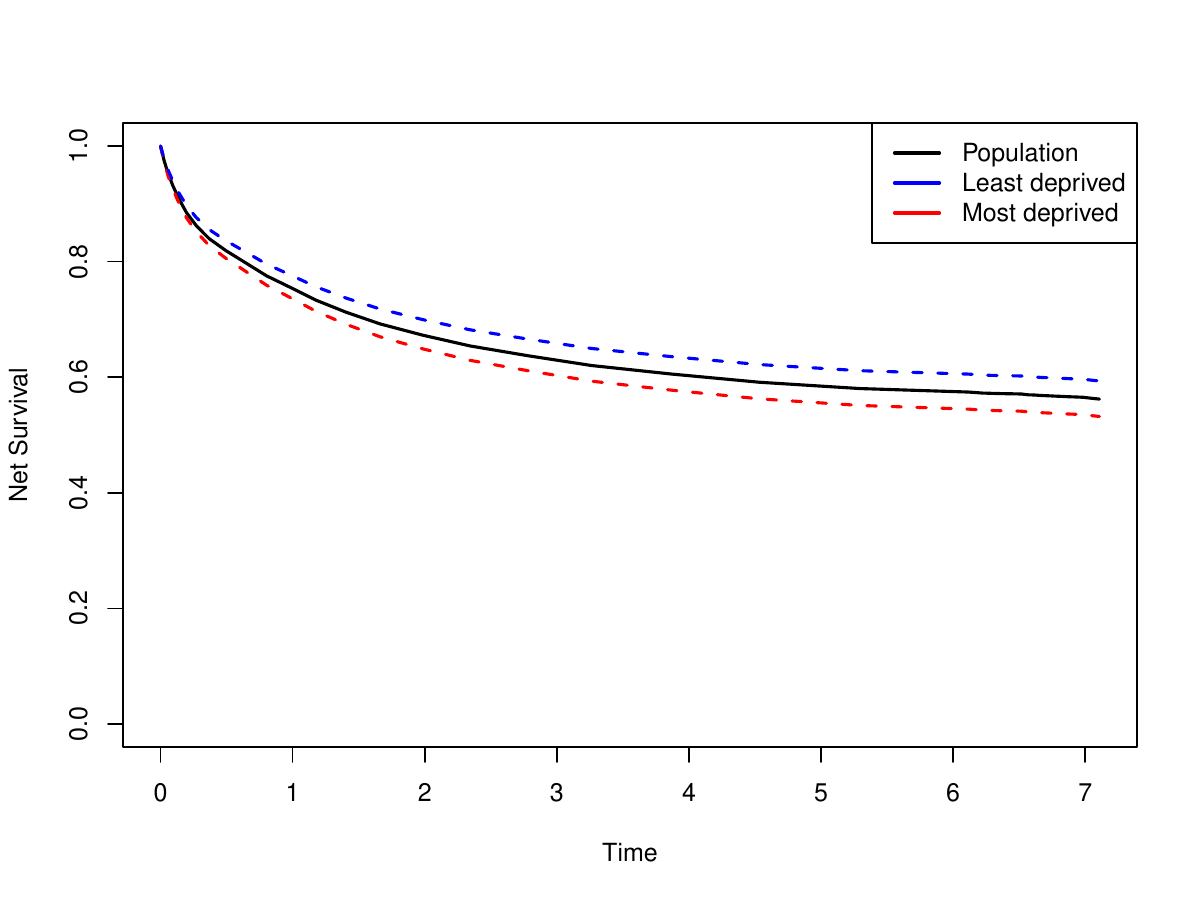} 
\caption{ Females colon cancer data. Net survival. Most deprived vs least deprived vs population}
\label{fig:nsdep_females}
\end{figure}

\begin{figure}[h!]
\centering
\begin{tabular}{c c c}
\includegraphics[width=0.3\textwidth]{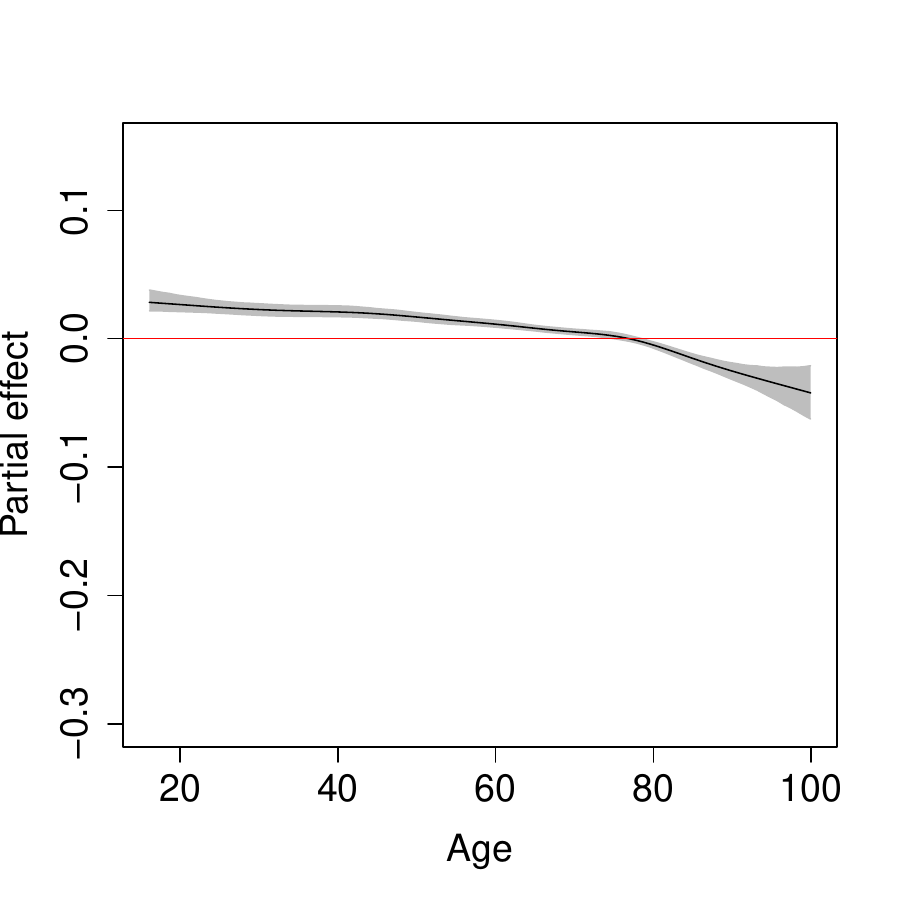} & 
\includegraphics[width=0.3\textwidth]{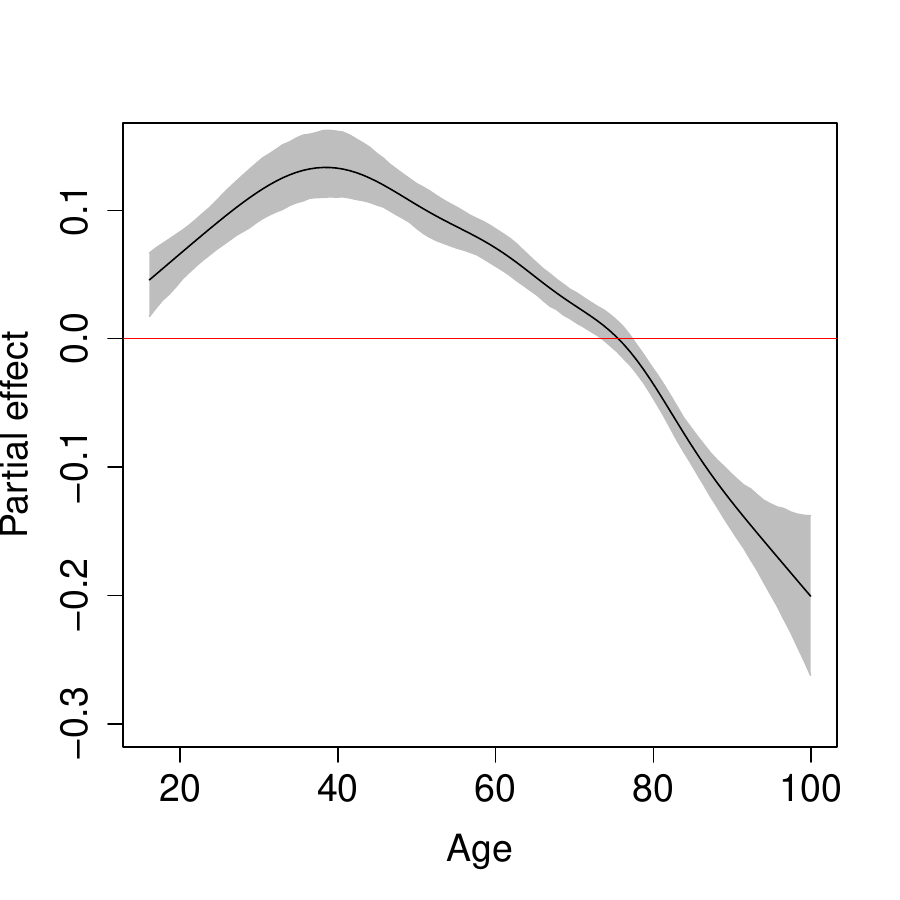} & 
\includegraphics[width=0.3\textwidth]{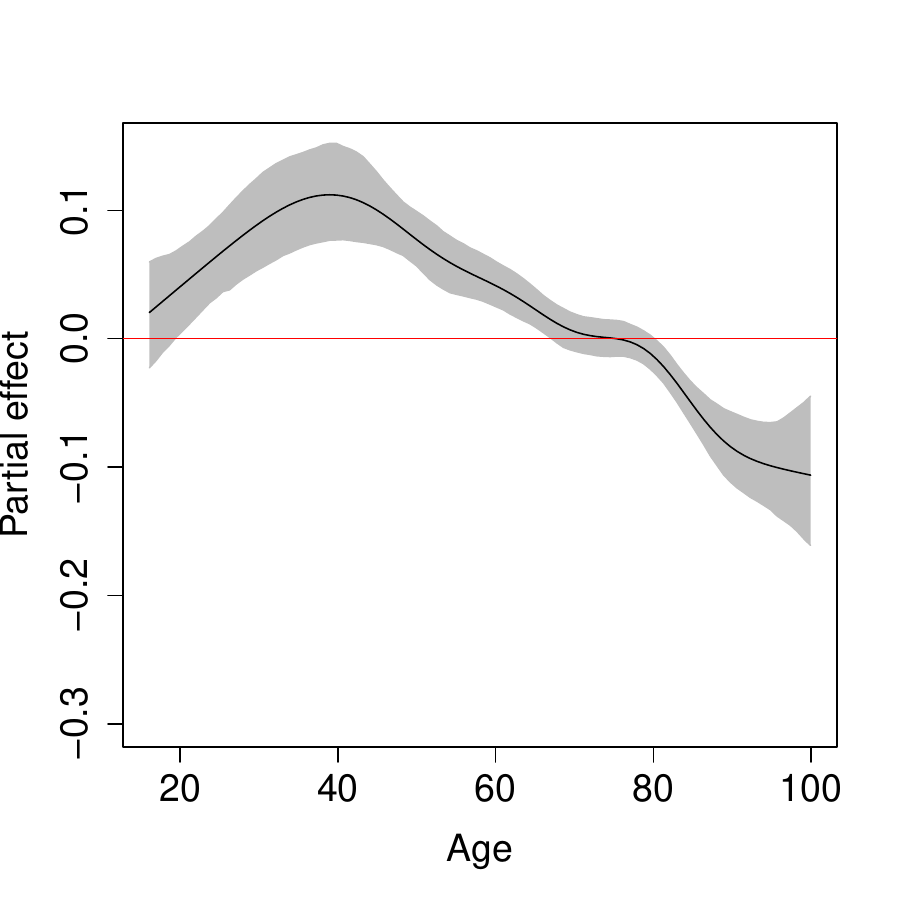}\\
 (a) & (b) & (c) \\
 \includegraphics[width=0.3\textwidth]{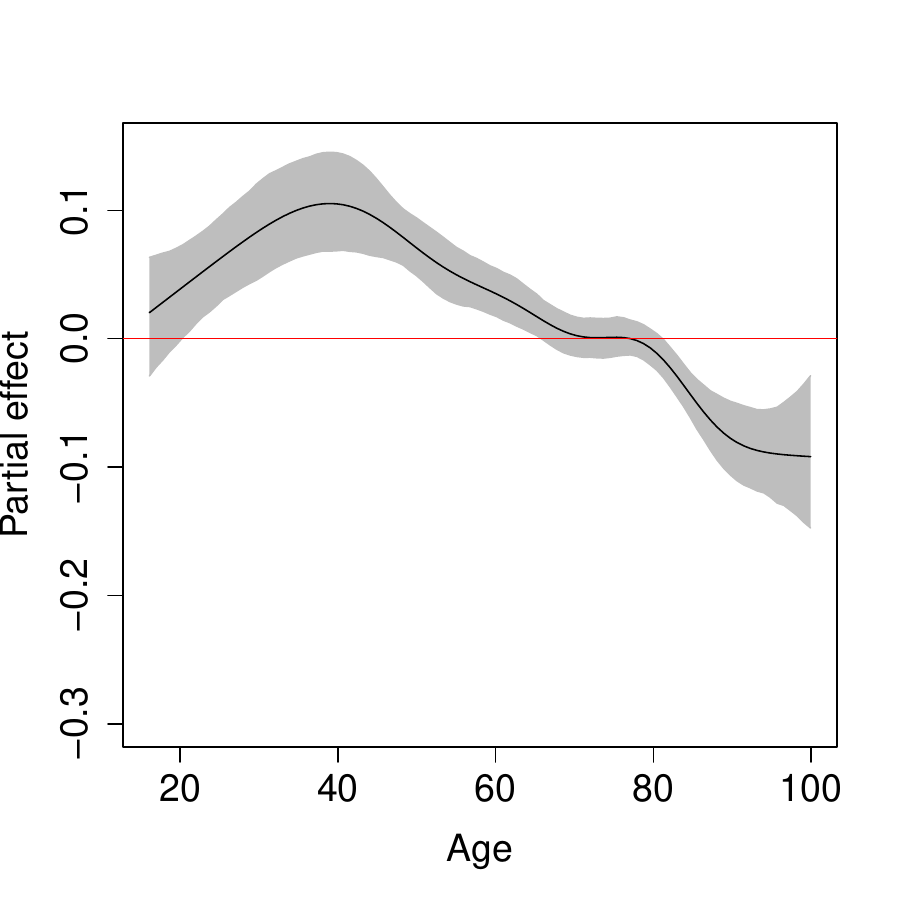} & 
\includegraphics[width=0.3\textwidth]{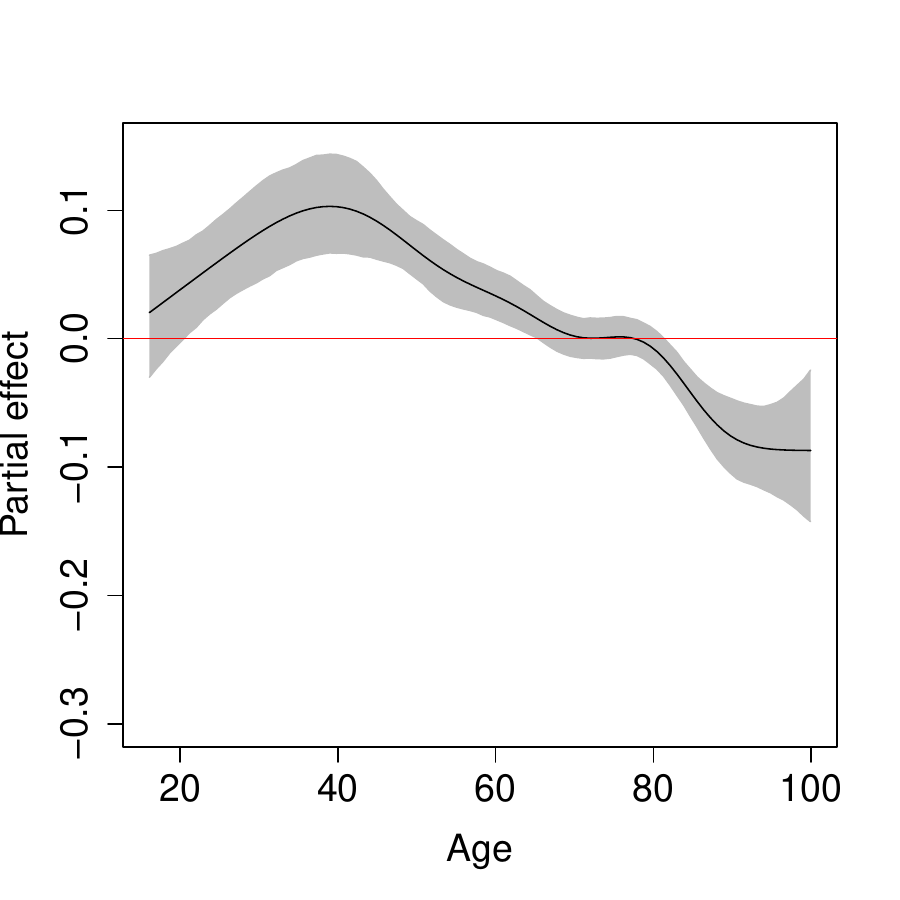} & 
\\
 (d) & (e) & 
 \end{tabular}
\caption{ Females colon cancer data: Partial effect of age at 2, 20, 40, 60, and 80 months.}
\label{fig:peff_age_females}
\end{figure}

Figure \ref{fig:nsdep_females} also reveals inequalities in net survival by deprivation for female cancer patients.
Figure \ref{fig:peff_age_females} shows the partial effect of age (as recorded at diagnosis) on net survival at different time points (bins 2, 20, 40, 60 and 80 months). The conclusions are similar to those drawn for male patients, with age having a non-linear and time-varying effect, and the largest amplitude of the partial effect of age observed at 20 months after diagnosis, and then reducing slightly as follow-up time elapses.

\begin{figure}[h!]
\centering
\begin{tabular}{c c c}
\includegraphics[width=0.3\textwidth]{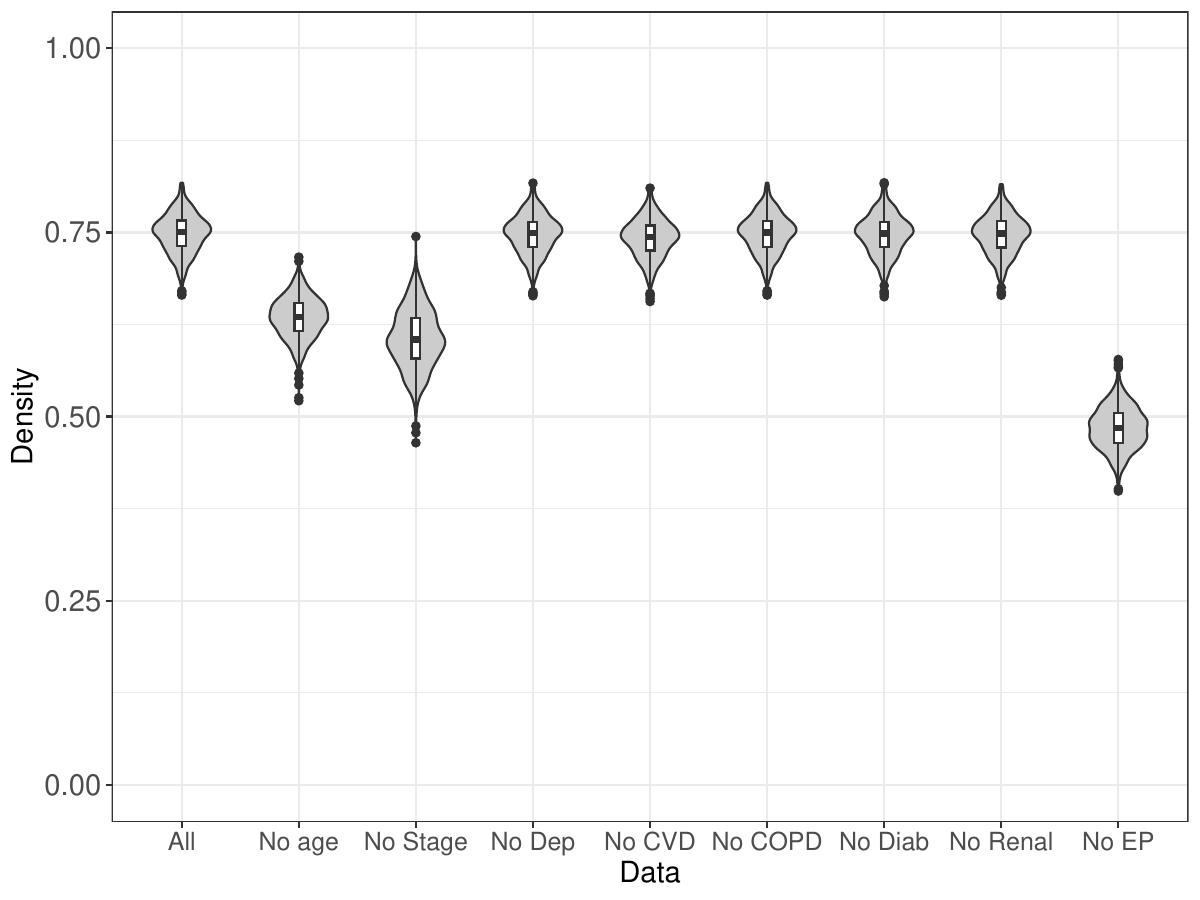} & 
\includegraphics[width=0.3\textwidth]{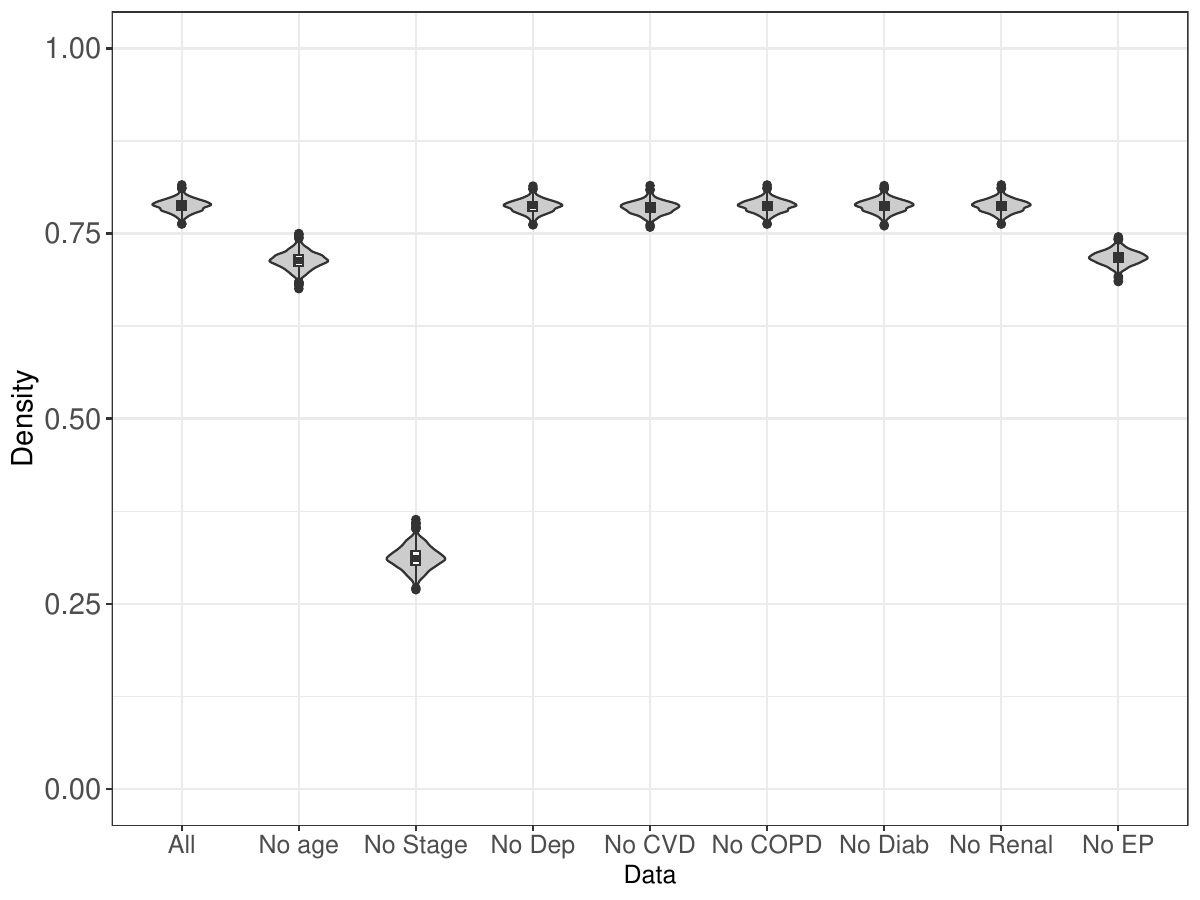} & 
\includegraphics[width=0.3\textwidth]{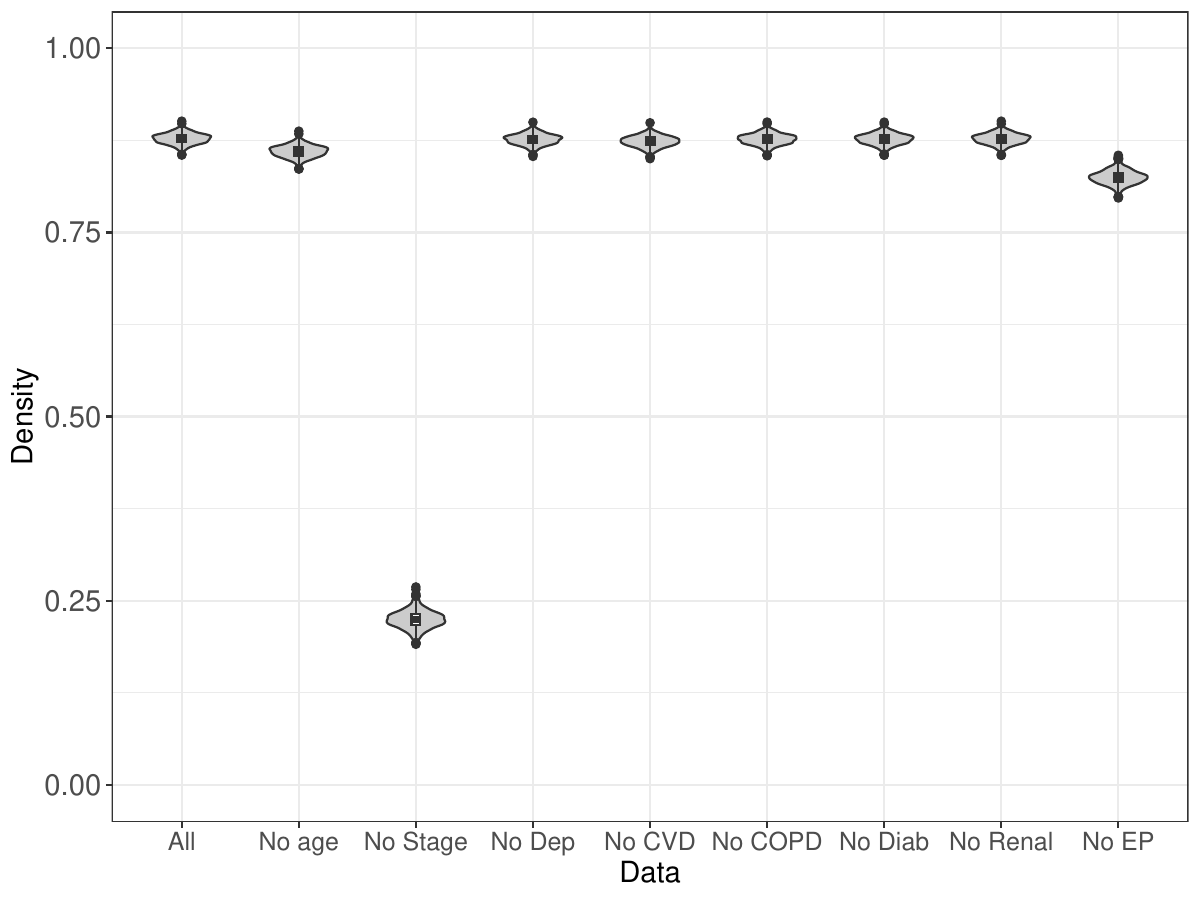}\\
 (a) & (b) & (c) \\
 \includegraphics[width=0.3\textwidth]{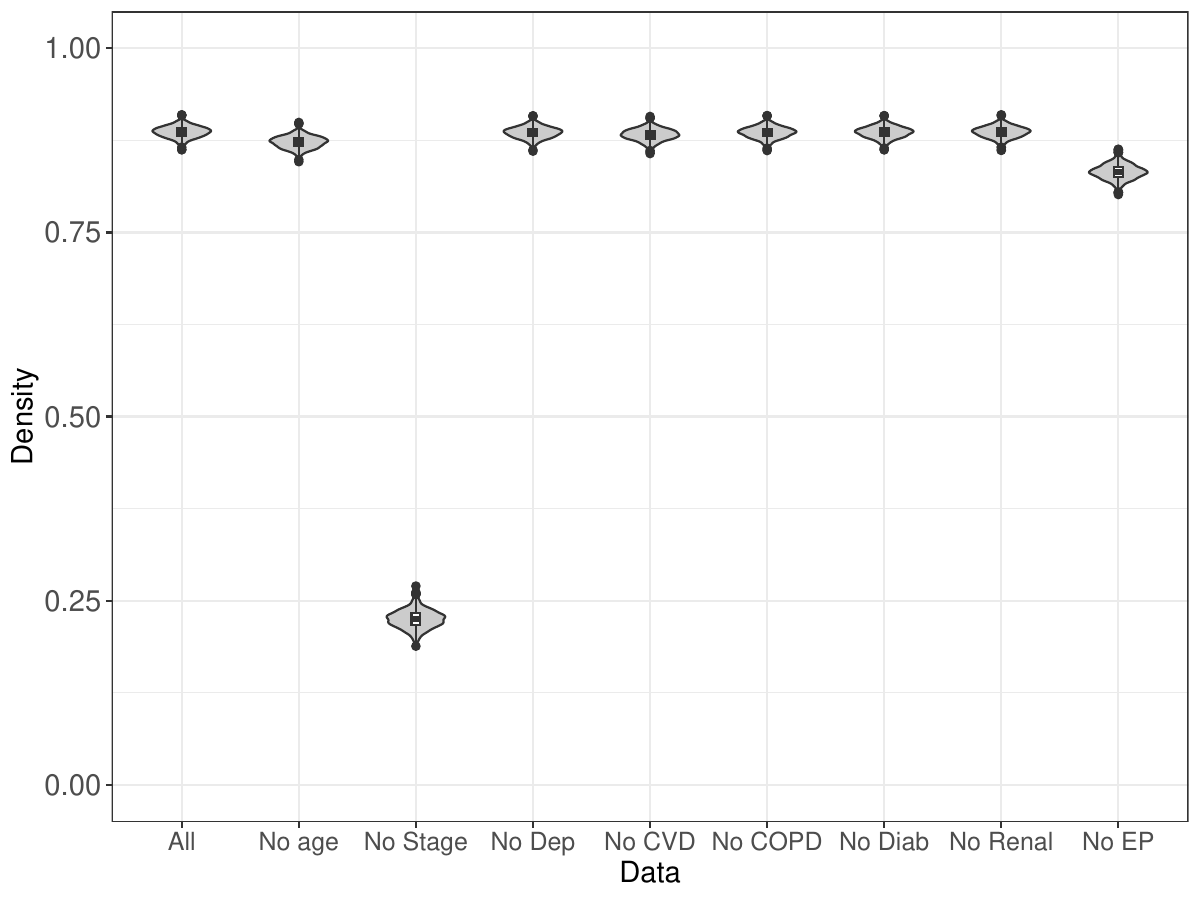} & 
\includegraphics[width=0.3\textwidth]{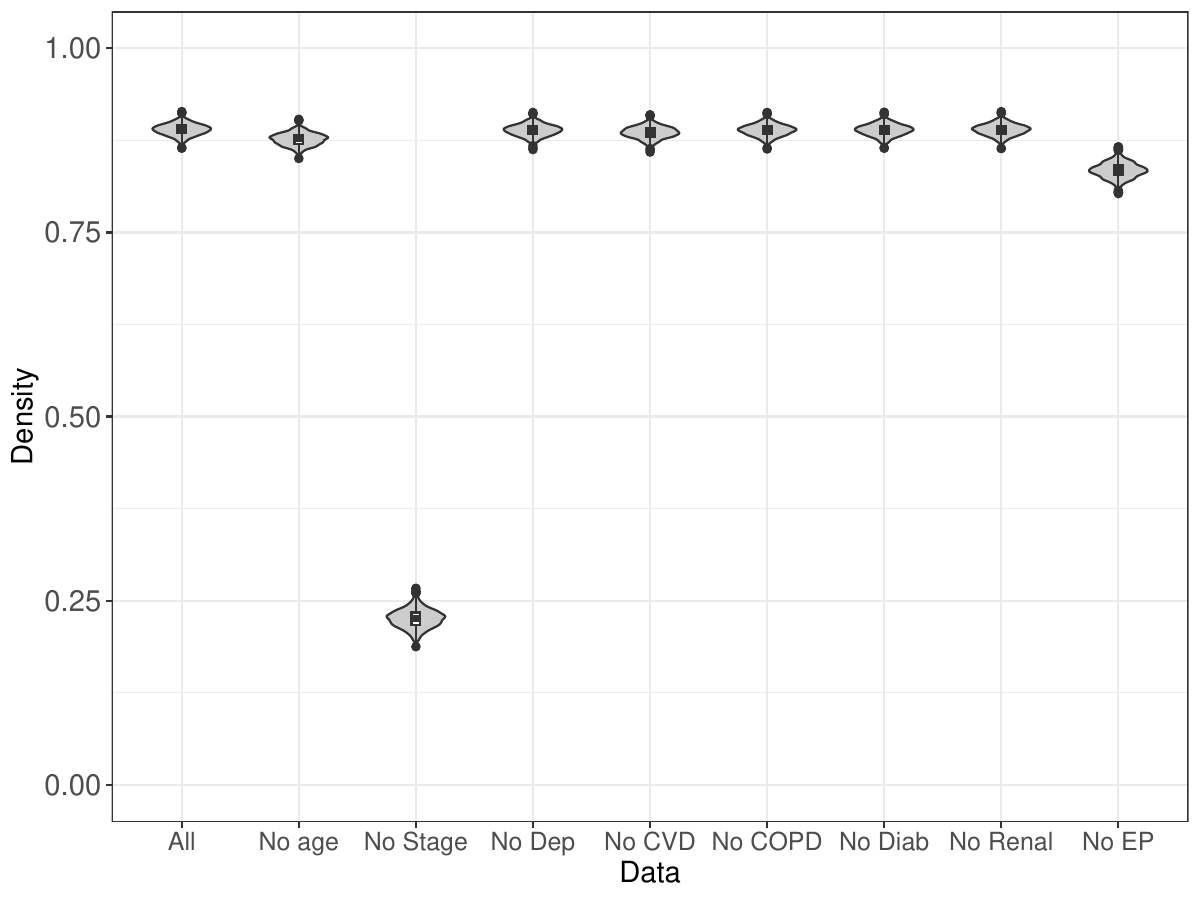} & 
\\
 (d) & (e) & 
 \end{tabular}
\caption{ Females colon cancer data. Variable importance ($R^2$) distributions at 2, 20, 40, 60, and 80 months.}
\label{fig:vimp_females}
\end{figure}

\begin{figure}[h!]
\centering
\begin{tabular}{c c c}
\includegraphics[width=0.3\textwidth]{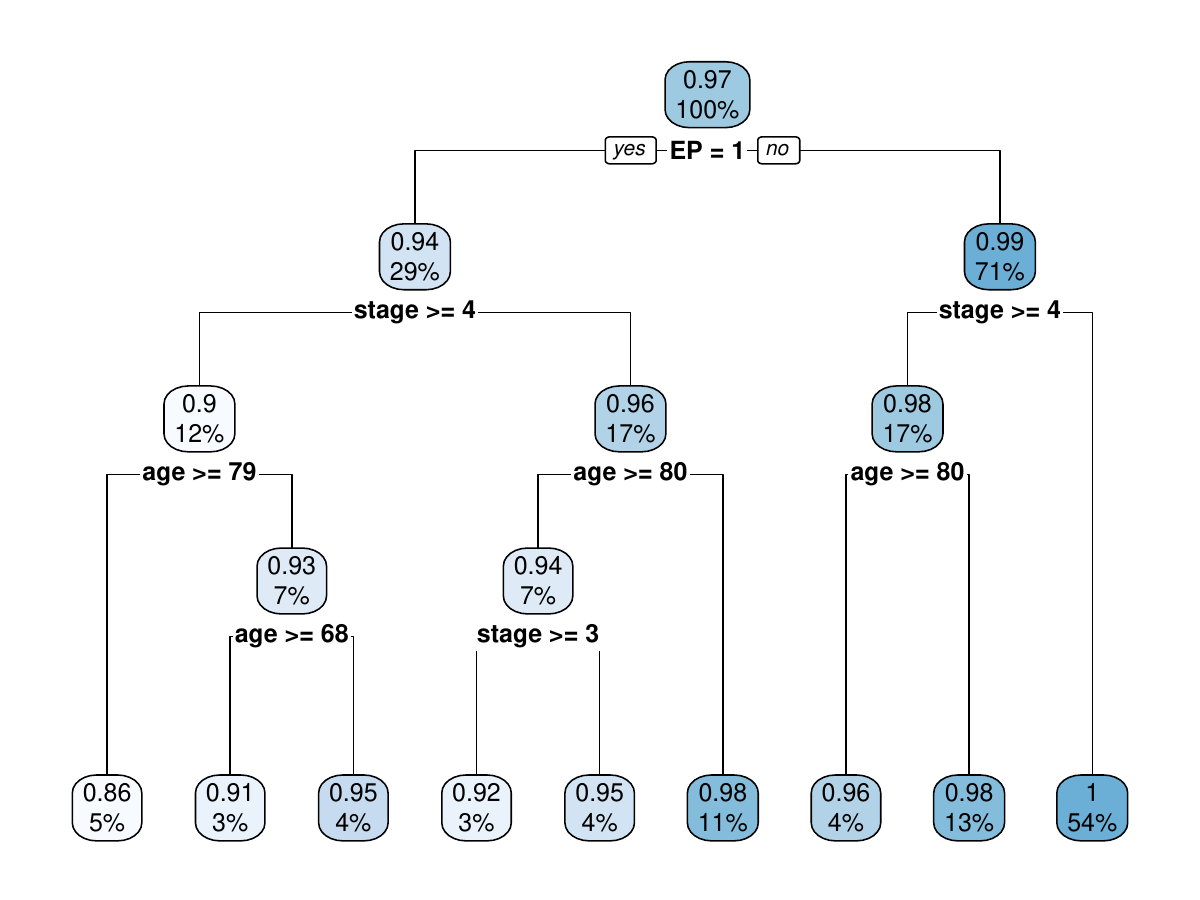} & 
\includegraphics[width=0.3\textwidth]{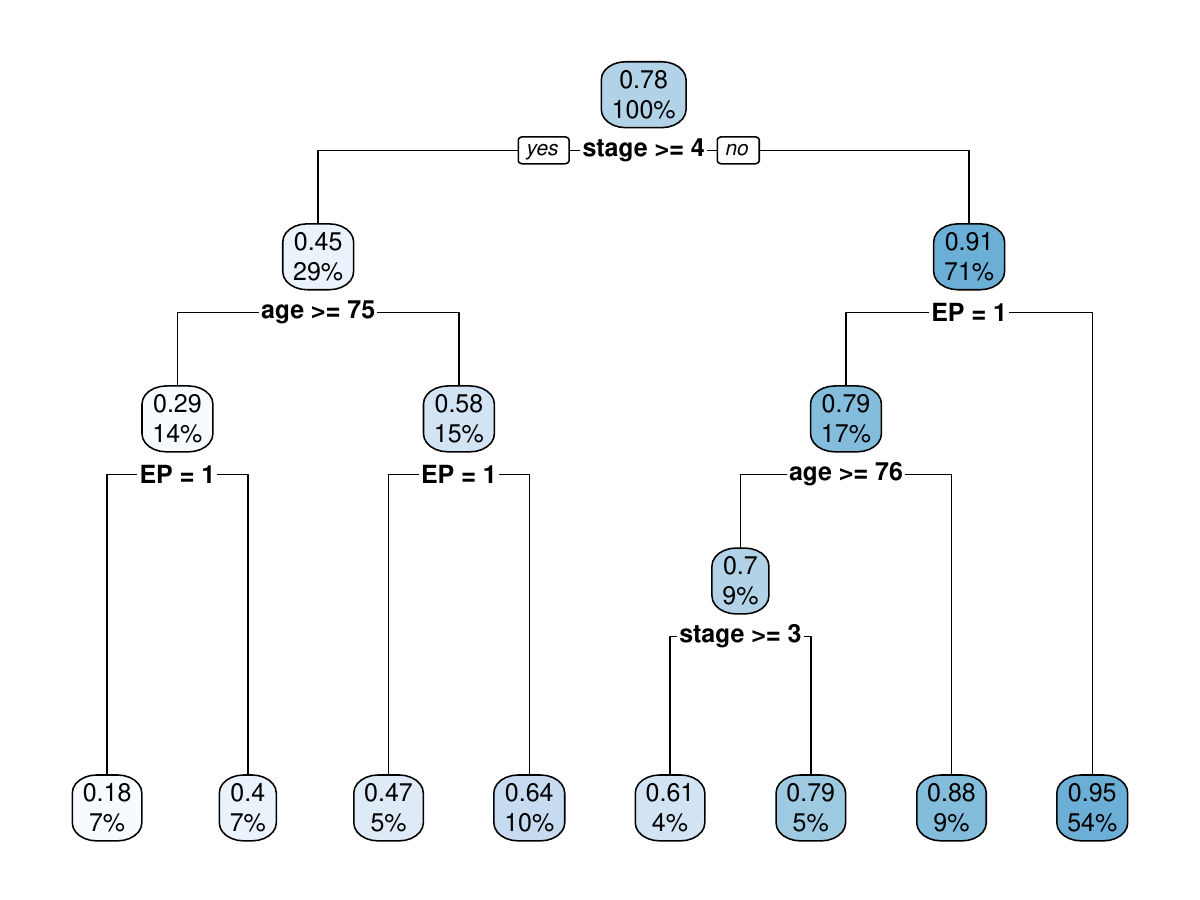} & 
\includegraphics[width=0.3\textwidth]{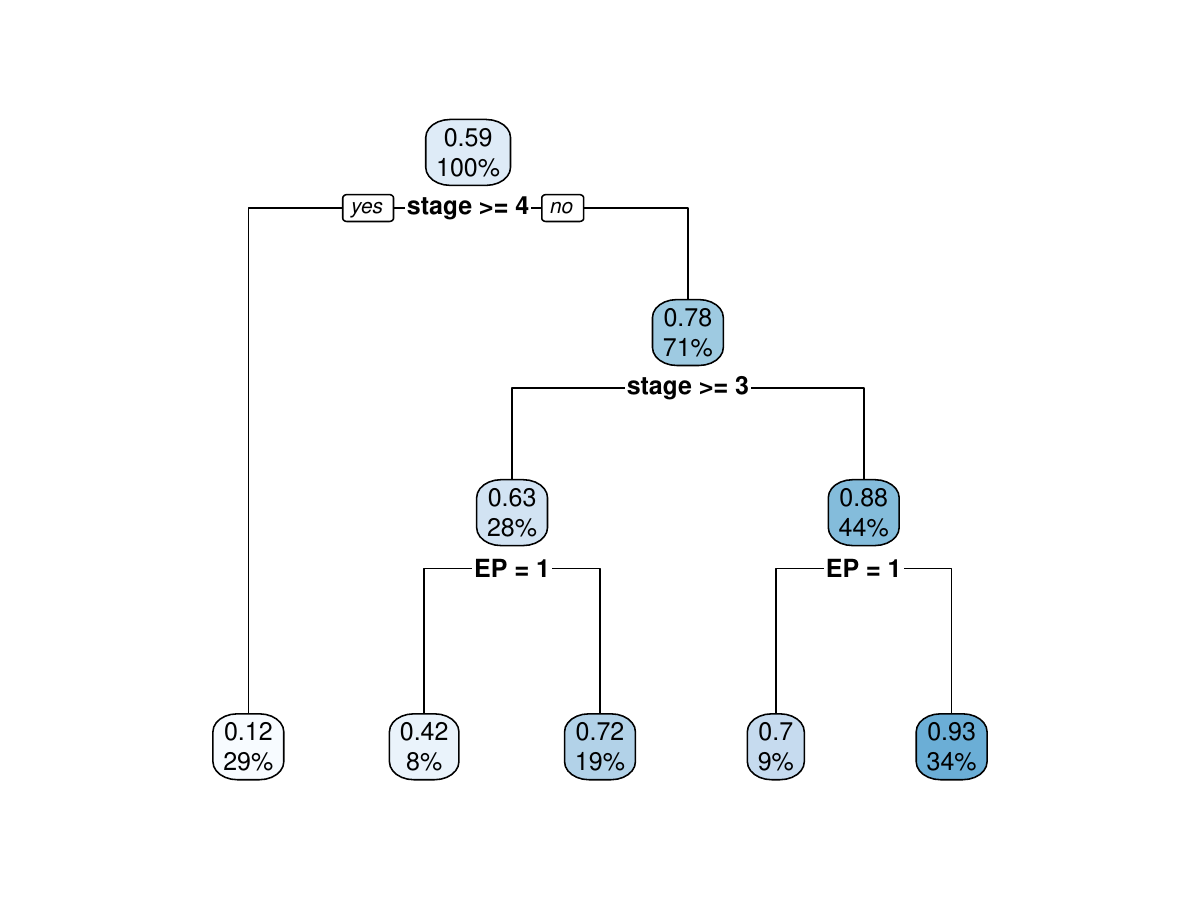}\\
 (a) & (b) & (c) \\
 \includegraphics[width=0.3\textwidth]{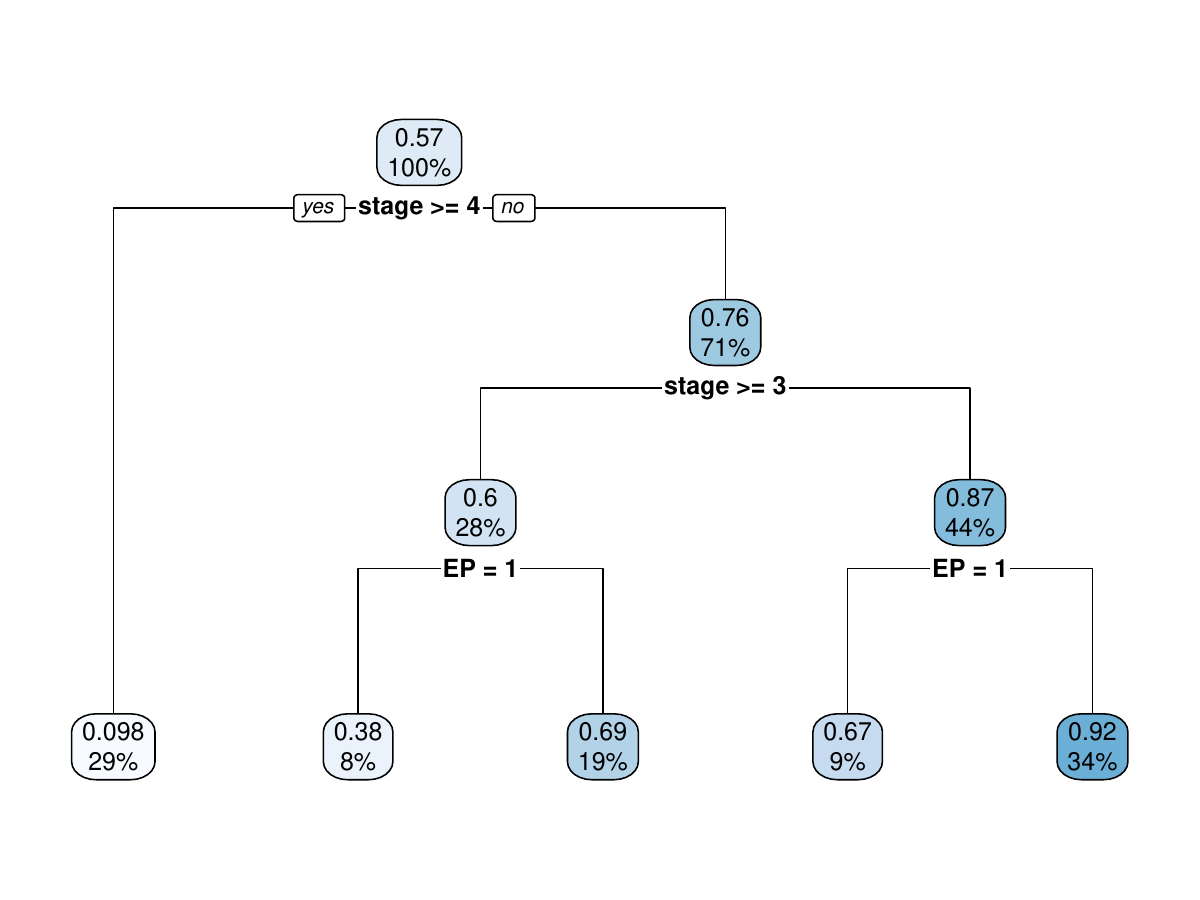} & 
\includegraphics[width=0.3\textwidth]{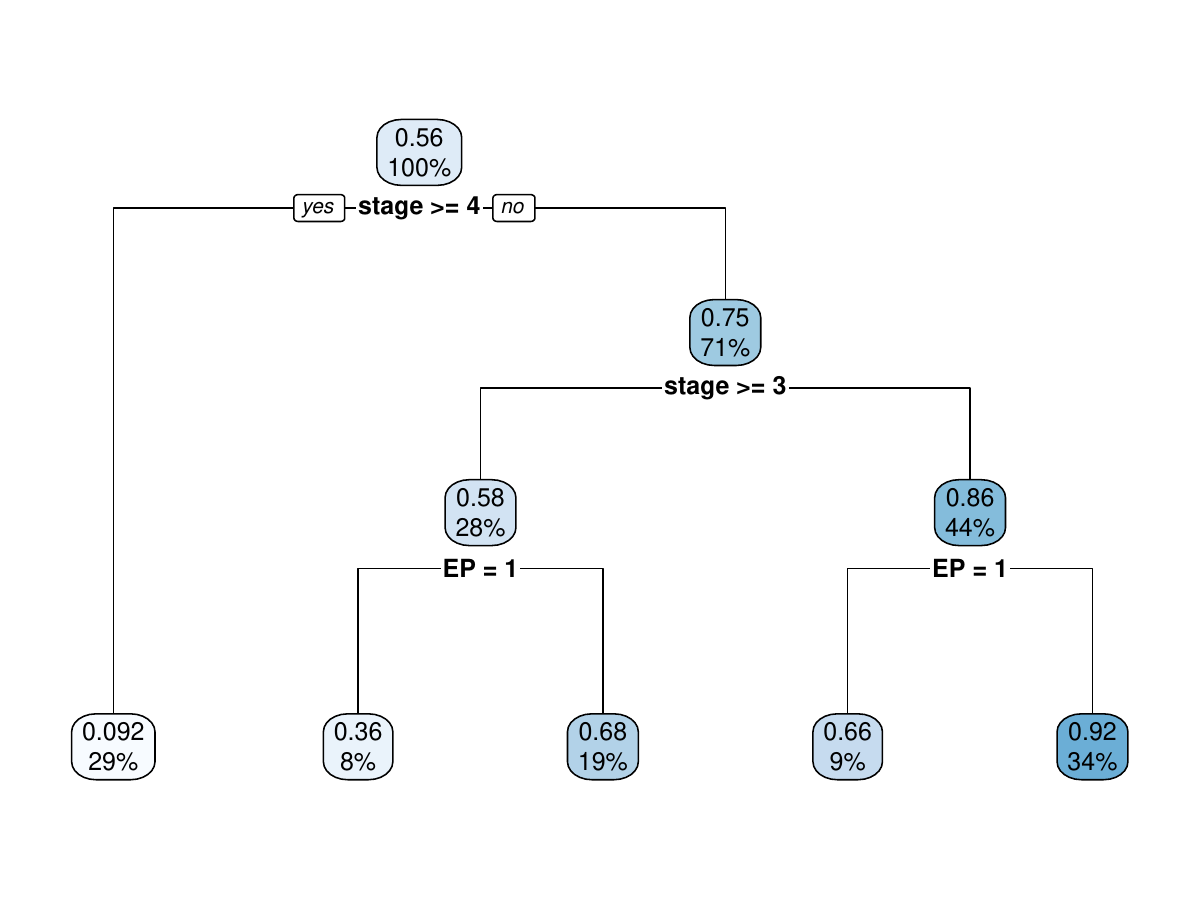} & 
\\
 (d) & (e) & 
 \end{tabular}
\caption{ Females colon cancer data. Variable importance trees (net survival) at 2, 20, 40, 60, and 80 months.}
\label{fig:tree_females}
\end{figure}

The variable importance, measured through $R^2$ based on net survival,  in presented in Figure \ref{fig:vimp_females}. At 2 months after diagnosis, the full model explains approximately $75\%$ of the variability in net survival, with the largest contributions coming from the EP status variable, followed by stage and age at diagnosis. Deprivation and comorbidities do not hold much variable importance in addition to the three aforementioned variables. After the acute diagnostic phase, the full-model's importance increases. The importance of stage at diagnosis is increased, and while emergency presentation and age at diagnosis remain important prognostic factors, they are not as discriminant as they are shortly after diagnosis.
A decision tree for the variable importance on explaining net survival is shown in Figure \ref{fig:tree_females}. At 2 months post-diagnosis, EP status is the first variable to divide the cohort, with 29\% of patients having received an emergency diagnosis. The next division is based on stage, distinguishing 12\% of late-stage (IV, metastatic) from 17\% of early-stage (I–III) patients with $\text{EP}=1$. For patients with $\text{EP} = 1$, age at diagnosis further differentiates the late-stage group showing some variation in net survival (from 86\% to 95\%). Earlier stages are further divided by age (at 80 years), with the older group subdivided by stages III versus I-II, exhibiting non-negligible differences in net survival.
In contrast, patients with $\text{EP} = 0$ are divided by stage (IV vs. I--III), with late stage further divided by two age groups at 80 years. In this branch net survival probabilities remain close to 1. At later intervals (40, 60, and 80 months post-diagnosis), the decision trees simplify, with only metastatic disease and EP status strongly influencing prognosis. Overall, we obtain similar conclusions for female colon cancer patients than for male patients, which provide further evidence about the usefulness of EP as a proxy for healthcare access, all through follow up time.

\renewcommand{\arraystretch}{1.1}
\begin{table}[ht!]
    \centering
    \resizebox{\textwidth}{!}{
    \begin{tabular}{lllllllll}
    \hline
        & \multicolumn{7}{c}{\textbf{Stage at diagnosis}}    \\ \hline
        & \multicolumn{2}{c}{\textbf{I}}  & \multicolumn{2}{c}{\textbf{II}} & \multicolumn{2}{c}{\textbf{III}} & \multicolumn{2}{c}{\textbf{IV}} \\ \hline
         & N  & \% &  N  & \% &  N  & \% &  N  & \% \\
        \textbf{N} &   1,114 & & 2,616 & & 2,369 & & 2,451 
 &  \\
        \textbf{Mean age (sd)} &   69.4 & (13.6) & 73.5 & (12.0) & 72.1 & (12.5) & 71.6 & (13.8) \\ 
        \textbf{Emergency presentation} &  121 & \textit{10.9} & 686 & \textit{26.2} & 712 & \textit{30.1} & 1,035 & \textit{42.2} \\ 
        \textbf{Comorbidities} &  &  &  &  &  &  &  &  \\ 
        Cardiovascular disease &  73 & \textit{6.6} & 245 & \textit{9.4} & 194 & \textit{8.2} & 211 & \textit{8.6} \\ 
        COPD &   104 & \textit{9.3} & 270 & \textit{10.3 }& 208 & \textit{8.8} & 256 & \textit{10.4} \\ 
        Diabetes &   73 & \textit{6.6} & 240 & \textit{9.2} & 212 & \textit{8.9} & 218 & \textit{8.9} \\ 
        Renal disease &   25 & \textit{2.2} & 85 & \textit{3.2} & 93 & \textit{3.9} & 100 & \textit{4.1} \\ 
        \textbf{Deprivation (in quintiles)} &  &  &  &  &  &  &  &  \\ 
        Least deprived &   259 & \textit{23.2} & 528 & \textit{20.2} & 517 & \textit{21.8} & 485 & \textit{19.8} \\ 
        2 &   244 & \textit{21.9} & 583 & \textit{22.3} & 529 & \textit{22.3 }& 540 & \textit{22.0} \\ 
        3 &   221 & \textit{19.8} & 542 & \textit{20.7} & 489 & \textit{20.6} & 508 & \textit{20.7} \\ 
        4 &   201 & \textit{18.0} & 513 & \textit{19.6} & 441 & \textit{18.6} & 476 & \textit{19.4} \\ 
        Most deprived &   189 & \textit{17.0} & 450 & \textit{17.2} & 393 & \textit{16.6} & 442 & \textit{18.0} \\ 
        \textbf{Number of deaths} &   257 & \textit{23.1} & 933 &\textit{ 35.7} & 1,250 & \textit{52.8} & 2,243 & \textit{91.5} \\ 
        \textbf{Mean follow-up time, years (sd)} &   5.8 & (1.8) & 5.2 & (2.3) & 4.2 & (2.6) & 1.4 & (1.9)  \\ \hline
    \end{tabular}
    }
    \caption{Characteristics of females with colon cancer}
    \label{table:table1_women}
\end{table}

\pagebreak

\begin{figure}[h!]
\centering
\begin{tabular}{c c c}
\includegraphics[width=0.3\textwidth]{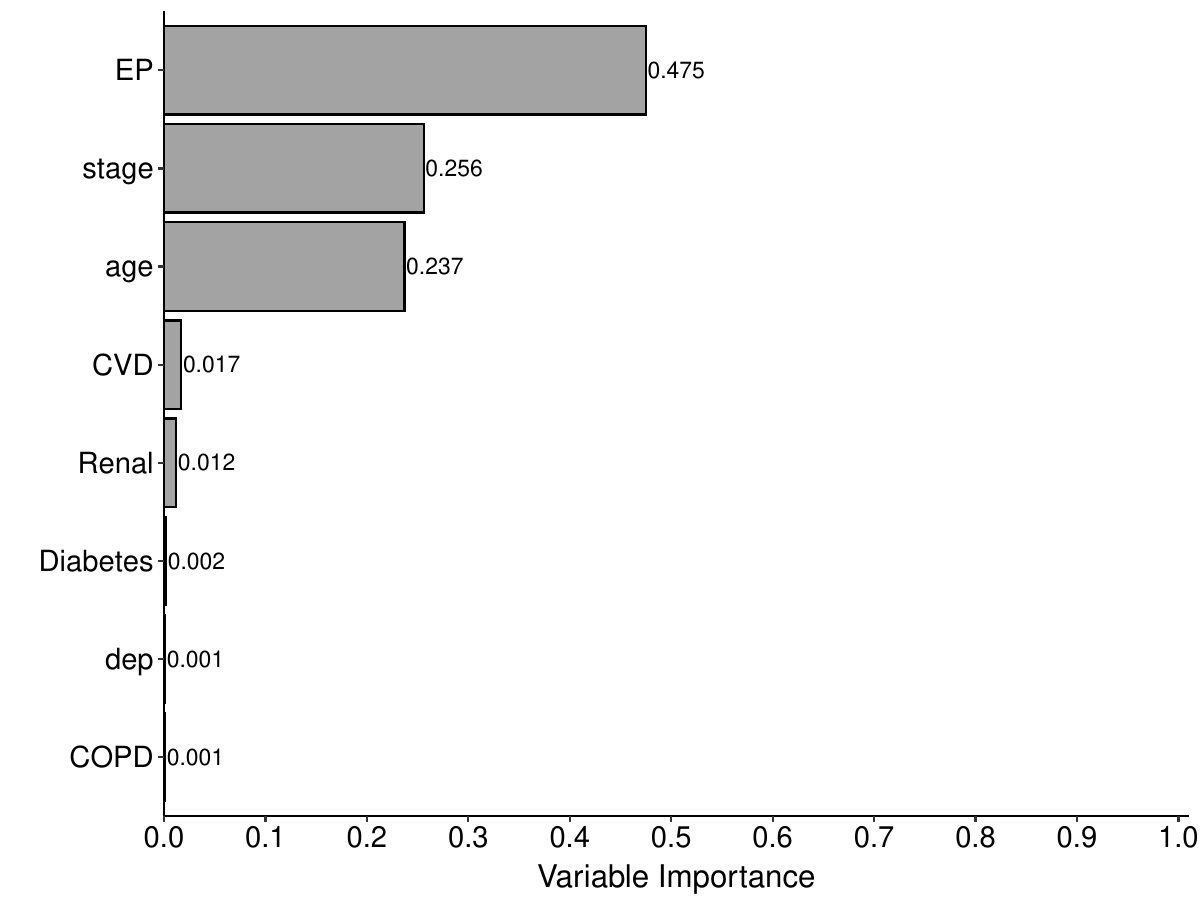} & 
\includegraphics[width=0.3\textwidth]{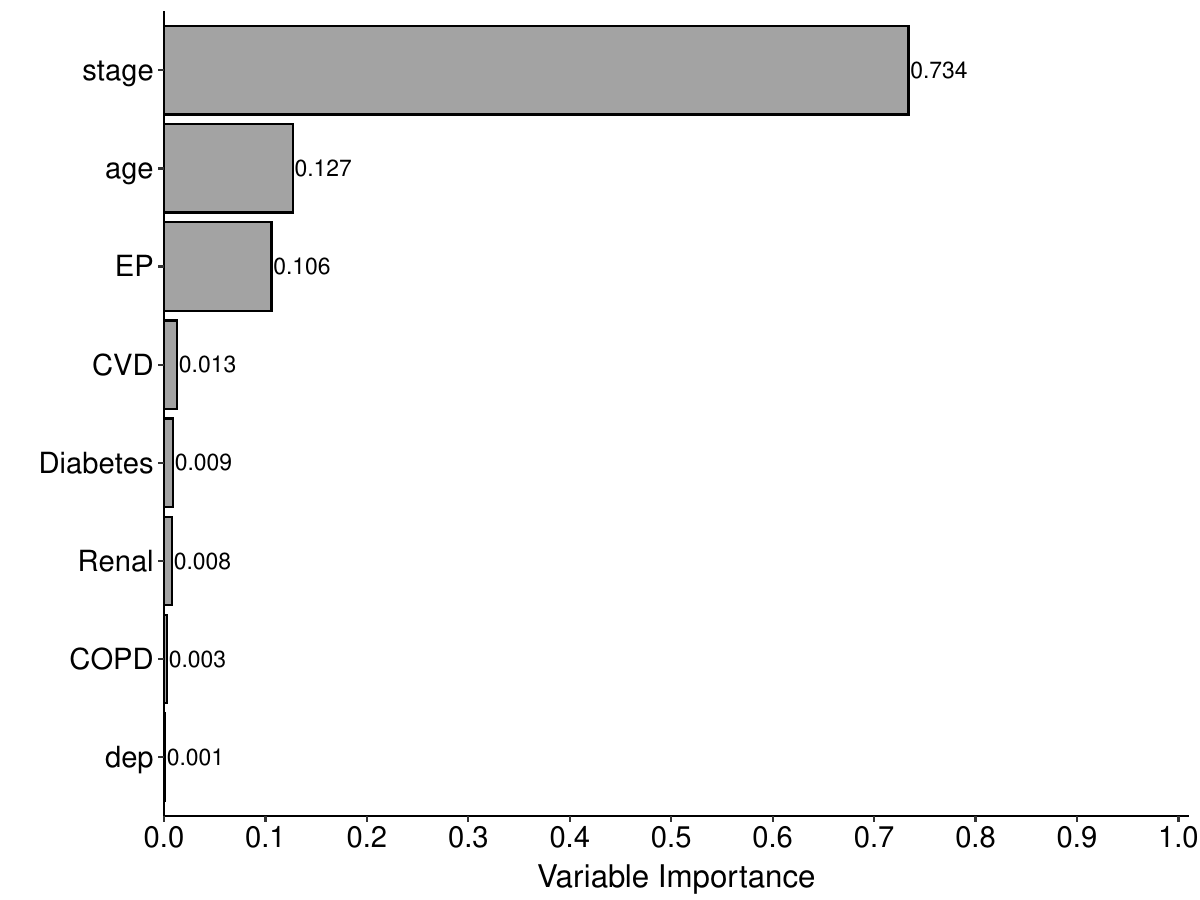} & 
\includegraphics[width=0.3\textwidth]{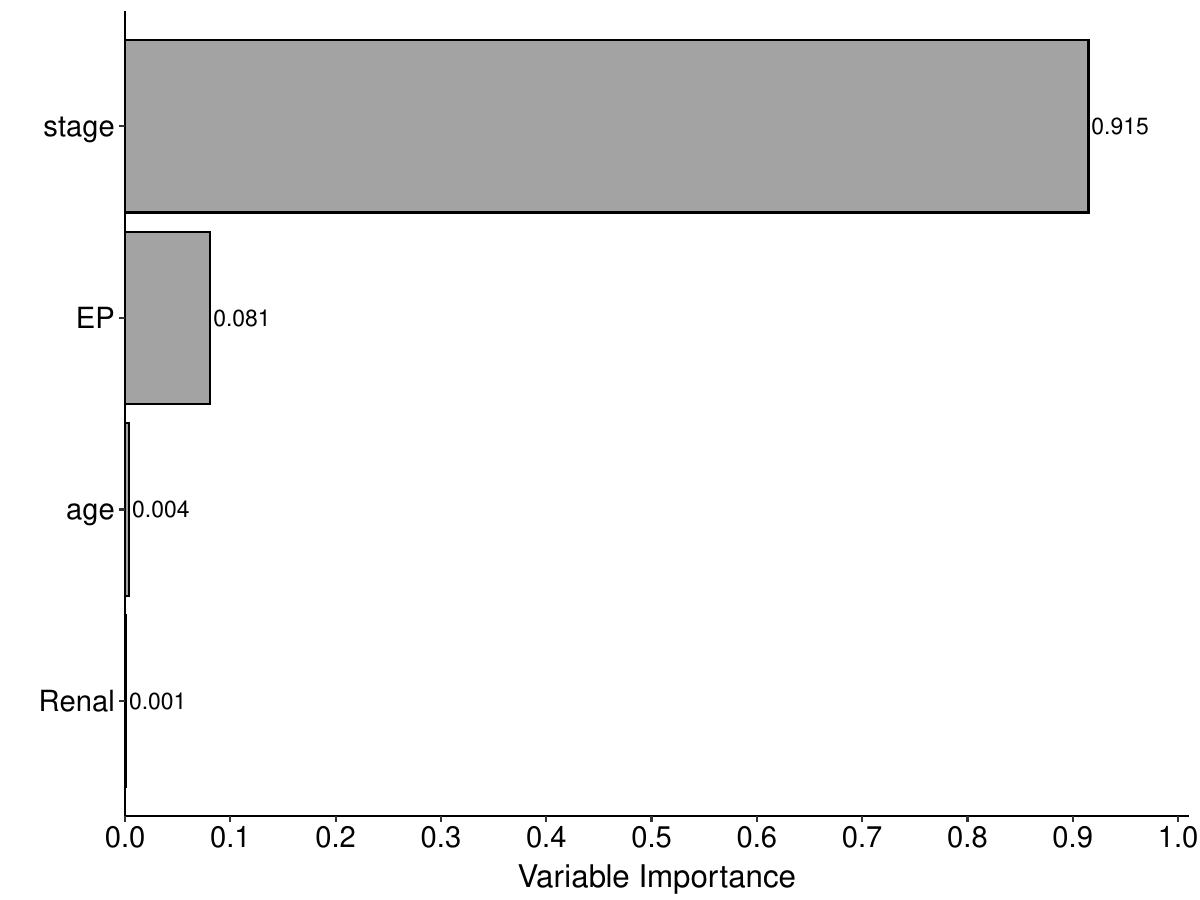}\\
 (a) & (b) & (c) \\
 \includegraphics[width=0.3\textwidth]{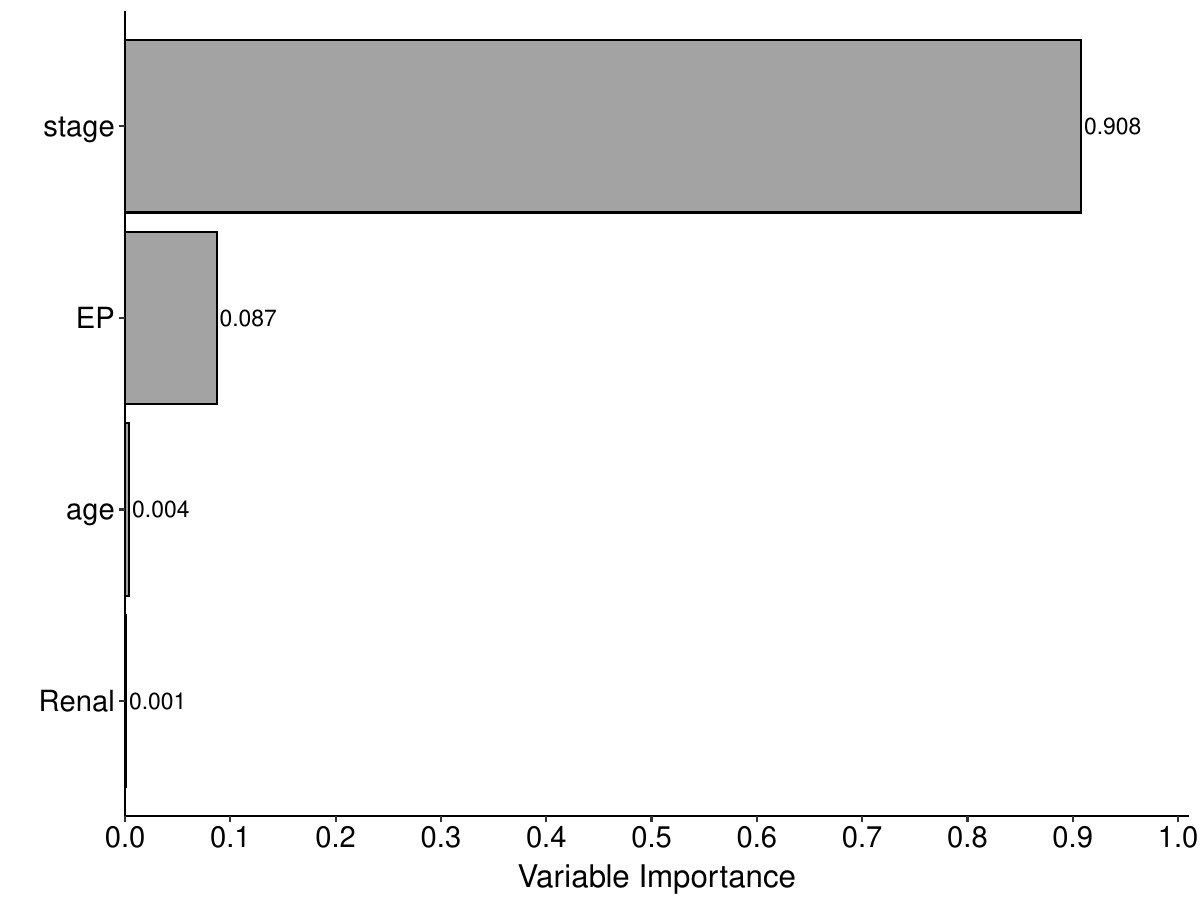} & 
\includegraphics[width=0.3\textwidth]{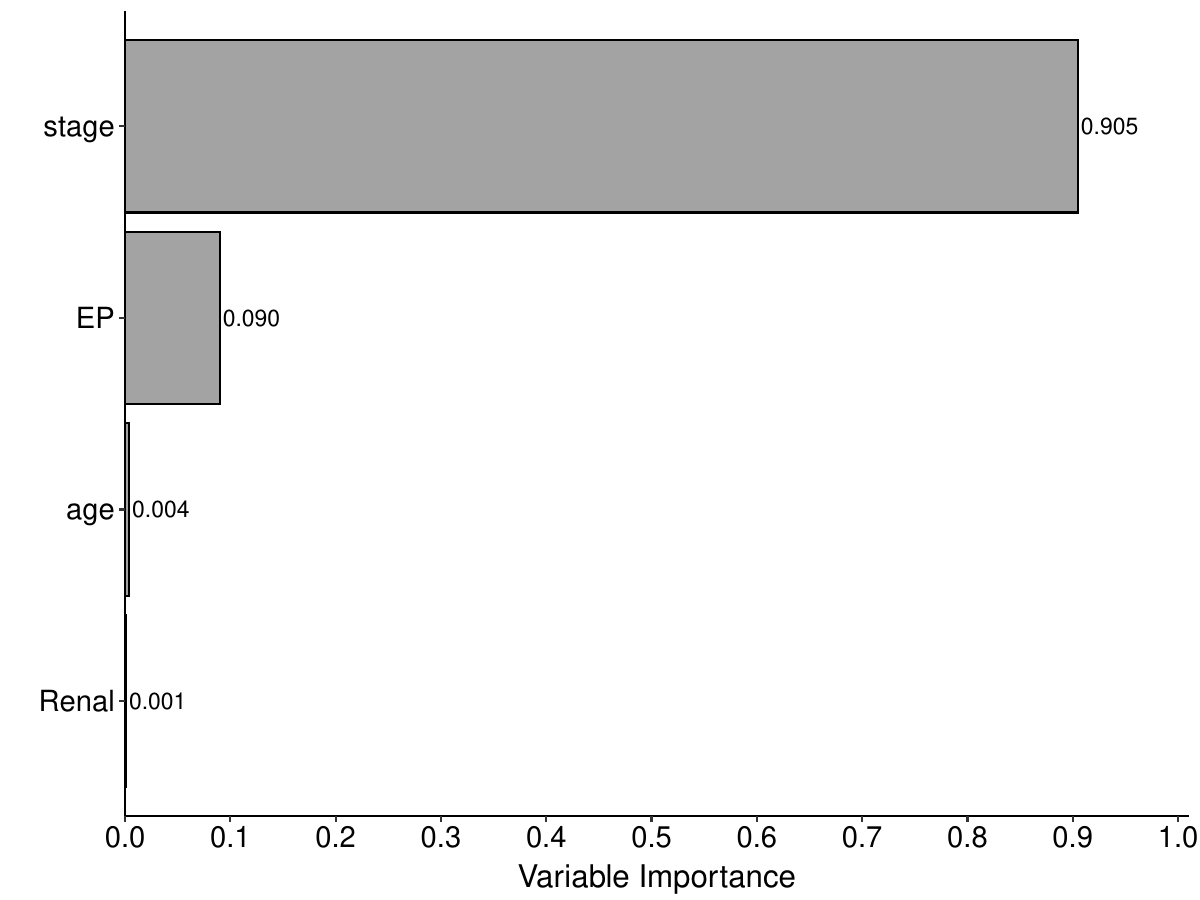} & 
\\
 (d) & (e) & 
 \end{tabular}
\caption{ Females colon cancer data. Variable importance from \texttt{rpart} at 2, 20, 40, 60, and 80 months.}
\label{fig:rpartvimp_females}
\end{figure}

%%%%%%%%%%%%%%%%%%%%%%%%%%%%%%%%%%%%%%%%%%%%%%%%%%%%%%%%%%%%%%%%%%%%%%%%%%%%%%%%
%%%%%%%%%%%%%%%%%%%%%%%%%%%%%%%%%%%%%%%%%%%%%%%%%%%%%%%%%%%%%%%%%%%%%%%%%%%%%%%%

\end{document}